\documentclass[a4paper,12pt]{article}
\pdfoutput=1

\usepackage{color}
\usepackage{multirow}
\usepackage{longtable}
\usepackage{xspace}
\usepackage{listings}
\usepackage{changepage}
\usepackage{amsmath,amssymb}
\usepackage{slashed}
\usepackage{xcolor,colortbl}
\usepackage{hyperref}

\RequirePackage[T1]{fontenc}
\RequirePackage{graphicx}
\RequirePackage{mathptmx}      
\RequirePackage[numbers,sort&compress]{natbib}

\hypersetup{colorlinks,bookmarksopen,bookmarksnumbered,linkcolor=blus,pdfstartview=FitH,urlcolor=blus,citecolor=verde}

\definecolor{tit}{rgb}{0.1,0.2,0.4}
\definecolor{blus}{cmyk}{1,1,0,0.6}
\definecolor{verde}{cmyk}{0.92,0,0.59,0.25}

\newcommand{\Dlr}{\overset\leftrightarrow{D}}
\newcommand{\DlrImu}{\Dlr_\mu \hspace*{-0.16cm}{}^I}

\newcommand{\vp}{\varphi}
\newcommand{\gc}{g^2}
\newcommand{\gpc}{g^{\prime 2}}
\newcommand{\gsc}{g_s^2}
\newcommand{\hc}{\mathrm{h.c.}}
\newcommand{\cc}{\mathrm{c.c.}}
\newcommand{\Tr}{\mathrm{Tr}}
\newcommand{\nn}{\nonumber}

\newcommand{\myv}[1]{{{\rule{0cm}{0.35cm}#1}}}
\newcommand{ \mysmall}[1]{\scriptscriptstyle #1} 

\newenvironment{myroutine}{
\vspace*{0.2cm}
\begin{adjustwidth}{0.5cm}{}
\setlength{\parindent}{0pt}}
{\end{adjustwidth}
\vspace*{0.5cm}}

\definecolor{shaded}{RGB}{245,245,245}

\lstdefinestyle{mathematica}{
  basicstyle=\ttfamily\mdseries,
  backgroundcolor=\color{shaded},
  language=Mathematica,
  frame=false	
}

\lstdefinestyle{WCsInput}{
  basicstyle=\ttfamily\mdseries,
  language=bash,
  frame=single,
  title=\hspace{14cm}{\tt WCsInput.dat}	
}

\lstdefinestyle{SMInput}{
  basicstyle=\ttfamily\mdseries,
  language=bash,
  frame=single,
  title=\hspace{14cm}{\tt SMInput.dat}	
}

\lstdefinestyle{Options}{
  basicstyle=\ttfamily\mdseries,
  language=bash,
  frame=single,
  title=\hspace{14cm}{\tt Options.dat}	
}

\definecolor{Gray}{gray}{0.95}
\definecolor{RGray}{gray}{0.85}
\definecolor{CGray}{gray}{0.92}
\definecolor{dgray}{gray}{0.4}

\definecolor{color1}{rgb}{0.9,.4,.2}
\definecolor{color2}{rgb}{0.3,.6,.7}
\definecolor{color3}{rgb}{0.7,.2,.7}

\newcommand{\pkg}[1]{{\tt #1}\xspace}
\newcommand{\dsix}{\pkg{DsixTools}}
\newcommand{\dsixbf}{\pkg{\bf DsixTools}}
\newcommand{\SMEFTrunner}{\pkg{SMEFTrunner}}
\newcommand{\SMEFTrunnerbf}{\pkg{\bf SMEFTrunner}}
\newcommand{\WETrunner}{\pkg{WETrunner}}
\newcommand{\WETrunnerbf}{\pkg{\bf WETrunner}}
\newcommand{\EWmatcher}{\pkg{EWmatcher}}
\newcommand{\EWmatcherbf}{\pkg{\bf EWmatcher}}
\newcommand{\mathe}{\pkg{Mathematica}}
\newcommand{\real}[1]{{\color{red} #1}}

\allowdisplaybreaks

\begin{document}

\vspace*{-1cm}

\hspace{-4mm}
\begin{minipage}{16cm}

\begin{flushright}
{\small
LMU-ASC 24/17\\
IFIC/17-18
}
\end{flushright}

\vspace{-1.3cm}
\noindent\includegraphics[width=4cm]{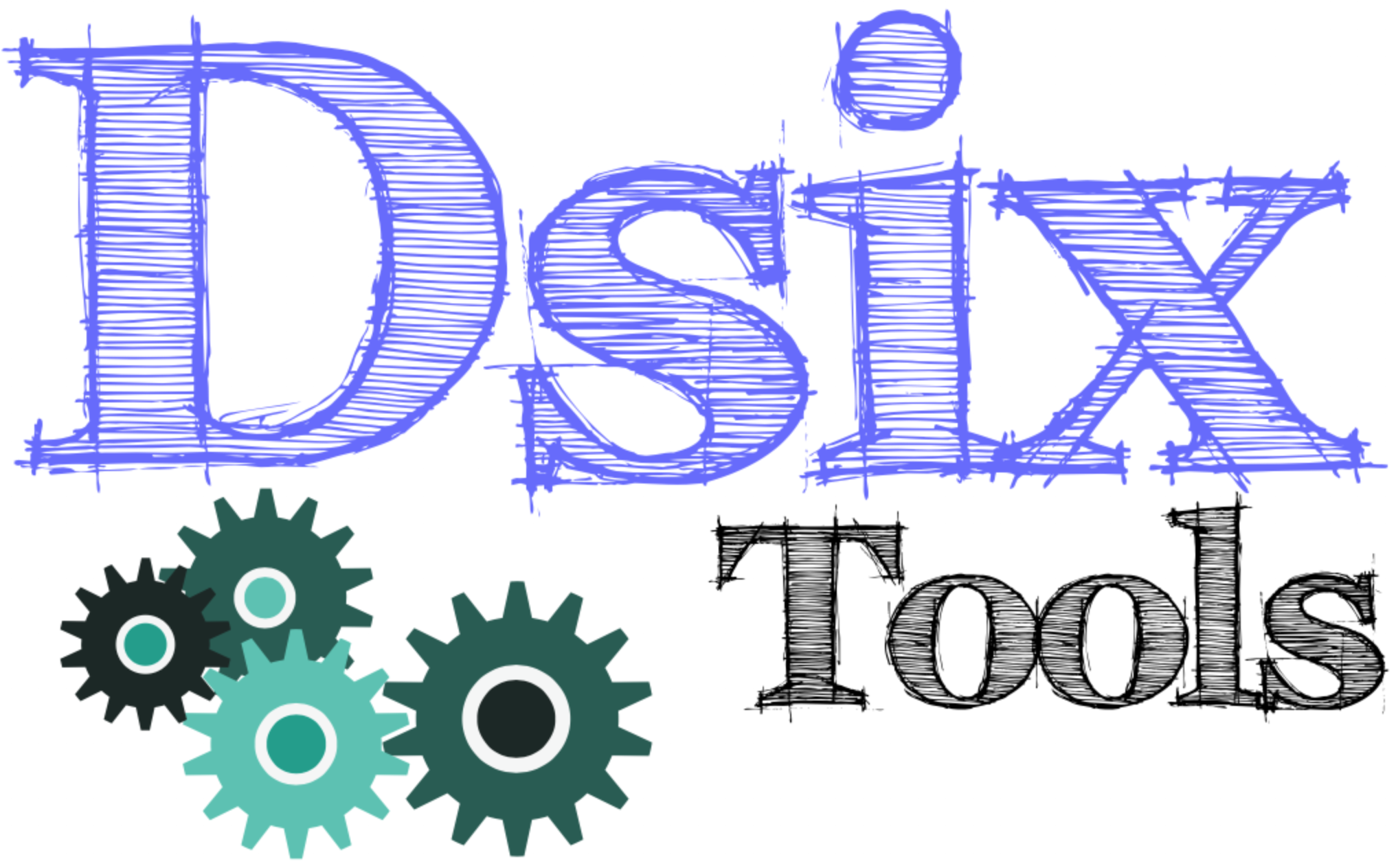}

\vspace{2cm}

\hrule
\vspace{5mm}
\begin{flushleft}
{\bf 
\huge{\color{dgray} DsixTools}\\[3mm]
\vspace*{0.3cm}
\LARGE{The Standard Model Effective Field Theory Toolkit}}\\[6mm]
\hrule

\vspace{1cm}

{\fontsize{14.3}{60}\selectfont Alejandro Celis$^{\ast,}$\footnote{Email:~Alejandro.Celis@physik.uni-muenchen.de}, Javier Fuentes-Mart\'in$^{\dagger,}$\footnote{Email:~Javier.Fuentes@ific.uv.es}, Avelino Vicente$^{\dagger,}$\footnote{Email:~Avelino.Vicente@ific.uv.es}, Javier Virto$^{\ddag,}$\footnote{Email:~jvirto@mit.edu}}  
\\[7mm]
{\it $^\ast$ Ludwig-Maximilians-Universit\"at M\"unchen, 
   Fakult\"at f\"ur Physik,\\
   \hspace{2.1mm} Arnold Sommerfeld Center for Theoretical Physics, 
   80333 M\"unchen, Germany.}\\[3mm]
{\it $^\dagger$ Instituto de F\'{\i}sica Corpuscular, Universitat de Val\`encia - CSIC, E-46071 Val\`encia, Spain.}\\[3mm]
{\it $^\ddag$ Albert Einstein Center for Fundamental Physics, Institute for Theoretical Physics,\\
\hspace{1.5mm} University of Bern, CH-3012 Bern, Switzerland.}
\end{flushleft}

\vspace{1cm}

\noindent  {\large\bf\color{blus} Abstract}\\
We present \dsix, a \mathe package for the handling of the dimension-six
Standard Model Effective Field Theory. Among other features, \dsix
allows the user to perform the full one-loop Renormalization Group
Evolution of the Wilson coefficients in the Warsaw
basis. This is achieved thanks to the \SMEFTrunner module, which
implements the full one-loop anomalous dimension matrix previously
derived in the literature. In addition, \dsix also contains modules
devoted to the matching to the $\Delta B = \Delta S = 1,2$ and $\Delta
B = \Delta C = 1$ operators of the Weak Effective Theory at the
electroweak scale, and their QCD and QED Renormalization Group
Evolution below the electroweak scale.

\vspace{25pt}
\end{minipage}


\thispagestyle{empty}

\begin{quote}
{\large\noindent\color{blus} 
}

\numberwithin{equation}{section}

\end{quote}

\tableofcontents

\setcounter{footnote}{0}

\section{Introduction}
\label{sec:intro}

The experimental success of the Standard Model (SM) of particle
physics and the absence of new physics (NP) signals so far seem to
hint at the presence of a mass gap between the SM degrees of freedom
and the new dynamics.  In this case, departures from the SM at
energies much smaller than the new physics scale can be described
using effective field theory (EFT) methods.  The so-called Standard
Model EFT (SMEFT) parametrizes possible deviations from the SM caused
by heavy degrees of freedom in a model independent way.

The SMEFT Lagrangian is organized as an expansion in powers of
$1/\Lambda$, where $\Lambda$ represents the new physics scale and
compensates the canonical dimension of the effective operators.  The
leading order of the SMEFT corresponds to the renormalizable SM
Lagrangian.  Dominant new physics contributions to most processes of
interest are expected to be encoded in the Wilson Coefficients (WC) of effective operators of
canonical dimension six~\cite{Buchmuller:1985jz}.

In recent years, a non-redundant basis for the dimension-six SMEFT
operators was derived~\cite{Grzadkowski:2010es}. This basis is
currently known as the \textit{Warsaw basis}. The complete one-loop
anomalous dimension matrix (ADM) for the dimension-six operators has
been calculated very recently in this
basis~\cite{Jenkins:2013zja,Jenkins:2013wua,Alonso:2013hga,Alonso:2014zka}. 
These advances, together with simultaneous theoretical developments
occurring in the field (such as in the matching of specific models to
the SMEFT at one loop~\cite{Henning:2014wua,Drozd:2015rsp,delAguila:2016zcb,Boggia:2016asg,Henning:2016lyp,Ellis:2016enq,Fuentes-Martin:2016uol,Zhang:2016pja} and in the automatization of calculations within the SMEFT~\cite{Dedes:2017zog}), 
pave the way to the systematic use of EFT methods in the
analysis of new physics models. The power of the SMEFT approach is
that it allows to relate physics at disparate energy scales, in our
case properties of the high-energy dynamics at the scale~$\Lambda$,
with measurements that take place at low energies, while performing an
expansion in $1/\Lambda$ that allows to keep leading new physics
effects in a consistent manner. Here we present \dsix, a
\mathe\footnote{\mathe~is a product from Wolfram Research,
  Inc.~\cite{wolfram}.}~package that aims to facilitate such
enterprise.

Given some initial conditions for the Warsaw basis at the high-energy
scale $\Lambda$, obtained from the matching of a UV model to the
SMEFT, \dsix~allows the user to perform the full one-loop
Renormalization Group Evolution (RGE) down to the electroweak
scale. Furthermore, when the physics of interest lies well below the
electroweak scale, it is useful to perform a matching of the dimension-six
basis to a low-energy EFT in the broken electroweak phase. In the
so-called Weak Effective Theory (WET), the SM heavy degrees of freedom
(top quark, Higgs, $W^{\pm}$ and $Z$) have been integrated
out. \dsix~implements the tree-level matching of the Warsaw basis to
$\Delta B = \Delta S = 1,2$ and $\Delta B = \Delta C = 1$ operators of
the WET based on the results obtained
in~\cite{Aebischer:2015fzz}. Last but not least, the relevant effects
due to QCD and QED running from the electroweak scale down to the
$b$-quark mass scale in the WET is also implemented in \dsix, using the
results in~\cite{Aebischer:2017gaw}.
  
\dsix~is structured into different modules, each of them taking care
of a specific task. Their functionality is described in
Sec.~\ref{sec:nutshell}. Sec.~\ref{sec:down} explains how to download
and load \dsix whereas a detailed review of \dsix and its modules is
given in Sec.~\ref{sec:use}. A summary is given in
Sec.~\ref{mysum}. Finally, this manual includes several appendices to
document all the features present in \dsix: \ref{sec:SMEFT} contains
details of the conventions used for the implementation of the SMEFT,
\ref{ap:rges} contains the SMEFT RGEs, \ref{sec:WET} describes
relevant aspects of the WET, \ref{ap:routines} contains a detailed
description of the \dsix routines, \ref{ap:parameters} is devoted to
compile the SMEFT and WET parameters list and \ref{ap:BVinMB} gives
results for the Baryon-number-violating operators in the fermion mass basis.

\begin{figure}
  \centering
  \includegraphics[width=9cm]{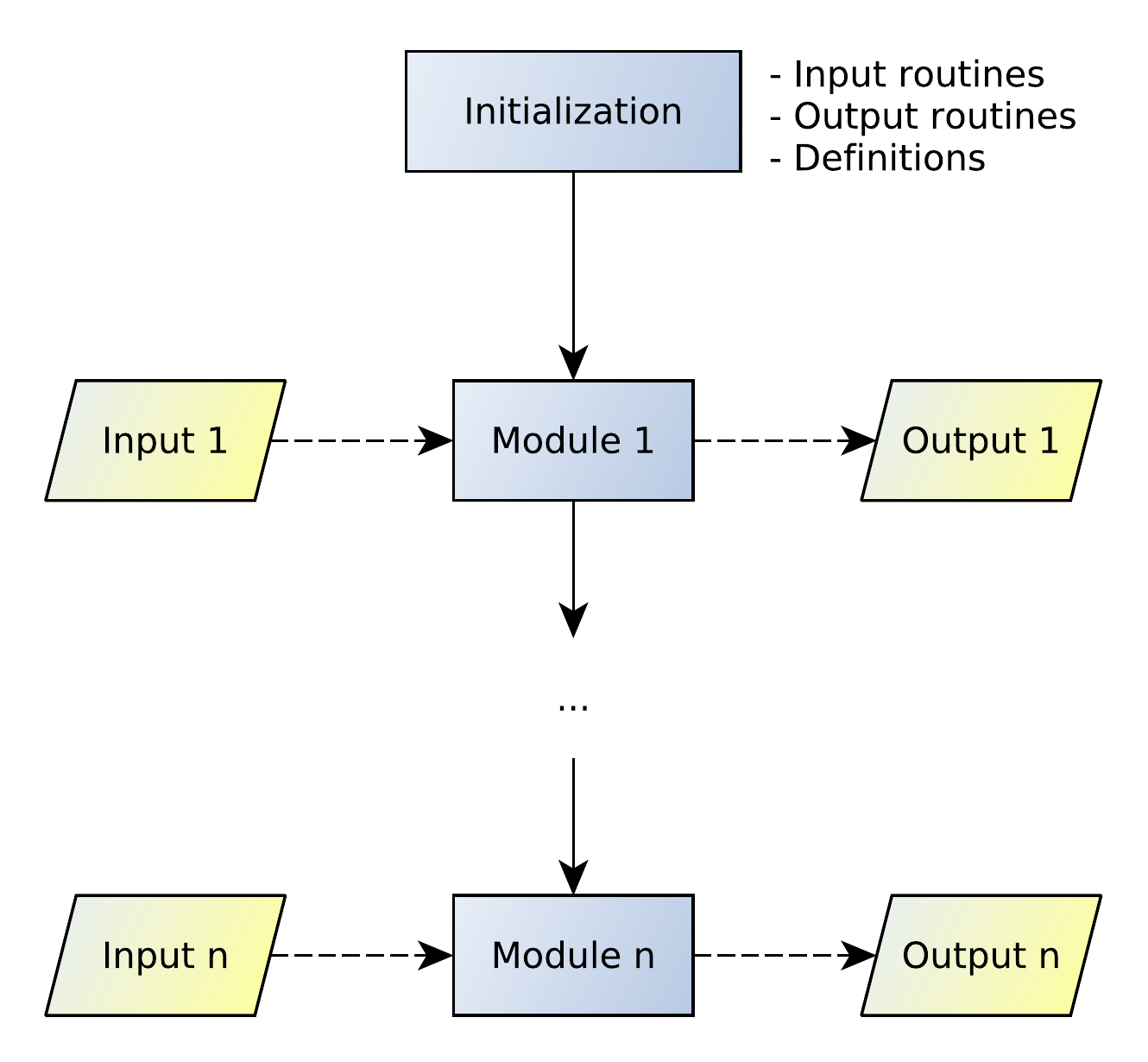}
  \caption{\small \dsix' global flowchart. The user can provide input
    to each module or use the results obtained with the previous one.}
  \label{fig:globalflow}
\end{figure}

\section{DsixTools in a nutshell}
\label{sec:nutshell}

\dsix is a modular package, with each module performing an independent
task related to dimension-six effective operators.  The different
modules can be easily communicated with each other, since one module's
output can be used as input for another module. In practice, this
means that a particular project with \dsix can involve all modules or
just a few, depending on the user's goals. Fig. \ref{fig:globalflow}
shows the global flowchart of \dsix.

The current version of \dsix contains three independent modules, called:
\SMEFTrunner, \EWmatcher and \WETrunner. The \SMEFTrunnerbf module
implements:

\begin{itemize}
\item The SM contribution to the one-loop RGEs of the SM
  parameters~\cite{Machacek:1983tz,Machacek:1983fi,Machacek:1984zw,Luo:2002ey}. \vspace{0.1cm}

\item The one-loop RGEs for the dimension-six operators in the Warsaw
  basis from Refs.~\cite{Jenkins:2013zja,Jenkins:2013wua,Alonso:2013hga}.\footnote{We
    have taken into account the errata published
    in~\href{http://einstein.ucsd.edu/smeft/}{http://einstein.ucsd.edu/smeft/}.}  \vspace{0.1cm}

\item The one-loop RGEs for the dimension-six Baryon-number-violating operators
  using the results in~\cite{Alonso:2014zka}. 
\end{itemize} 

\vspace{0.1cm}
\noindent The \EWmatcherbf module implements:
\begin{itemize}
\item The tree-level matching of the Warsaw basis to the $\Delta B =
  \Delta S = 1,2$ and $\Delta B = \Delta C = 1$ operators of the WET
  at the electroweak scale, using the results
  in~\cite{Aebischer:2015fzz}.
\end{itemize}

\vspace{0.1cm}
\noindent The \WETrunnerbf module implements:
\begin{itemize}
\item QCD and QED RGEs in the WET from the electroweak scale down to
  the low-energy scale $\Lambda_{\rm IR}$. The \WETrunner
  module fixes by default $\Lambda_{\rm IR} = m_b$, the $b$-quark mass scale, and
  uses the RGEs derived in~\cite{Aebischer:2017gaw}.
\end{itemize}

\begin{figure}
  \centering
  \includegraphics[width=9cm,height=6.5cm]{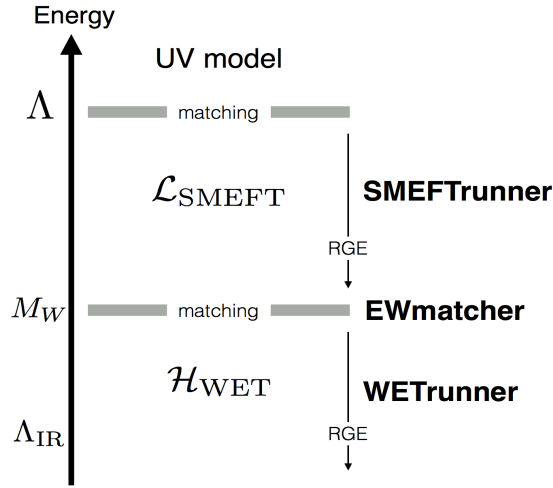}
  \caption{\small Functionality of \dsix~modules.}
  \label{fig:diagram}
\end{figure}

The functionality of the \dsix~modules is illustrated in
Fig.~\ref{fig:diagram}, where one can also see how they relate to the
different energy ranges and effective theories. Relevant details of
the SMEFT and WET implementations are given in \ref{sec:SMEFT} to
\ref{sec:WET}, where our conventions are also presented.

\section{Downloading and loading \dsixbf}
\label{sec:down}

\dsix~is free software under the copyleft of the \href{https://www.gnu.org/copyleft/gpl.html}{GNU General Public License}. It can be downloaded from the web page~\cite{dsixweb}:
\begin{center}
\href{https://dsixtools.github.io}{https://dsixtools.github.io}
\end{center}
After placing the \dsix~folder in the {\textit{Applications}} folder of the \mathe~base directory, the user can load \dsix~with the command
\begin{lstlisting}[style=mathematica]
Needs["DsixTools`"]
\end{lstlisting}
Alternatively, the user can place the \dsix~folder in a given directory and call the package by specifying its location via
\begin{lstlisting}[style=mathematica]
pathtoDsixTools = "<directory>";
AppendTo[$Path, pathtoDsixTools];
\end{lstlisting}
before executing the {\tt Needs} command. We also recommend to use
\begin{lstlisting}[style=mathematica]
SetDirectory[NotebookDirectory[]];
\end{lstlisting}
at the beginning of all projects with \dsix.

\section{Using \dsixbf}
\label{sec:use}

In this Section we describe how to use \dsix in detail. Once the
package has been loaded, the user can already execute some basic I/O
functions and routines. Several global variables are also introduced
at this stage. Here we list the general \dsix routines:

\begin{itemize}
\item {\tt LoadModule[moduleName]}: Loads the \dsix module {\tt moduleName}. \vspace{0.1cm}
\item {\tt MyPrint[string]}: Prints the message {\tt string}. It can be switched on and off by using the \dsix routines {\tt TurnOnMessages} and {\tt TurnOffMessages}. \vspace{0.1cm}
\item {\tt TurnOnMessages}: Turns on the messages written by \dsix. \vspace{0.1cm}
\item {\tt TurnOffMessages}: Turns off the messages written by \dsix. \vspace{0.1cm}
\item {\tt ReadInputFiles[options\_file,\,WCsInput\_file,\,\{SMInput\_file\}]}: Reads all the input files. The third argument is optional and should be absent when the routine is used to read WET input.\vspace{0.1cm}
\item {\tt WriteInputFiles[options\_file,\,WCsInput\_file,\,\{SMInput\_file\},data]}: Creates input files with the parameter values in {\tt data}. The third argument is optional and should be absent when the routine is used to write WET input. \vspace{0.1cm}
\item {\tt WriteAndReadInputFiles[options\_file,WCsInput\_file,\{SMInput\_file\}]}:
Writes data into new input files and then reads them. The third argument is optional and should be absent when the routine is used to write and read WET input.\vspace{0.1cm}
\item {\tt NewInput[parameter,newvalue,dispatch]}: Replaces the current input (contained in the \mathe dispatch {\tt dispatch}) by a new one in which {\tt parameter} takes the value {\tt newvalue}.\footnote{For those not familiar with \mathe dispatch tables, we clarify that these are optimized representations of lists of replacement rules. In practice they work in exactly the same way as replacement rules, but their execution time is much lower when the list of replacements is long.}\vspace{0.1cm}
\item {\tt NewScale[scale,newvalue]}: Replaces the current value of {\tt scale} by {\tt newvalue}. Here {\tt scale} can be either {\tt "high"} or {\tt "low"}. \vspace{0.1cm}
\item {\tt H[mat]}: Returns the Hermitian conjugate of the matrix {\tt mat}. If the option {\tt CPV} has been set to {\tt 0} then it returns the transpose. \vspace{0.1cm}
\item {\tt CC[x]}: Returns the complex conjugate of {\tt x}. If the option {\tt CPV} is set to {\tt 0} then it returns {\tt x}.
\end{itemize}  
A more detailed description of these \dsix routines can be found in
\ref{ap:routines-general}.   Once \dsix is loaded the user can already perform some basic
operations. However, the real power of \dsix resides in its modules. A
project with \dsix can involve one or many modules, depending on the
user's goals. For instance, if one is interested in the running of WET
operators between the electroweak scale and the $b$-quark mass scale,
the \WETrunner module suffices for the task. But if one wants to study
the RGE evolution of the SMEFT operators and their matching to WET
ones at the electroweak scale, both the \SMEFTrunner and \EWmatcher
modules must be included in the project. In the following we proceed
to describe the existing modules, the routines they include and how
they can be combined together in a practical \dsix project.

\subsection{\SMEFTrunner module}
\label{subsec:SMEFTrunner}

The \SMEFTrunner module takes care of the one-loop RGE of the
SMEFT between the high-energy scale $\Lambda$ and the electroweak
scale. The module is based on the anomalous dimension matrices
computed in \cite{Jenkins:2013zja,Jenkins:2013wua,Alonso:2013hga} (for
the Baryon-number-conserving operators) and \cite{Alonso:2014zka}
(for the Baryon-number-violating operators). This is the list of routines
implemented in this module:

\begin{itemize}
\item {\tt InitializeSMEFTrunnerInput}: Initializes the input for the \SMEFTrunner module. \vspace{0.1cm}
\item {\tt FindParameterSMEFT[parameter]}: Returns the position in which {\tt parameter} is located within the {\tt Parameters} list.\vspace{0.1cm}
\item {\tt GetBeta}: Computes the SMEFT $\beta$ functions.\vspace{0.1cm}
\item {\tt LoadBetaFunctions}: Constructs the SMEFT $\beta$ functions or reads
them from a file. \vspace{0.1cm}
\item {\tt RunRGEsSMEFT}: Runs the SMEFT RGEs. \vspace{0.1cm}
\item {\tt ExportSMEFTrunner}: Exports the \SMEFTrunner results to an output file. \vspace{0.1cm}
\item {\tt WriteSMEFTrunnerOutputFile[Output\_file,data]}: Exports the \SMEFTrunner results in {\tt data} to {\tt Output\_file}.
\end{itemize}
More details about these routines can be found in
\ref{ap:routines-SMEFTrunner}.

\begin{figure}
  \centering
  \includegraphics[width=10cm]{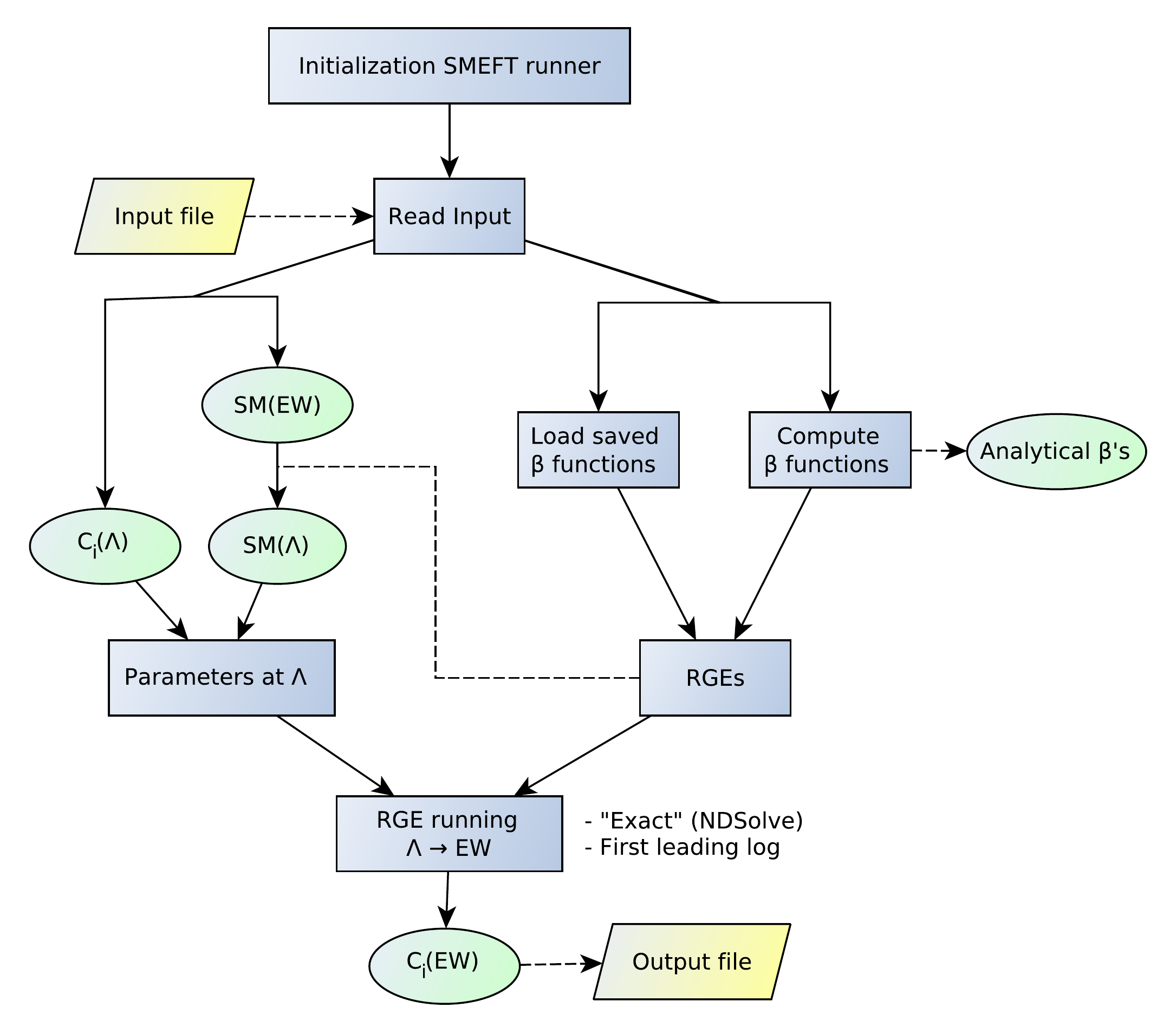}
  \caption{\small \SMEFTrunner module flowchart.}
  \label{fig:SMEFTrunner}
\end{figure}

The general flowchart of the \SMEFTrunner module can be seen in Fig. \ref{fig:SMEFTrunner}. The first step is to read the input. This not only includes the numerical values of the SMEFT WCs at the high-energy scale $\Lambda$, but also the numerical values of the SM parameters and some options. The input and output format of \dsix~is inspired by the Supersymmetry Les Houches Accord (SLHA)~\cite{Skands:2003cj,Allanach:2008qq}. A complete {\textit{options card}} would read:
\begin{lstlisting}[style=options]
Block SCALES
1 10000	  # UV scale [GeV]
2 80.385  # EW scale [GeV]
Block OPTIONS
1 0 	# CPV=0 : all parameters and WCs are assumed to be real
2 1	# ReadRGEs : 0 (RGEs reconstructed) or 1 (RGEs read from a file)
3 1	# Method to solve RGEs : 1 (NDSolve) or 2 (leading log)
4 0	# Export RGEs
5 1	# Use SM RGEs to compute SM parameters at the high-energy scale
6 0	# Export SMEFTrunner results
7 0	# Export EWmatcher results
8 0	# Export WETrunner results
9 1	# Type of input WCs: 1 (SMEFT) or 2 (WET)
\end{lstlisting}
\vspace{5mm}
For this particular example we have chosen the default options. In
fact, if a flag is not present in the options file, its value will be
taken as in this example. The user defines the high-energy and
electroweak scales in the first block. We see that in this example the
electroweak scale is fixed to $80.385$ GeV, the $W$-boson mass,
whereas the high-energy scale is $\Lambda = 10$ TeV. Then the user
defines several options. For instance, in this case CP-violation has
been switched off by setting {\tt CPV} to $0$, but the user can also
consider complex parameters setting {\tt CPV} to $1$. The comments
accompanying the other options are self-explanatory, but we will
review them below. A sample of a typical {\textit{input card}} for the
SM parameters is as follows:
\begin{lstlisting}[style=SMInput]
Block GAUGE
1 0.629787	# g
2 0.345367	# gp
3 1.218232	# gs
Block SCALAR
1 0.257736	# lambda
2 7812.500000	# m2 [GeV^2]
Block GU
1 1 1.23231E-5	# Gamma_u(1,1)
...
3 3 0.994858	# Gamma_u(3,3)
Block IMGU
1 1 0		# Gamma_u(1,1)
...
3 3 0		# Gamma_u(3,3)
...
Block THETA
1 0		# theta
2 0		# thetap
3 0		# thetas
\end{lstlisting}
\vspace{5mm}
Again, the input values are distributed in blocks, each devoted to a
set of parameters. For instance, the SM gauge couplings are given in
the block {\tt GAUGE}. We note that the Yukawa matrices (and in
general any complex parameter) are given in two blocks. For the
up-quark Yukawas these are {\tt GU} and {\tt IMGU}: the first one is
used for the real parts and the second for the imaginary ones. By
default, the SM parameters are assumed to be given at the electroweak
scale, where their experimental values are known. Then, before running
down from $\Lambda$ to the electroweak scale they must be computed at
$\Lambda$. This is done by running up from the electroweak scale using
pure SM RGEs, hence neglecting possible deviations caused by non-zero
SMEFT WCs.\footnote{The user can check the validity of this
  approximation by using the \SMEFTrunner routines, for instance by
  checking whether the resulting values for the SM parameters at the
  electroweak scale (after running down) do not match their initial
  values. This can be fixed by readjusting the SM parameters at
  $\Lambda$. We note, however, that one should take into account NP
  corrections to the standard electroweak parameters induced by
  non-zero SMEFT WCs.} However, in case the user prefers to give the
SM parameters directly at high-energy scale $\Lambda$, this can be
done by setting the {\tt UseRGEsSM} option to $0$ or, equivalently, by
introducing a $0$ in flag number 5 of the options file,
\begin{lstlisting}[style=options]
5 0	# Use SM RGEs to compute SM parameters at the high-energy scale
\end{lstlisting}
\vspace{5mm}
This choice is recommended when the user wants to use the \emph{First
  leading log} method to solve the RGEs, see below.  Finally, a
standard {\textit{input card}} for the SMEFT WCs reads:
\begin{lstlisting}[style=WCsInput]
Block WC1
1 0.000000	# G
2 0.000000	# G tilde
3 0.000000	# W
4 0.000000	# W tilde
Block WC2
1 0.000000	# phi
Block WC3
1 0.000000	# phiBox
2 0.000000	# phiD
Block WC4
1 0.000000	# phiG
...
8 0.000000	# phiWtildeB
Block WCUPHI
1 1 0.000000	# uphi(1,1)
1 2 0.000000	# uphi(1,2)
...
3 3 0.000000	# uphi(3,3)
...
\end{lstlisting}
\vspace{5mm}
The notation for the operators is self-explanatory:\\ 
\begin{align*}
&\mathtt{Block WC1}   \nonumber \\
&Q_G = f^{ABC} G_\mu^{A \nu} G_\nu^{B \rho} G_\rho^{C \mu}     \nonumber \\
&Q_{\widetilde G} = f^{ABC} \widetilde G_\mu^{A \nu} G_\nu^{B \rho} G_\rho^{C \mu}   \nonumber \\
&Q_W = \epsilon^{IJK} W_\mu^{I \nu} W_\nu^{J \rho} W_\rho^{K \mu}    \nonumber \\
&Q_{\widetilde W} = \epsilon^{IJK} \widetilde W_\mu^{I \nu} W_\nu^{J \rho} W_\rho^{K \mu}  \, \nonumber \\[0.3cm]
&\mathtt{Block WC2}   \nonumber \\
&Q_{\varphi} = \left( \varphi^\dagger \varphi \right)^3     \, \nonumber \\[0.3cm]
&\mathtt{Block WC3}   \nonumber \\
&Q_{\varphi \Box} = \left( \varphi^\dagger \varphi \right) \Box \left( \varphi^\dagger \varphi \right)      \nonumber \\
&Q_{\varphi D} = \left( \varphi^\dagger D^\mu \varphi \right)^\ast \left( \varphi^\dagger D_\mu \varphi \right)      \nonumber  \\[0.15cm]
&\cdots     \nonumber  \\[0.15cm]
&\mathtt{Block WCUPHI}   \nonumber \\
&Q_{u \varphi}[1,1] =  \left( \varphi^\dagger \varphi \right) \left( \bar q_1 u_1 \widetilde \varphi \right)�\nonumber \\
&Q_{u \varphi}[1,2] =  \left( \varphi^\dagger \varphi \right) \left( \bar q_1 u_2 \widetilde \varphi \right)�\nonumber \\
&\cdots \nonumber \\
&Q_{u \varphi}[3,3] =  \left( \varphi^\dagger \varphi \right) \left( \bar q_3 u_3 \widetilde \varphi \right)
\end{align*}\\[3mm]
It is important to note that all WCs are assumed to vanish by
default. Therefore, it suffices to include the non-zero WCs (and only
these) in the input card. The rest can be absent.

In addition to getting input values at $\Lambda$, the \SMEFTrunner
module constructs the SMEFT RGEs and prepares a routine for their
evaluation. There are two ways to do this: (i) to ``compute'' the
RGEs, or (ii) to read them from a file. In the first case, \dsix takes
their saved forms and computes all flavor traces and expands over all
possible indices. In the second case, \dsix simply reads them from a
file (already present in the \dsix folder). In both cases, the command
to give this step is {\tt LoadBetaFunctions}. The user can choose
between these two methods in the options file using flag number
$2$. Method (i) is more time consuming but provides nice-looking
$\beta$ functions in \mathe format. This would be very useful for
analytical studies. Moreover, the user can obtain these equations at
any moment by using the routine {\tt GetBeta}. Method (ii) is much
faster and produces the RGEs in the format required for their
immediate evaluation with the \SMEFTrunner routines (in particular
with {\tt RunRGEsSMEFT}).

Once both the initial conditions (input values at $\Lambda$) and the
RGEs are completely built, the user can apply the RGE to
obtain the goal of \SMEFTrunner: the SMEFT WCs at the electroweak
scale. This is done with the {\tt RunRGEsSMEFT} routine. We have
implemented two different methods for the resolution of the RGEs:

\begin{itemize}
\item \emph{``Exact''}: This method applies the \mathe internal
  command {\tt NDSolve} for the numerical resolution of differential
  equations. Given the large number of differential equations involved
  in this case (several thousand), this might be time consuming, with
  each evaluation requiring a few (< 10) seconds, the exact number
  depending on the particular case and computer.
\item \emph{First leading log}: This approximate method might be
  sufficient for many phenomenological studies, in particular when
  $\Lambda$ is not too far from the electroweak scale. The solution of
  the RGEs is obtained as
\begin{equation}
C_i(\mu) = C_i(\Lambda) + \frac{\beta_i}{16 \pi^2} \log \left( \frac{\mu}{\Lambda} \right)\,,
\end{equation}
where $C_i$ is any of the running parameters, $\mu$ is the
renormalization scale and $\beta_i$ is the $\beta$ function for the
$C_i$ evaluated at $\mu = \Lambda$. This method is much faster but
neglects subleading effects.
\end{itemize}

The user chooses between these two methods by setting the global
option {\tt RGEsMethod} to $0$ (for the {\tt NDSolve} method) or $1$
(for the first leading log approximate method). This is also done via
flag number 3 in the options card. After running, the results are
saved in the array {\tt outSMEFTrunner} as a function of $t =
\log_{10} \mu$. The ordering of the parameters is the same as in the
    {\tt Parameters} global array, see
    \ref{ap:parameters}.\footnote{The position of a specific
      parameter can also be obtained by using the {\tt
        FindParameterSMEFT} function, see \ref{ap:routines}.}
    Moreover, two important values of $t$ are predefined: {\tt tLOW}
    ($=\log_{10} \mu_{\rm EW}$) and {\tt tHIGH} ($=\log_{10}
    \Lambda$). Therefore, for instance, one can obtain the value of
    the $\left[ C_{e \vp} \right]_{12}$ at the electroweak scale by
    evaluating
\begin{lstlisting}[style=mathematica]
outSMEFTrunner[[70]]/.t->tLOW
\end{lstlisting}

Finally, the output of \SMEFTrunner can be exported to a text
file. This is done by running {\tt ExportSMEFTrunner}. The file {\tt
  Output\_SMEFTrunner.dat} is then generated, also with an SLHA
format, completely analogous to the SMEFT WCs input card.

\subsection{\EWmatcher module}
\label{subsec:EWmatcher}

The \EWmatcher module applies the tree-level matching of the Warsaw
basis of the SMEFT to the $\Delta B = \Delta S = 1,2$ and $\Delta B =
\Delta C = 1$ operators of the WET. For this purpose we make use of
the analytical results obtained in~\cite{Aebischer:2015fzz}. This is
the list of routines implemented in this module:

\begin{itemize}
\item {\tt InitializeEWmatcherInput}: Initializes the input for the \EWmatcher module. \vspace{0.1cm}
\item {\tt FindParameterWET[parameter]}: Returns the position where {\tt parameter} is located within the {\tt WETParameters} list.\vspace{0.1cm}
\item {\tt Biunitary[mat,dim]}: Applies a biunitary transformation that diagonalizes the {\tt dim} $\times$ {\tt dim} matrix {\tt mat}. \vspace{0.1cm}
\item {\tt RotateToMassBasis}: Transforms the SMEFT WCs to the fermion
  mass basis.  \vspace{0.1cm}
\item {\tt ApplyEWmatching}: Matches the SMEFT WCs onto the WET WCs. \vspace{0.1cm}
\item {\tt Match[WC]}: Returns the value of the WET Wilson coefficient {\tt WC} after matching it to the SMEFT. \vspace{0.1cm}
\item {\tt MatchAnalytical[WC]}: Returns the analytical expression of the WET Wilson coefficient {\tt WC} after matching it to the SMEFT. \vspace{0.1cm}
\item {\tt WriteWCsMassBasisOutputFile[Output\_file]}: Exports the SMEFT 2- and 4-fermion WCs in the fermion mass basis to {\tt Output\_file}. \vspace{0.1cm}
\item {\tt ExportEWmatcher}: exports the \EWmatcher results to an output file.   \vspace{0.1cm}
\item {\tt WriteEWmatcherOutputFile[Output\_file,data]}: Exports the \EWmatcher results in {\tt data} to {\tt Output\_file}.
\end{itemize}
Appendix~\ref{ap:routines-EWmatcher} contains more details about these routines
and functions.\\

\begin{figure}
  \centering
  \includegraphics[width=13cm,height=9.5cm]{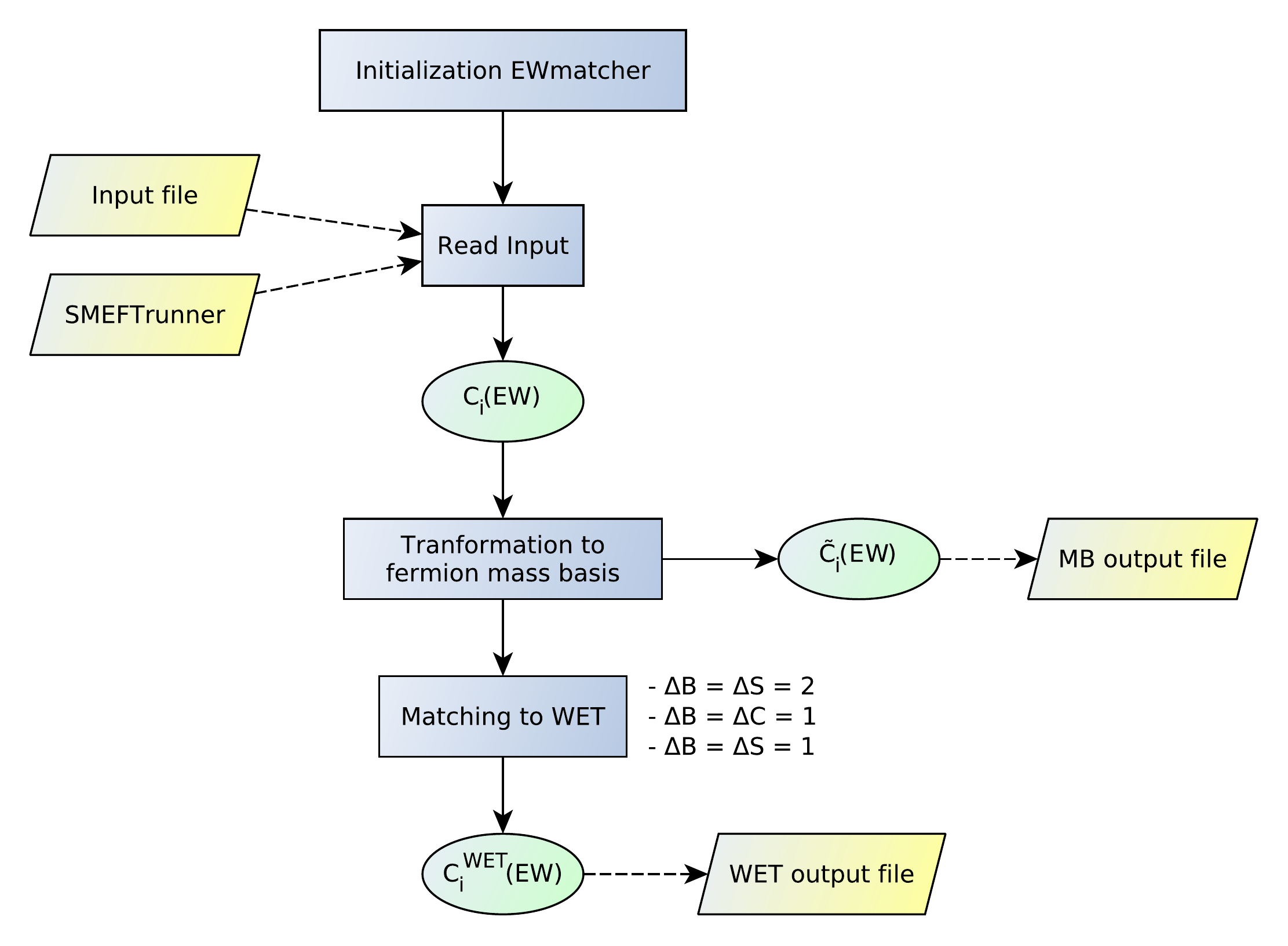}
  \caption{\small \EWmatcher module flowchart.}
  \label{fig:EWmatcher}
\end{figure}

The \EWmatcher module flowchart can be seen in
Fig. \ref{fig:EWmatcher}. The first step is, as usual, to determine
the input of the module. If the \SMEFTrunner module has preceded the
\EWmatcher, the input is automatically created using the output of the
former. Otherwise, the user can also provide an input file and start
the \dsix project with the \EWmatcher module as first step.

In what concerns the input card, the same {\tt WCsInput.dat} file that
is read by the \SMEFTrunner module can also be used as WCs input for
the \EWmatcher module. Instead of interpreting the input values as
given at $\mu = \Lambda$, the \EWmatcher module will interpret them as
given at $\mu = \mu_{\rm EW}$.

Next, the \EWmatcher transforms the SMEFT WCs to the fermion mass
basis, applying the required biunitary transformations to the fermion
mass matrices.\footnote{In the current version of \dsix, the
  dimension five Weinberg operator is absent and neutrinos remain
  massless. For this reason, we have not considered rotations in the
  lepton sector, which is assumed to be diagonal.} This necessary step
is done by means of the {\tt RotateToMassBasis} routine and gives, as
a result, the SMEFT WCs in the fermion mass basis, generally denoted
as $\widetilde C_i$.\footnote{The Baryon-number-conserving $\widetilde C_i$
  coefficients in the mass basis were computed in
  \cite{Aebischer:2015fzz} whereas the Baryon-number-violating ones can be found
  in \ref{ap:BVinMB}.} In case the user is particularly interested in
these coefficients, they can be exported to an external text file by
running\\[-4mm]
\begin{lstlisting}[style=mathematica]
WriteWCsMassBasisOutputFile[Output_file]
\end{lstlisting}
\vspace{1mm}
Furthermore, {\tt RotateToMassBasis} also creates the replacements array {\tt
  ToMassBasis}, which can be used to print the numerical values of the
SMEFT WCs in the fermion mass basis. For example,\\[-4mm]
\begin{lstlisting}[style=mathematica]
LL[1, 1, 2, 2]/.ToMassBasis
\end{lstlisting}
\vspace{1mm}
would print the numerical value of the coefficient $\big[\widetilde{C}_{\ell \ell}\big]_{1122}$.\\[-1mm]

The next step is the matching to the WET operators. For this the user
must execute\\[-4mm]
\begin{lstlisting}[style=mathematica]
ApplyEWmatching
\end{lstlisting}
\vspace{1mm}
This creates the arrays {\tt BS2},
{\tt BC1}, {\tt BS1Hunprimed}, {\tt BS1Hprimed}, {\tt BS1GB}, {\tt
  BS1SLunprimed} and {\tt BS1SLprimed}, containing the numerical
values of the WET WCs at the electroweak scale. While {\tt BS2} and
{\tt BC1} contain the results for the $\Delta B = \Delta S = 2$ and
$\Delta B = \Delta C = 1$ coefficients, respectively, the other arrays
contain the results for the $\Delta B = \Delta S = 2$ ones, split into
several sub-arrays with hadronic (H), semileptonic (SL) and magnetic,
or gauge boson (GB) involving, WCs. They can also be accessed
individually thanks to the function {\tt Match}. For example, {\tt
  CBS1[e][1] // Match} would print the numerical value of
$C_1^{bsee}$. If one is interested in the analytical expression after
matching the function to use is {\tt MatchAnalytical}, which only
replaces the energy scales and the SM parameters by their numerical
values.

Finally, the output of \EWmatcher can be exported to a text file. This
is done by running {\tt ExportEWmatcher}, which generates the file
{\tt Output\_EWmatcher.dat} with a compilation of all results obtained
with this module.

\subsection{\WETrunner module}
\label{subsec:WETrunner}

The \WETrunner module is devoted to the RGE of the WET WCs
between the electroweak scale and the infrared scale $\Lambda_{\rm
  IR}$. In the current version of \dsix, $\Lambda_{\rm IR}$ is set by default at
$b$-quark mass scale, $m_b = 4.18$ GeV, and the RGE is based on
the analytical results obtained in Ref.~\cite{Aebischer:2017gaw}. This is the list of routines
implemented in this module:

\begin{itemize}
\item {\tt InitializeWETrunnerInput}: Initializes the input for the \WETrunner module. \vspace{0.1cm}
\item {\tt RunRGEsWET}: Runs the WET RGEs.  \vspace{0.1cm}
\item {\tt ExportWETrunner}: Exports the \WETrunner results to an output file.  \vspace{0.1cm}
\item {\tt WriteWETrunnerOutputFile[Output\_file,data,scale]}: Exports the \WETrunner results in data to {\tt Output\_file} after evaluating them at $\mu =$ {\tt scale}.  
\end{itemize}
For more details about these routines see Appendix~\ref{ap:routines-WETrunner}.\\[-1mm]

\begin{figure}
  \centering
  \includegraphics[width=12cm,height=9cm]{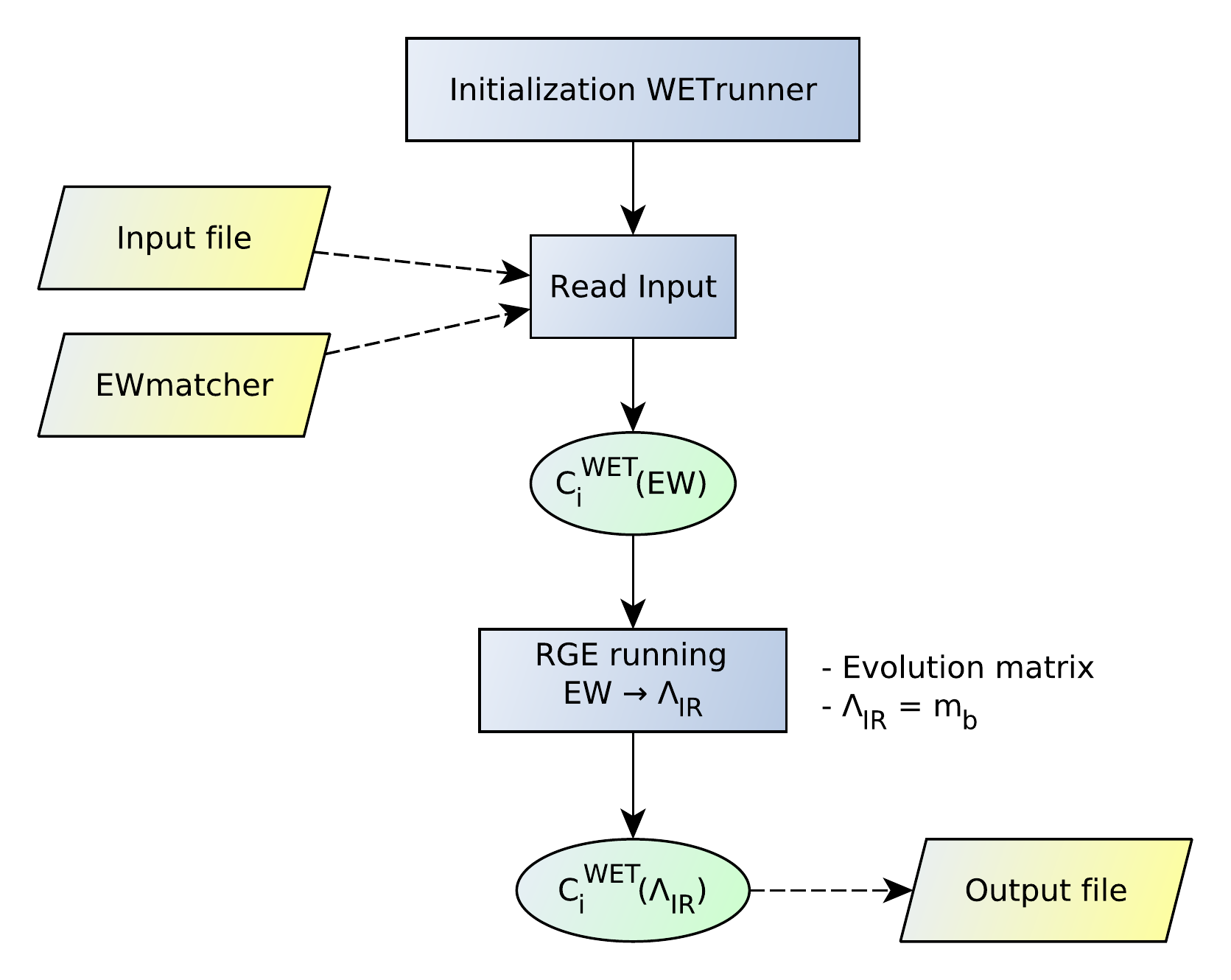}
  \caption{\small \WETrunner module flowchart.}
  \label{fig:WETrunner}
\end{figure}

The general flowchart of the \WETrunner module can be seen in
Fig. \ref{fig:WETrunner}. As for the \EWmatcher module, the input for
the \WETrunner can be obtained directly (and automatically) from the
previous module (\EWmatcher in this case) if this is run
before. Otherwise, the user can also provide new input values and
begin a \dsix project in this step of the chain. In this case, the
input card for the WET WCs, which we will call {\tt WCsInput.dat} as
for the SMEFT ones, also follows a simple SLHA-inspired format:
\begin{lstlisting}[style=WCsInput]
Block SCALES
2 80.385  # EW scale [GeV]
Block BS2
1 0.000000	# C1sb
2 0.000000	# C2sb
3 0.000000	# C3sb
4 1.000000	# C4sb
5 0.000000	# C5sb
6 0.000000	# C1sb'
7 0.000000	# C2sb'
8 0.000000	# C3sb'
Block BC1
1 0.000000	# CV(e)
2 0.000000	# CV(mu)
3 0.000000	# CV(tau)
4 0.000000	# CS(e)
5 0.000000	# CS(mu)
6 0.000000	# CS(tau)
7 0.000000	# CT(e)
8 0.000000	# CT(mu)
9 0.000000	# CT(tau)
10 0.000000	# CV(e)'
11 0.000000	# CV(mu)'
12 0.000000	# CV(tau)'
13 0.000000	# CS(e)'
14 0.000000	# CS(mu)'
15 0.000000	# CS(tau)'
Block BS1H
1 0.000000	# C1(sbuu)
...
\end{lstlisting}
\vspace{3mm}
We see that the first block gives the input value for the electroweak
scale while the other blocks give the input values for the WET WCs. In
this particular example the only non-vanishing WET WC is $C_4^{sbsb} =
1$. It is important to note, however, that \dsix must know that the
input WCs are of WET type. This is done via the option {\tt
  inputWCsType}, which must be set to $2$ (the default value is $1$,
which indicates SMEFT WCs) before reading the input file with the {\tt
  ReadInputFiles} routine. This option choice can also be achieved
by using the flag number $9$ in the options card:
\begin{lstlisting}[style=options]
9 2	# Type of input WCs: 1 (SMEFT) or 2 (WET)
\end{lstlisting}
\vspace{3mm}
Once the input has been read (or taken from the \EWmatcher module) one
can proceed to the RGE of the WET WCs. The \WETrunner module
implements an evolution matrix formalism for this step, based on the
results of Ref.~\cite{Aebischer:2017gaw}. This can be performed with the {\tt RunRGEsWET}
which creates the numerical arrays {\tt BS2Low}, {\tt BC1Low}, {\tt
  BS1unprimedLow} and {\tt BS1primedLow}, all of them functions of $t
= \log_{10} \mu$. Therefore, in order to find the values of the WET
WCs at the $b$-quark mass scale one can evaluate commands like\\[-4mm]
\begin{lstlisting}[style=mathematica]
BS2Low[[4]] /. t -> Log[10, mb]
\end{lstlisting}
where {\tt mb} is the global \dsix variable containing the $b$-quark
mass (in GeV). This command would print the numerical value of
$C_4^{sbsb}(m_b)$. Finally, the results can be exported to an output
file with the routine {\tt ExportWETrunner}. This creates the text file
{\tt Output\_WETrunner.dat} with the values of all WET WCs at the infrared scale.

\section{Summary}
\label{mysum}

We have presented \dsix, a \mathe package for the handling of the
dimension-six SMEFT and WET. \dsix facilitates the treatment of these 
two effective theories in a systematic and complete manner.

\dsix is a modular code. In the current version, it includes three
independent modules, designed for specific tasks related to the SMEFT
and WET. The \SMEFTrunner performs the one-loop RGE from the
UV scale $\Lambda$ to the electroweak scale, the \EWmatcher matches
the SMEFT Wilson coefficients to the $\Delta B = \Delta S = 1,2$ and $\Delta B = \Delta C = 1$ 
operators of the WET, and the \WETrunner runs
these down to an IR scale, in this case the $b$-quark mass scale.

The structure of \dsix allows for an easy implementation of new
modules. Therefore, the current content of the package is expected to
grow substantially with future improvements, including additional
tools and features. The final outcome of this endevour will be a
complete and powerful framework for the systematic exploration of new
physics models using the language of Effective Field Theories.

\section*{Acknowledgements}
We thank Aneesh Manohar for comments on the manuscript.
The work of A.C. is supported by the Alexander von Humboldt
Foundation. The work of J.F. is supported in part by the Spanish
Government, by Generalitat Valenciana and by ERDF funds from the EU
Commission [grants FPA2011-23778,FPA2014-53631-C2-1-P,
  PROMETEOII/2013/007, SEV-2014-0398]. J.F. also acknowledges
VLC-CAMPUS for an ``Atracci\'o de Talent''
scholarship. A.V. acknowledges financial support from the ``Juan de la
Cierva'' program (27-13-463B-731) funded by the Spanish MINECO as well
as from the Spanish grants FPA2014-58183-P, Multidark CSD2009-00064,
SEV-2014-0398 and PROMETEOII/ 2014/084 (Generalitat Valenciana).
J.V. is funded by the Swiss National Science Foundation and
acknowledges support from Explora project FPA2014-61478-EXP.    A.C. thanks the Albert Einstein Center for Fundamental Physics of the University of Bern for their hospitality and support while part of this work was performed. 

\newpage

\appendix

\section{Standard Model Effective Field Theory}
\label{sec:SMEFT}

The Lagrangian for the SMEFT can be written as
\begin{equation} \label{eq:SMEFT}
\mathcal{L} = \mathcal{L}_{\rm SM}^{(4)} + \frac{1}{\Lambda} \sum_k C_k^{(5)} Q_k^{(5)}  + \frac{1}{\Lambda^2} \sum_k C_k^{(6)} Q_k^{(6)} + \mathcal{O}\left( \frac{1}{\Lambda^3} \right) \, .
\end{equation}
Here $\Lambda$ is the new physics scale suppressing higher dimensional operators, assumed to be much larger than the electroweak scale.  There is only one operator of dimension five, the so-called Weinberg operator that gives a Majorana mass term for the neutrinos~\cite{Weinberg:1979sa}.  A non-redundant basis of dimension-six operators was introduced in \cite{Grzadkowski:2010es} and is known as the \textit{Warsaw basis}. 
We list these operators in Tables~\ref{pbsot},~\ref{mixt} and \ref{fft}.   
Barring flavor structure, and
assuming Baryon number conservation, there are 59 operators, some of which
are non-Hermitian, yielding in total 76 real coefficients.
Taking into account flavour indices, the dimension-six Lagrangian
contains 1350 CP-even and 1149 CP-odd operators, for a total of 2499
hermitian operators~\cite{Alonso:2013hga}.    Finally, the complete set
of independent dimension-6 Baryon number violating operators were identified in \cite{Abbott:1980zj}.  Barring flavor structure, there are only 4 Baryon-number-violating operators.   These are listed in Table~\ref{bvt}.

The implementation of the SMEFT in \dsix~follows the conventions used in
\cite{Grzadkowski:2010es}.\footnote{The reader should keep in mind that these conventions differ from those used in~\cite{Jenkins:2013zja,Jenkins:2013wua,Alonso:2013hga}. The differences appear in the normalization of $\lambda$ and $m$, the definition of the Yukawa matrices, the name of the gauge couplings, and in whether the NP scale is introduced into the definition of the WCs.  } The SM renormalizable Lagrangian
$\mathcal{L}_{\rm SM}^{(4)}$ is given by
\begin{align} 
\mathcal{L}_{\rm SM}^{(4)} =& -\frac{1}{4} G_{\mu \nu}^A G^{A \mu \nu} -\frac{1}{4} W_{\mu \nu}^I W^{I \mu \nu} -\frac{1}{4} B_{\mu \nu} B^{\mu \nu} + \left( D_\mu \varphi \right)^\dagger \left( D^\mu \varphi \right) + m^2 \varphi^\dagger \varphi -   \frac{\lambda}{2} \left( \varphi^\dagger \varphi \right)^2 \nonumber \\
&+ i \left( \bar \ell \slashed{D} \ell + \bar e \slashed{D} e + \bar q \slashed{D} q + \bar u \slashed{D} u + \bar d \slashed{D} d \right) - \left( \bar \ell \Gamma_e e \varphi + \bar q \Gamma_u u \widetilde \varphi + \bar q \Gamma_d d \varphi + \hc \right) \, . \label{eq:SMLag}
\end{align}
Here $A=1\dots8$ and $I=1\dots3$ denote gauge indices associated to $SU(3)_c$ and $SU(2)_L$.  The fields $\ell$ and $q$ correspond to the lepton and quark $SU(2)_L$ doublets of the SM, while $e,u,d$ are the right-handed fields.  The Higgs $SU(2)_L$ doublet is denoted by $\varphi$. The Yukawa couplings $\Gamma_{e,u,d}$ are $3 \times 3$ matrices in
flavor space. The
covariant derivative is generically defined as
\begin{equation}
D_\mu = \partial_\mu + i g_s T^A G_\mu^A + i g T^I W_\mu^I + i g^\prime Y B_\mu \, ,
\end{equation}
where $\{g_s,g,g^\prime\}$ and $\{G,W,B\}$ are, respectively, the $SU(3)_c$, $SU(2)_L$ and
$U(1)_Y$ gauge couplings and gauge fields. $T^A$ and $T^I$ are the corresponding gauge group generators.    The hypercharge assignments for the matter fields are given in Table~\ref{hypercharge}.
%
\begin{table}[ht]
\renewcommand{\arraystretch}{1.6}
\begin{center}
\caption{Hypercharge assignments.}
\label{hypercharge}
\begin{tabular}{|ccccccc|}
\hline 
Field & $\ell_L$ & $e_R$ & $q_L$ & $u_R$ & $d_R$ & $\vp$ \\
\hline
$Y$ & $-\frac{1}{2}$ & $-1$ & $\frac{1}{6}$ & $\frac{2}{3}$ & $-\frac{1}{3}$ & $\frac{1}{2}$ \\
\hline
\end{tabular}
\end{center}
\end{table}
\dsix~also implements the running of the $\theta$ terms
\begin{align}
\mathcal{L}_{\theta} =  \frac{\theta^{\prime} g^{\prime 2 }}{  32 \pi^2 }  \widetilde B_{\mu \nu}  B^{\mu \nu} + 
 \frac{\theta  g^{2}}{  32 \pi^2 }  \widetilde W_{\mu \nu}^{I}  W^{\mu \nu}_{I}  +  \frac{\theta_s g_s^{2}}{  32 \pi^2 }  \widetilde G_{\mu \nu}^{A}  G^{\mu \nu}_{A}   \,,
\end{align}
calculated in~\cite{Jenkins:2013zja}.  The dual tensors are defined as $\widetilde X = \frac{1}{2}
\epsilon_{\mu \nu \rho \sigma} X^{\rho \sigma}$ (with $\epsilon_{0123}
= +1$).

\begin{table} 
\renewcommand{\arraystretch}{1.6}
\caption{Purely bosonic operators. \label{pbsot}}
\begin{center}
\begin{tabular}{|c|c||c|c|}
\hline  
\multicolumn{2}{|c||}{$X^3$} & \multicolumn{2}{|c|}{$X^2 \varphi^2$} \\
\hline
$Q_G$  & $f^{ABC} G_\mu^{A \nu} G_\nu^{B \rho} G_\rho^{C \mu}$ & $Q_{\varphi G}$ & $\varphi^\dagger \varphi G_{\mu \nu}^A G^{A \mu \nu}$ \\
$Q_{\widetilde G}$ & $f^{ABC} \widetilde G_\mu^{A \nu} G_\nu^{B \rho} G_\rho^{C \mu}$ & $Q_{\varphi B}$ & $\varphi^\dagger \varphi B_{\mu \nu} B^{\mu \nu}$ \\
$Q_W$ & $\epsilon^{IJK} W_\mu^{I \nu} W_\nu^{J \rho} W_\rho^{K \mu}$ & $Q_{\varphi W}$  & $\varphi^\dagger \varphi W_{\mu \nu}^I W^{I \mu \nu}$ \\
$Q_{\widetilde W}$ & $\epsilon^{IJK} \widetilde W_\mu^{I \nu} W_\nu^{J \rho} W_\rho^{K \mu}$ &  $Q_{\varphi W B}$  & $\varphi^\dagger \tau^I \varphi W_{\mu \nu}^I B^{\mu \nu}$  \\
\cline{1-2}  
\multicolumn{2}{|c||}{$\myv{\varphi^6}$} & $Q_{\varphi \widetilde G}$   & $\varphi^\dagger \varphi \widetilde G_{\mu \nu}^A G^{A \mu \nu}$  \\
\cline{1-2}
 $Q_{\varphi}$ &$\myv{\left( \varphi^\dagger \varphi \right)^3}$ & $Q_{\varphi \widetilde B}$ & $\varphi^\dagger \varphi \widetilde B_{\mu \nu} B^{\mu \nu}$ \\
 \cline{1-2}
\multicolumn{2}{|c||}{$\varphi^4 D^2$} & $Q_{\varphi \widetilde W}$ &  $\varphi^\dagger \varphi \widetilde W_{\mu \nu}^I W^{I \mu \nu}$ \\
\cline{1-2}
$Q_{\varphi \Box}$ & $\myv{\left( \varphi^\dagger \varphi \right) \Box \left( \varphi^\dagger \varphi \right)}$ & $Q_{\varphi \widetilde W B}$ & $\varphi^\dagger \tau^I \varphi \widetilde W_{\mu \nu}^I B^{\mu \nu}$ \\
$Q_{\varphi D}$ & $\left( \varphi^\dagger D^\mu \varphi \right)^\ast \left( \varphi^\dagger D_\mu \varphi \right)$ &  &  \\
\hline
\end{tabular}
\end{center}
\end{table}

\begin{table}
\renewcommand{\arraystretch}{1.6}
\caption{Mixed operators involving bosons and fermions. \label{mixt}}
\begin{center}
\begin{tabular}{|c|c||c|c|}
\hline
\multicolumn{2}{|c||}{$\psi^2 \varphi^3$} & \multicolumn{2}{|c|}{$\psi^2 \varphi^2 D$} \\
\hline
$Q_{u \varphi}$  & $\left( \varphi^\dagger \varphi \right) \left( \bar q u \widetilde \varphi \right)$  & $Q_{\varphi \ell}^{(1)}$ & $\left( \varphi^\dagger i \Dlr_\mu \varphi \right) \left( \bar \ell \gamma^\mu \ell \right)$ \\
$Q_{d \varphi}$  & $\left( \varphi^\dagger \varphi \right) \left( \bar q d \varphi \right)$  & $Q_{\varphi \ell}^{(3)}$ & $\left( \varphi^\dagger i \DlrImu \varphi \right) \left( \bar \ell \tau^I \gamma^\mu \ell \right)$ \\
$Q_{e \varphi}$ & $\left( \varphi^\dagger \varphi \right) \left( \bar \ell e \varphi \right)$ & $Q_{\varphi e}$ & $\left( \varphi^\dagger i \Dlr_\mu \varphi \right) \left( \bar e \gamma^\mu e \right)$ \\  \cline{1-2}
\multicolumn{2}{|c||}{$\myv{\psi^2 X \varphi}$} & $Q_{\varphi q}^{(1)}$ & $\left( \varphi^\dagger i \Dlr_\mu \varphi \right) \left( \bar q \gamma^\mu q \right)$ \\
\cline{1-2} 
 $Q_{e W}$ &  ${\rule{0cm}{0.3cm}\left( \bar \ell \sigma^{\mu \nu} e \right) \tau^I \varphi W_{\mu \nu}^I}$ &  $Q_{\varphi q}^{(3)}$& $\left( \varphi^\dagger i \DlrImu \varphi \right) \left( \bar q \tau^I \gamma^\mu q \right)$ \\
  $Q_{e B}$ & $\left( \bar \ell \sigma^{\mu \nu} e \right) \varphi B_{\mu \nu}$  & $Q_{\varphi u}$ & $\left( \varphi^\dagger i \Dlr_\mu \varphi \right) \left( \bar u \gamma^\mu u \right)$ \\
  $Q_{u G}$  &  $\left( \bar q \sigma^{\mu \nu} T^A u \right) \widetilde \varphi G_{\mu \nu}^A$  &  $Q_{\varphi d}$  & $\left( \varphi^\dagger i \Dlr_\mu \varphi \right) \left( \bar d \gamma^\mu d \right)$ \\
$Q_{u W}$ & $\left( \bar q \sigma^{\mu \nu} u \right) \tau^I \widetilde \varphi W_{\mu \nu}^I$ & $Q_{\varphi u d}$ & $\left( \widetilde \varphi^\dagger i D_\mu \varphi \right) \left( \bar u \gamma^\mu d \right)$ \\[0.1cm]
$Q_{u B}$ & $\left( \bar q \sigma^{\mu \nu} u \right) \widetilde \varphi B_{\mu \nu}$ &  &  \\[0.1cm]
$Q_{d G}$ & $\left( \bar q \sigma^{\mu \nu} T^A d \right) \varphi G_{\mu \nu}^A$ &  &  \\[0.1cm]
$Q_{d W}$ & $\left( \bar q \sigma^{\mu \nu} d \right) \tau^I \varphi W_{\mu \nu}^I$ &  &  \\[0.1cm]
$Q_{d B}$ & $\left( \bar q \sigma^{\mu \nu} d \right) \varphi B_{\mu \nu}$ &  &  \\
\hline
\end{tabular}
\end{center}
\end{table}

\begin{table}
\renewcommand{\arraystretch}{1.6}
\caption{Purely fermionic operators which preserve Baryon number. \label{fft}}
\begin{center}
\begin{tabular}{|c|c||c|c|}
\hline
\multicolumn{2}{|c||}{$\left( \bar L L \right) \left( \bar L L \right)$} & \multicolumn{2}{|c|}{$\left( \bar L L \right) \left( \bar R R \right)$} \\
\hline
$Q_{\ell \ell}$ & $\left( \bar \ell \gamma_\mu \ell \right) \left( \bar \ell \gamma^\mu \ell \right)$ & $Q_{\ell e}$ & $\left( \bar \ell \gamma_\mu \ell \right) \left( \bar e \gamma^\mu e \right)$ \\
$Q_{q q}^{(1)}$ & $\left( \bar q \gamma_\mu q \right) \left( \bar q \gamma^\mu q \right)$ & $Q_{\ell u}$ & $\left( \bar \ell \gamma_\mu \ell \right) \left( \bar u \gamma^\mu u \right)$ \\
$Q_{q q}^{(3)}$ & $\left( \bar q \gamma_\mu \tau^I q \right) \left( \bar q \gamma^\mu \tau^I q \right)$ & $Q_{\ell d}$ & $\left( \bar \ell \gamma_\mu \ell \right) \left( \bar d \gamma^\mu d \right)$ \\
$Q_{\ell q}^{(1)}$ & $\left( \bar \ell \gamma_\mu \ell \right) \left( \bar q \gamma^\mu q \right)$ & $Q_{q e}$ & $\left( \bar q \gamma_\mu q \right) \left( \bar e \gamma^\mu e \right)$ \\
$Q_{\ell q}^{(3)}$ & $\left( \bar \ell \gamma_\mu \tau^I \ell \right) \left( \bar q \gamma^\mu \tau^I q \right)$ & $Q_{q u}^{(1)}$ & $\left( \bar q \gamma_\mu q \right) \left( \bar u \gamma^\mu u \right)$ \\
\cline{1-2}
\multicolumn{2}{|c||}{$\left( \bar R R \right) \left( \bar R R \right)$} & $Q_{q u}^{(8)}$ & $\left( \bar q \gamma_\mu T^A q \right) \left( \bar u \gamma^\mu T^A u \right)$ \\
\cline{1-2}
$Q_{ee}$ & $\left( \bar e \gamma_\mu e \right) \left( \bar e \gamma^\mu e \right)$ & $Q_{q d}^{(1)}$ & $\left( \bar q \gamma_\mu q \right) \left( \bar d \gamma^\mu d \right)$ \\
$Q_{uu}$ & $\left( \bar u \gamma_\mu u \right) \left( \bar u \gamma^\mu u \right)$ & $Q_{q d}^{(8)}$ & $\left( \bar q \gamma_\mu T^A q \right) \left( \bar d \gamma^\mu T^A d \right)$ \\
\cline{3-4}
$Q_{dd}$ & $\left( \bar d \gamma_\mu d \right) \left( \bar d \gamma^\mu d \right)$ & \multicolumn{2}{|c|}{$\myv{\left( \bar L R \right) \left( \bar R L \right)}$} \\
\cline{3-4}
$Q_{eu}$ & $\left( \bar e \gamma_\mu e \right) \left( \bar u \gamma^\mu u \right)$ & $Q_{\ell e d q}$ & $\myv{\left( \bar \ell^j e \right) \left( \bar d q^j \right)}$ \\
\cline{3-4}
$Q_{ed}$ & $\left( \bar e \gamma_\mu e \right) \left( \bar d \gamma^\mu d \right)$ & \multicolumn{2}{|c|}{$\myv{\left( \bar L R \right) \left( \bar L R \right)}$} \\
\cline{3-4}
$\myv{Q_{ud}^{(1)}}$ & $\left( \bar u \gamma_\mu u \right) \left( \bar d \gamma^\mu d \right)$ & $Q_{q u q d}^{(1)}$ & $\left( \bar q^j u \right) \epsilon_{jk} \left( \bar q^k d \right)$ \\
$Q_{ud}^{(8)}$ & $\left( \bar u \gamma_\mu T^A u \right) \left( \bar d \gamma^\mu T^A d \right)$ & $Q_{q u q d}^{(8)}$ & $\left( \bar q^j T^A u \right) \epsilon_{jk} \left( \bar q^k T^A d \right)$ \\
 & & $Q_{\ell e q u}^{(1)}$ & $\left( \bar \ell^j e \right) \epsilon_{jk} \left( \bar q^k u \right)$ \\
 & & $Q_{\ell e q u}^{(3)}$ & $\left( \bar \ell^j \sigma_{\mu \nu} e \right) \epsilon_{jk} \left( \bar q^k \sigma^{\mu \nu} u \right)$ \\
\hline
\end{tabular}
\end{center}
\end{table}

\begin{table}
\renewcommand{\arraystretch}{1.6}
\caption{Baryon-number-violating operators. We use $C$ to denote the Dirac charge conjugation matrix.   \label{bvt}}
\begin{center}
\begin{tabular}{|c|c|}
\hline
\multicolumn{2}{|c|}{Baryon-number-violating} \\
\hline
$Q_{duq\ell}$ & $\left( d^T C u \right) \left( q^T C \ell \right)$ \\
$Q_{qque}$ & $\left( q^T C q \right) \left( u^T C e \right)$ \\
$Q_{qqq\ell}$ & $\epsilon_{il} \epsilon_{jk} \left( q_i^T C q_j \right) \left( q_k^T C \ell_l \right)$ \\
$Q_{duue}$ & $\left( d^T C u \right) \left( u^T C e \right)$ \\
\hline
\end{tabular}
\end{center}
\end{table}

\section{SMEFT Renormalization Group Equations}
\label{ap:rges}

At leading order, the RGEs governing the energy evolution of the SMEFT
Wilson coefficients $C_i$ can be written as
\begin{equation}
\frac{dC_i}{d \log \mu} = \frac{1}{16 \pi^2} \sum_j \gamma_{ij} C_j \equiv \frac{1}{16 \pi^2} \, \beta_i \, .
\end{equation}
Here $\mu$ is the renormalization scale, $\gamma$ is the anomalous
dimensions matrix and $\beta_i$ are the $\beta$ functions. The
complete anomalous dimension matrix for the dimension-six SMEFT
operators has been recently computed in
\cite{Jenkins:2013zja,Jenkins:2013wua,Alonso:2013hga,Alonso:2014zka}. We
collect here the resulting $\beta$ functions, adapted to our notation
and conventions.

\newpage
First, we give some definitions that turn out to be useful in order to
simplify the $\beta$ functions:
\begin{align}
\eta_1 & = \frac{1}{2} \left[ 3 \, \Tr \left( C_{u \vp} \Gamma_u^\dagger \right) + 3 \, \Tr \left( C_{d \vp} \Gamma_d^\dagger \right) + \, \Tr \left( C_{e \vp} \Gamma_e^\dagger \right) + \cc \right] \, , \\[2mm]
\eta_2 & = - 6 \, \Tr \left( C_{\vp q}^{(3)} \Gamma_u \Gamma_u^\dagger \right) - 6 \, \Tr \left( C_{\vp q}^{(3)} \Gamma_d \Gamma_d^\dagger \right) - 2 \, \Tr \left( C_{\vp \ell}^{(3)} \Gamma_e \Gamma_e^\dagger \right) + 3 \left[ \Tr \left( C_{\vp u d} \Gamma_d^\dagger \Gamma_u \right) + \cc \right] \, , \\[2mm]
\eta_3 & = 3 \, \Tr \left( C_{\vp q}^{(1)} \Gamma_d \Gamma_d^\dagger \right) - 3 \, \Tr \left( C_{\vp q}^{(1)} \Gamma_u \Gamma_u^\dagger \right) + 9 \, \Tr \left( C_{\vp q}^{(3)} \Gamma_d \Gamma_d^\dagger \right) + 9 \, \Tr \left( C_{\vp q}^{(3)} \Gamma_u \Gamma_u^\dagger \right) \nn \\
& + 3 \, \Tr \left( C_{\vp u} \Gamma_u^\dagger \Gamma_u \right) - 3 \, \Tr \left( C_{\vp d} \Gamma_d^\dagger \Gamma_d \right) - 3 \left[ \Tr \left( C_{\vp u d} \Gamma_d^\dagger \Gamma_u \right) + \cc \right] \nn \\
& + \Tr \left( C_{\vp \ell}^{(1)} \Gamma_e \Gamma_e^\dagger \right) + 3 \, \Tr \left( C_{\vp \ell}^{(3)} \Gamma_e \Gamma_e^\dagger \right) - \Tr \left( C_{\vp e} \Gamma_e^\dagger \Gamma_e \right) \, , \\[2mm]
\eta_4 & = 12 \, \Tr \left( C_{\vp q}^{(1)} \Gamma_d \Gamma_d^\dagger \right) - 12 \, \Tr \left( C_{\vp q}^{(1)} \Gamma_u \Gamma_u^\dagger \right) \nn \\
& + 12 \, \Tr \left( C_{\vp u} \Gamma_u^\dagger \Gamma_u \right) - 12 \, \Tr \left( C_{\vp d} \Gamma_d^\dagger \Gamma_d \right) + 6 \left[ \Tr \left( C_{\vp u d} \Gamma_d^\dagger \Gamma_u \right) + \cc \right] \nn \\
& + 4 \, \Tr \left( C_{\vp \ell}^{(1)} \Gamma_e \Gamma_e^\dagger \right) - 4 \, \Tr \left( C_{\vp e} \Gamma_e^\dagger \Gamma_e \right) \, , \\[2mm]
\eta_5 & = \frac{3}{2} i \left[ \, \Tr \left( \Gamma_d C_{d \vp}^\dagger \right) - \cc \right] - \frac{3}{2} i \left[ \, \Tr \left( \Gamma_u C_{u \vp}^\dagger \right) - \cc \right] + \frac{1}{2} i \left[ \, \Tr \left( \Gamma_e C_{e \vp}^\dagger \right) - \cc \right] \, ,
\end{align}
as well as
\begin{align}
\xi_B & = \frac{2}{3} \left( C_{\vp \Box} + C_{\vp D} \right) + \frac{8}{3} \left( - \Tr \, C_{\vp \ell}^{(1)} + \Tr \, C_{\vp q}^{(1)} - \Tr \, C_{\vp e} + 2 \Tr \, C_{\vp u} - \Tr \, C_{\vp d} \right) \, ,
\end{align}
and the wavefunction renormalization terms
\begin{align}
\gamma_H^{(Y)} & = \Tr \left( 3 \Gamma_u \Gamma_u^\dagger + 3 \Gamma_d \Gamma_d^\dagger + \Gamma_e \Gamma_e^\dagger \right) \, ,
\end{align}
and
\begin{align}
\left[ \gamma_q^{(Y)} \right]_{rs} & = \frac{1}{2} \left[ \Gamma_u \Gamma_u^\dagger + \Gamma_d \Gamma_d^\dagger \right]_{rs} \, , \\[2mm]
\left[ \gamma_u^{(Y)} \right]_{rs} & = \left[ \Gamma_u^\dagger \Gamma_u \right]_{rs} \, , \\[2mm]
\left[ \gamma_d^{(Y)} \right]_{rs} & = \left[ \Gamma_d^\dagger \Gamma_d \right]_{rs} \, , \\[2mm]
\left[ \gamma_\ell^{(Y)} \right]_{rs} & = \frac{1}{2} \left[ \Gamma_e \Gamma_e^\dagger \right]_{rs} \, , \\[2mm]
\left[ \gamma_e^{(Y)} \right]_{rs} & = \left[ \Gamma_e^\dagger \Gamma_e \right]_{rs} \, .
\end{align}
Finally we also use the following definitions
\begin{align}
\left[\xi_e\right]_{pt}&= 2\left[C_{\ell e}\right]_{prst}\left[\Gamma_e\right]_{rs}-3\left[C_{\ell edq}\right]_{ptsr}\left[\Gamma_d\right]_{rs}+3\left[C_{lequ}^{(1)}\right]_{ptsr}\left[\Gamma_u\right]^\ast_{sr}\,,\\[2mm]
\left[\xi_d\right]_{pt}&= 2\left[C_{qd}^{(1)}+\frac{4}{3}C_{qd}^{(8)}\right]_{prst}\left[\Gamma_d\right]_{rs}-\left(3\left[C_{quqd}^{(1)}\right]_{srpt}+\frac{1}{2}\left[C_{quqd}^{(1)}+\frac{4}{3}C_{quqd}^{(8)}\right]_{prst}\right)\left[\Gamma_u\right]^\ast_{sr}\nn\\
&\quad-\left[C_{ledq}\right]^\ast_{srtp}\left[\Gamma_e\right]_{sr}\,,\\[2mm]
\left[\xi_u\right]_{pt}&= 2\left[C_{qu}^{(1)}+\frac{4}{3}C_{qu}^{(8)}\right]_{prst}\left[\Gamma_u\right]_{rs}-\left(3\left[C_{quqd}^{(1)}\right]_{ptsr}+\frac{1}{2}\left[C_{quqd}^{(1)}+\frac{4}{3}C_{quqd}^{(8)}\right]_{stpr}\right)\left[\Gamma_d\right]^\ast_{sr}\nn\\
&\quad+\left[C_{lequ}^{(1)}\right]_{srpt}\left[\Gamma_e\right]^\ast_{sr}\,.
\end{align}
  
\subsubsection*{$\boxed{\boldsymbol{X^3}}$}

\begin{align}
\beta_G & = 15 \, \gsc \, C_G  \, , \\[2mm]
\beta_{\widetilde G} & = 15 \, \gsc \, C_{\widetilde G}  \, , \\[2mm]
\beta_W & = \frac{29}{2} \, \gc \, C_W  \, , \\[2mm]
\beta_{\widetilde W} & = \frac{29}{2} \, \gc \, C_{\widetilde W}  \, .
\end{align}

\subsubsection*{$\boxed{\boldsymbol{\vp^6}}$}

\begin{align}
\beta_\vp & = - \frac{9}{2} \left( 3 \gc + \gpc \right) C_\vp + \lambda \left[ \frac{20}{3} \gc C_{\vp \Box} + 3 \left( \gpc - \gc \right) C_{\vp D} \right] - \frac{3}{4} \left( \gc + \gpc \right)^2 C_{\vp D} \nn \\
& + 6 \lambda \left( 3 \gc C_{\vp W} + \gpc C_{\vp B} + g g^\prime C_{\varphi W B} \right) - 3 \left( \gc \gpc + 3 g^4 \right) C_{\vp W} - 3 \left( g^{\prime 4} + \gc \gpc \right) C_{\vp B} \nn \\
& - 3 \left( g g^{\prime 3} + g^3 g^\prime \right) C_{\vp W B} + \frac{8}{3} \lambda \gc \left( \Tr \, C_{\vp \ell}^{(3)} + 3 \, \Tr \, C_{\vp q}^{(3)} \right) + 54 \lambda C_\vp - 40 \lambda^2 C_{\vp \Box} + 12 \lambda^2 C_{\vp D} \nn \\
& + 4 \lambda \left(\eta_1 + \eta_2 \right) - 4 \left[ 3 \, \Tr \left( C_{u \vp} \Gamma_u^\dagger \Gamma_u \Gamma_u^\dagger \right) + 3 \, \Tr \left( C_{d \vp} \Gamma_d^\dagger \Gamma_d \Gamma_d^\dagger \right) + \, \Tr \left( C_{e \vp} \Gamma_e^\dagger \Gamma_e \Gamma_e^\dagger \right) + \cc \right] \nn \\
& + 6 \gamma_H^{(Y)} \, C_\vp \, .
\end{align}

\subsubsection*{$\boxed{\boldsymbol{\vp^4 D^2}}$}

\begin{align}
\beta_{\vp \Box} & = - \left( 4 \gc + \frac{4}{3} \gpc \right) C_{\vp \Box} + \frac{5}{3} \gpc C_{\vp D} + 2 \gc \left( \Tr \, C_{\vp \ell}^{(3)} + 3 \, \Tr \, C_{\vp q}^{(3)} \right) \nn \\
& + \frac{2}{3} \gpc \left( 2 \, \Tr \, C_{\vp u} - \Tr \, C_{\vp d} - \Tr \, C_{\vp e} + \Tr \, C_{\vp q}^{(1)} - \Tr \, C_{\vp \ell}^{(1)} \right) + 12 \lambda \, C_{\vp \Box} -2  \, \eta_3 + 4 \gamma_H^{(Y)} \, C_{\vp \Box} \, , \\[4mm]
\beta_{\vp D} & = \frac{20}{3} \gpc C_{\vp \Box} + \left( \frac{9}{2} \gc - \frac{5}{6} \gpc \right) C_{\vp D} + \frac{8}{3} \gpc \left( 2 \, \Tr \, C_{\vp u} - \Tr \, C_{\vp d} - \Tr \, C_{\vp e} + \Tr \, C_{\vp q}^{(1)} - \Tr \, C_{\vp \ell}^{(1)} \right) \nn \\
& + 6 \lambda \, C_{\vp D} -2  \, \eta_4 + 4 \gamma_H^{(Y)} \, C_{\vp D} \, .
\end{align}

\subsubsection*{$\boxed{\boldsymbol{X^2 \vp^2}}$}

\begin{align}
\beta_{\vp G} & = \left( - \frac{3}{2} \gpc - \frac{9}{2} \gc - 14 \gsc \right) C_{\vp G} + 6 \, \lambda C_{\vp G} \nn \\
& - 2 g_s \left[ \Tr \left( C_{u G} \Gamma_u^\dagger \right) + \Tr \left( C_{d G} \Gamma_d^\dagger \right) + \cc \right] + 2 \gamma_H^{(Y)} C_{\vp G} \, , \\[3mm]
\beta_{\vp B} & = \left( \frac{85}{6} \gpc - \frac{9}{2} \gc \right) C_{\vp B} + 3 g g^\prime C_{\vp W B} + 6 \, \lambda C_{\vp B} \nn \\
& + g^\prime \left[ - 5 \, \Tr \left( C_{u B} \Gamma_u^\dagger \right) + \Tr \left( C_{d B} \Gamma_d^\dagger \right) + 3 \, \Tr \left( C_{e B} \Gamma_e^\dagger \right) + \cc \right] + 2 \gamma_H^{(Y)} C_{\vp B} \, , \\[3mm]
\beta_{\vp W} & = \left( - \frac{3}{2} \gpc - \frac{53}{6} \gc \right) C_{\vp W} + g g^\prime C_{\vp W B} - 15 g^3 C_W + 6 \, \lambda C_{\vp W} \nn \\
& - g \left[ 3 \, \Tr \left( C_{u W} \Gamma_u^\dagger \right) + 3 \, \Tr \left( C_{d W} \Gamma_d^\dagger \right) + \Tr \left( C_{e W} \Gamma_e^\dagger \right) + \cc \right] + 2 \gamma_H^{(Y)} C_{\vp W} \, , \\[3mm]
\beta_{\vp W B} & = \left( \frac{19}{3} \gpc + \frac{4}{3} \gc \right) C_{\vp W B} + 2 g g^\prime \left( C_{\vp B} + C_{\vp W} \right) + 3 g^2 g^\prime C_W + 2 \, \lambda C_{\vp W B} \nn \\
& + g \left[ 3 \, \Tr \left( C_{u B} \Gamma_u^\dagger \right) - 3 \, \Tr \left( C_{d B} \Gamma_d^\dagger \right) - \Tr \left( C_{e B} \Gamma_e^\dagger \right) + \cc \right] \nn \\
& + g^\prime \left[ 5 \, \Tr \left( C_{u W} \Gamma_u^\dagger \right) + \Tr \left( C_{d W} \Gamma_d^\dagger \right) + 3 \, \Tr \left( C_{e W} \Gamma_e^\dagger \right) + \cc \right] + 2 \gamma_H^{(Y)} C_{\vp W B} \, , \\[3mm]
\beta_{\vp \widetilde G} & = \left( - \frac{3}{2} \gpc - \frac{9}{2} \gc - 14 \gsc \right) C_{\vp \widetilde  G} + 6 \, \lambda C_{\vp \widetilde G} \nn \\
& + 2 i g_s \left[ \Tr \left( C_{u G} \Gamma_u^\dagger \right) + \Tr \left( C_{d G} \Gamma_d^\dagger \right) - \cc \right] + 2 \gamma_H^{(Y)} C_{\vp \widetilde G} \, , \\[3mm]
\beta_{\vp \widetilde  B} & = \left( \frac{85}{6} \gpc - \frac{9}{2} \gc \right) C_{\vp \widetilde  B} + 3 g g^\prime C_{\vp \widetilde  W B} + 6 \, \lambda C_{\vp \widetilde B} \nn \\
& - i g^\prime \left[ - 5 \, \Tr \left( C_{u B} \Gamma_u^\dagger \right) + \Tr \left( C_{d B} \Gamma_d^\dagger \right) + 3 \, \Tr \left( C_{e B} \Gamma_e^\dagger \right) - \cc \right] + 2 \gamma_H^{(Y)} C_{\vp \widetilde B} \, , \\[3mm]
\beta_{\vp \widetilde  W} & = \left( - \frac{3}{2} \gpc - \frac{53}{6} \gc \right) C_{\vp \widetilde  W} + g g^\prime C_{\vp \widetilde  W B} - 15 g^3 C_{\widetilde  W} + 6 \, \lambda C_{\vp \widetilde W} \nn \\
& + i g \left[ 3 \, \Tr \left( C_{u W} \Gamma_u^\dagger \right) + 3 \, \Tr \left( C_{d W} \Gamma_d^\dagger \right) + \Tr \left( C_{e W} \Gamma_e^\dagger \right) - \cc \right] + 2 \gamma_H^{(Y)} C_{\vp \widetilde W} \, , \\[3mm]
\beta_{\vp \widetilde W B} & = \left( \frac{19}{3} \gpc + \frac{4}{3} \gc \right) C_{\vp \widetilde W B} + 2 g g^\prime \left( C_{\vp \widetilde  B} + C_{\vp \widetilde  W} \right) + 3 g^2 g^\prime C_{\widetilde W} + 2 \, \lambda C_{\vp \widetilde W B} \nn \\
& - i g \left[ 3 \, \Tr \left( C_{u B} \Gamma_u^\dagger \right) - 3 \, \Tr \left( C_{d B} \Gamma_d^\dagger \right) - \Tr \left( C_{e B} \Gamma_e^\dagger \right) - \cc \right] \nn \\
& - i g^\prime \left[ 5 \, \Tr \left( C_{u W} \Gamma_u^\dagger \right) + \Tr \left( C_{d W} \Gamma_d^\dagger \right) + 3 \, \Tr \left( C_{e W} \Gamma_e^\dagger \right) - \cc \right] + 2 \gamma_H^{(Y)} C_{\vp \widetilde W B} \, .
\end{align}

\subsubsection*{$\boxed{\boldsymbol{\psi^2 \vp^3}}$}

\begin{align}
\left[ \beta_{u \vp} \right]_{rs} & = \left[ \frac{10}{3} \gc C_{\vp \Box} + \frac{3}{2} \left( \gpc - \gc \right) C_{\vp D} + 32 \gsc \left( C_{\vp G} + i C_{\vp \widetilde G} \right) + 9 \gc \left( C_{\vp W} + i C_{\vp \widetilde W} \right) \right. \nn \\
& \left. + \frac{17}{3} \gpc \left( C_{\vp B} + i C_{\vp \widetilde B} \right) - g g^\prime \left( C_{\vp W B} + i C_{\vp \widetilde W B} \right) + \frac{4}{3} \gc \left( \Tr \, C_{\vp \ell}^{(3)} + 3 \, \Tr \, C_{\vp q}^{(3)} \right) \right] \left[ \Gamma_u \right]_{rs} \nn \\
& - \left( \frac{35}{12} \gpc + \frac{27}{4} \gc + 8 \gsc \right) \left[ C_{u \vp} \right]_{rs} - g^\prime \left( 5 \gpc - 3 \gc \right) \left[ C_{u B} \right]_{rs} + g \left( 5 \gpc - 9 \gc \right) \left[ C_{u W} \right]_{rs} \nn \\
& - \left( 3 \gc - \gpc \right) \left[ \Gamma_u C_{\vp u} \right]_{rs} + 3 \gc \left[ \Gamma_d C_{\vp u d}^\dagger \right]_{rs} + 4 \gpc \left[ C_{\vp q}^{(1)} \Gamma_u \right]_{rs} - 4 \gpc \left[ C_{\vp q}^{(3)} \Gamma_u \right]_{rs} \nn \\
& - 5 g^\prime \left[ C_{u B} \Gamma_u^\dagger \Gamma_u + \Gamma_u \Gamma_u^\dagger C_{u B} \right]_{rs} - 3 g \left[ C_{u W} \Gamma_u^\dagger \Gamma_u - \Gamma_u \Gamma_u^\dagger C_{u W} \right]_{rs} \nn \\
& - 16 g_s \left[ C_{u G} \Gamma_u^\dagger \Gamma_u + \Gamma_u \Gamma_u^\dagger C_{u G} \right]_{rs} - 12 g \left[ \Gamma_d \Gamma_d^\dagger C_{u W} \right]_{rs} - 6 g \left[ C_{d W} \Gamma_d^\dagger \Gamma_u \right]_{rs} \nn \\
& + \lambda \, \Big( 12 \left[ C_{u \vp} \right]_{rs} - 2 \left[ C_{\vp q}^{(1)} \Gamma_u \right]_{rs} + 6 \left[ C_{\vp q}^{(3)} \Gamma_u \right]_{rs} + 2 \left[ \Gamma_u C_{\vp u} \right]_{rs} - 2 \left[ \Gamma_d C_{\vp u d}^\dagger \right]_{rs} \nn \\
& - 2 C_{\vp \Box} \left[ \Gamma_u \right]_{rs} + C_{\vp D} \left[ \Gamma_u \right]_{rs} - 4 \left[ C_{qu}^{(1)} \right]_{rpts} \left[ \Gamma_u \right]_{pt} - \frac{16}{3} \left[ C_{qu}^{(8)} \right]_{rpts} \left[ \Gamma_u \right]_{pt} \nn \\
& - 2 \left[ C_{\ell e q u}^{(1)} \right]_{ptrs} \left[ \Gamma_e \right]_{pt}^\ast + 6 \left[ C_{quqd}^{(1)} \right]_{rspt} \left[ \Gamma_d \right]_{pt}^\ast + \left[ C_{quqd}^{(1)} \right]_{psrt} \left[ \Gamma_d \right]_{pt}^\ast + \frac{4}{3} \left[ C_{quqd}^{(8)} \right]_{psrt} \left[ \Gamma_d \right]_{pt}^\ast \Big) \nn \\
& + 2 \left( \eta_1 + \eta_2 - i \eta_5 \right) \left[ \Gamma_u \right]_{rs} + \left( C_{\vp D} - 6 \, C_{\vp \Box} \right) \left[ \Gamma_u \Gamma_u^\dagger \Gamma_u \right]_{rs} - 2 \left[ C_{\vp q}^{(1)} \Gamma_u \Gamma_u^\dagger \Gamma_u \right]_{rs} \nn \\
& + 6 \left[ C_{\vp q}^{(3)} \Gamma_d \Gamma_d^\dagger \Gamma_u \right]_{rs} + 2 \left[ \Gamma_u \Gamma_u^\dagger \Gamma_u C_{\vp u} \right]_{rs} - 2 \left[ \Gamma_d \Gamma_d^\dagger \Gamma_d C_{\vp u d}^\dagger \right]_{rs} \nn \\
& + 8 \left[ C_{q u}^{(1)} + \frac{4}{3} C_{q u}^{(8)} \right]_{rpts} \left[ \Gamma_u \Gamma_u^\dagger \Gamma_u \right]_{pt} - 2 \left[ C_{q u q d}^{(1)} + \frac{4}{3} C_{q u q d}^{(8)} \right]_{tsrp} \left[ \Gamma_d^\dagger \Gamma_d \Gamma_d^\dagger \right]_{pt} \nn \\
& - 12 \left[ C_{q u q d}^{(1)} \right]_{rstp} \left[ \Gamma_d^\dagger \Gamma_d \Gamma_d^\dagger \right]_{pt} + 4 \left[ C_{\ell e q u}^{(1)} \right]_{tprs} \left[ \Gamma_e \Gamma_e^\dagger \Gamma_e \right]_{pt} + 4 \left[ C_{u \vp} \Gamma_u^\dagger \Gamma_u \right]_{rs} \nn \\
& + 5 \left[ \Gamma_u \Gamma_u^\dagger C_{u \vp} \right]_{rs} - 2 \left[ \Gamma_d C_{d \vp}^\dagger \Gamma_u \right]_{rs} - \left[ C_{d \vp} \Gamma_d^\dagger \Gamma_u \right]_{rs} - 2 \left[ \Gamma_d \Gamma_d^\dagger C_{u \vp} \right]_{rs} \nn \\
& + 3 \gamma_H^{(Y)} \left[ C_{u \vp} \right]_{rs} + \left[ \gamma_q^{(Y)} C_{u \vp} \right]_{rs} + \left[ C_{u \vp} \gamma_u^{(Y)} \right]_{rs} \, , \\[4mm]
\left[ \beta_{d \vp} \right]_{rs} & = \left[ \frac{10}{3} \gc C_{\vp \Box} + \frac{3}{2} \left( \gpc - \gc \right) C_{\vp D} + 32 \gsc \left( C_{\vp G} + i C_{\vp \widetilde G} \right) + 9 \gc \left( C_{\vp W} + i C_{\vp \widetilde W} \right) \right. \nn \\
& \left. + \frac{5}{3} \gpc \left( C_{\vp B} + i C_{\vp \widetilde B} \right) + g g^\prime \left( C_{\vp W B} + i C_{\vp \widetilde W B} \right) + \frac{4}{3} \gc \left( \Tr \, C_{\vp \ell}^{(3)} + 3 \, \Tr \, C_{\vp q}^{(3)} \right) \right] \left[ \Gamma_d \right]_{rs} \nn \\
& - \left( \frac{23}{12} \gpc + \frac{27}{4} \gc + 8 \gsc \right) \left[ C_{d \vp} \right]_{rs} - g^\prime \left( 3 \gc - \gpc \right) \left[ C_{d B} \right]_{rs} - g \left( 9 \gc - \gpc \right) \left[ C_{d W} \right]_{rs} \nn \\
& + \left( 3 \gc + \gpc \right) \left[ \Gamma_d C_{\vp d} \right]_{rs} + 3 \gc \left[ \Gamma_u C_{\vp u d} \right]_{rs} - 2 \gpc \left[ C_{\vp q}^{(1)} \Gamma_d \right]_{rs} - 2 \gpc \left[ C_{\vp q}^{(3)} \Gamma_d \right]_{rs} \nn \\
& + g^\prime \left[ C_{d B} \Gamma_d^\dagger \Gamma_d + \Gamma_d \Gamma_d^\dagger C_{d B} \right]_{rs} - 3 g \left[ C_{d W} \Gamma_d^\dagger \Gamma_d - \Gamma_d \Gamma_d^\dagger C_{d W} \right]_{rs} \nn \\
& - 16 g_s \left[ C_{d G} \Gamma_d^\dagger \Gamma_d + \Gamma_d \Gamma_d^\dagger C_{d G} \right]_{rs} - 12 g \left[ \Gamma_u \Gamma_u^\dagger C_{d W} \right]_{rs} - 6 g \left[ C_{u W} \Gamma_u^\dagger \Gamma_d \right]_{rs} \nn \\
& + \lambda \, \Big( 12 \left[ C_{d \vp} \right]_{rs} + 2 \left[ C_{\vp q}^{(1)} \Gamma_d \right]_{rs} + 6 \left[ C_{\vp q}^{(3)} \Gamma_d \right]_{rs} - 2 \left[ \Gamma_d C_{\vp d} \right]_{rs} - 2 \left[ \Gamma_u C_{\vp u d} \right]_{rs} \nn \\
& - 2 C_{\vp \Box} \left[ \Gamma_d \right]_{rs} + C_{\vp D} \left[ \Gamma_d \right]_{rs} - 4 \left[ C_{qd}^{(1)} \right]_{rpts} \left[ \Gamma_d \right]_{pt} - \frac{16}{3} \left[ C_{qd}^{(8)} \right]_{rpts} \left[ \Gamma_d \right]_{pt} \nn \\
& + 2 \left[ C_{\ell e d q}^\ast \right]_{ptsr} \left[ \Gamma_e \right]_{tp}^\ast + 6 \left[ C_{quqd}^{(1)} \right]_{ptrs} \left[ \Gamma_u \right]_{pt}^\ast + \left[ C_{quqd}^{(1)} \right]_{rtps} \left[ \Gamma_u \right]_{pt}^\ast + \frac{4}{3} \left[ C_{quqd}^{(8)} \right]_{rtps} \left[ \Gamma_u \right]_{pt}^\ast \Big) \nn \\
& + 2 \left( \eta_1 + \eta_2 + i \eta_5 \right) \left[ \Gamma_d \right]_{rs} + \left( C_{\vp D} - 6 \, C_{\vp \Box} \right) \left[ \Gamma_d \Gamma_d^\dagger \Gamma_d \right]_{rs} + 2 \left[ C_{\vp q}^{(1)} \Gamma_d \Gamma_d^\dagger \Gamma_d \right]_{rs} \nn \\
& + 6 \left[ C_{\vp q}^{(3)} \Gamma_u \Gamma_u^\dagger \Gamma_d \right]_{rs} - 2 \left[ \Gamma_d \Gamma_d^\dagger \Gamma_d C_{\vp d} \right]_{rs} - 2 \left[ \Gamma_u \Gamma_u^\dagger \Gamma_u C_{\vp u d} \right]_{rs} \nn \\
& + 8 \left[ C_{q d}^{(1)} + \frac{4}{3} C_{q d}^{(8)} \right]_{rpts} \left[ \Gamma_d \Gamma_d^\dagger \Gamma_d \right]_{pt} - 2 \left[ C_{q u q d}^{(1)} + \frac{4}{3} C_{q u q d}^{(8)} \right]_{rpts} \left[ \Gamma_u^\dagger \Gamma_u \Gamma_u^\dagger \right]_{pt} \nn \\
& - 12 \left[ C_{q u q d}^{(1)} \right]_{tprs} \left[ \Gamma_u^\dagger \Gamma_u \Gamma_u^\dagger \right]_{pt} - 4 \left[ C_{\ell e d q}^\ast \right]_{ptsr} \left[ \Gamma_e \Gamma_e^\dagger \Gamma_e \right]_{pt} + 4 \left[ C_{d \vp} \Gamma_d^\dagger \Gamma_d \right]_{rs} \nn \\
& + 5 \left[ \Gamma_d \Gamma_d^\dagger C_{d \vp} \right]_{rs} - 2 \left[ \Gamma_u C_{u \vp}^\dagger \Gamma_d \right]_{rs} - \left[ C_{u \vp} \Gamma_u^\dagger \Gamma_d \right]_{rs} - 2 \left[ \Gamma_u \Gamma_u^\dagger C_{d \vp} \right]_{rs} \nn \\
& + 3 \gamma_H^{(Y)} \left[ C_{d \vp} \right]_{rs} + \left[ \gamma_q^{(Y)} C_{d \vp} \right]_{rs} + \left[ C_{d \vp} \gamma_d^{(Y)} \right]_{rs} \, ,\\[4mm]
\left[ \beta_{e \vp} \right]_{rs} & = \left[ \frac{10}{3} \gc C_{\vp \Box} + \frac{3}{2} \left( \gpc - \gc \right) C_{\vp D} + 9 \gc \left( C_{\vp W} + i C_{\vp \widetilde W} \right) + 15 \gpc \left( C_{\vp B} + i C_{\vp \widetilde B} \right) \right. \nn \\
& \left. - 3 g g^\prime \left( C_{\vp W B} + i C_{\vp \widetilde W B} \right) + \frac{4}{3} \gc \left( \Tr \, C_{\vp \ell}^{(3)} + 3 \, \Tr \, C_{\vp q}^{(3)} \right) \right] \left[ \Gamma_e \right]_{rs} - \frac{3}{4} \left( 7 \gpc + 9 \gc \right) \left[ C_{e \vp} \right]_{rs} \nn \\
& - 3 g^\prime \left( \gc - 3 \gpc \right) \left[ C_{e B} \right]_{rs} - 9 g \left( \gc - \gpc \right) \left[ C_{e W} \right]_{rs} + 3 \left( \gc - \gpc \right) \left[ \Gamma_e C_{\vp e} \right]_{rs} - 6 \gpc \left[ C_{\vp \ell}^{(1)} \Gamma_e \right]_{rs} \nn \\
& - 6 \gpc \left[ C_{\vp \ell}^{(3)} \Gamma_e \right]_{rs} + 9 g^\prime \left[ C_{e B} \Gamma_e^\dagger \Gamma_e + \Gamma_e \Gamma_e^\dagger C_{e B} \right]_{rs} - 3 g \left[ C_{e W} \Gamma_e^\dagger \Gamma_e - \Gamma_e \Gamma_e^\dagger C_{e W} \right]_{rs} \nn \\
& + \lambda \, \Big( 12 \left[ C_{e \vp} \right]_{rs} + 2 \left[ C_{\vp \ell}^{(1)} \Gamma_e \right]_{rs} + 6 \left[ C_{\vp \ell}^{(3)} \Gamma_e \right]_{rs} - 2 \left[ \Gamma_e C_{\vp e} \right]_{rs} - 2 C_{\vp \Box} \left[ \Gamma_e \right]_{rs} + C_{\vp D} \left[ \Gamma_e \right]_{rs} \nn \\
& - 4 \left[ C_{\ell e} \right]_{rpts} \left[ \Gamma_e \right]_{pt} + 6 \left[ C_{\ell e d q} \right]_{rspt} \left[ \Gamma_d \right]_{pt} - 6 \left[ C_{\ell e q u}^{(1)} \right]_{rspt} \left[ \Gamma_u \right]_{pt}^\ast \Big) + 2 \left( \eta_1 + \eta_2 + i \eta_5 \right) \left[ \Gamma_e \right]_{rs} \nn \\
& + \left( C_{\vp D} - 6 \, C_{\vp \Box} \right) \left[ \Gamma_e \Gamma_e^\dagger \Gamma_e \right]_{rs} + 2 \left[ C_{\vp \ell}^{(1)} \Gamma_e \Gamma_e^\dagger \Gamma_e \right]_{rs} - 2 \left[ \Gamma_e \Gamma_e^\dagger \Gamma_e C_{\vp e} \right]_{rs} \nn \\
& + 8 \left[ C_{\ell e} \right]_{rpts} \left[ \Gamma_e \Gamma_e^\dagger \Gamma_e \right]_{pt} - 12 \left[ C_{\ell e d q} \right]_{rspt} \left[ \Gamma_d \Gamma_d^\dagger \Gamma_d \right]_{tp} + 12 \left[ C_{\ell e q u}^{(1)} \right]_{rstp} \left[ \Gamma_u^\dagger \Gamma_u \Gamma_u^\dagger \right]_{pt} \nn \\
& + 4 \left[ C_{e \vp} \Gamma_e^\dagger \Gamma_e \right]_{rs} + 5 \left[ \Gamma_e \Gamma_e^\dagger C_{e \vp} \right]_{rs} + 3 \gamma_H^{(Y)} \left[ C_{e \vp} \right]_{rs} + \left[ \gamma_\ell^{(Y)} C_{e \vp} \right]_{rs} + \left[ C_{e \vp} \gamma_e^{(Y)} \right]_{rs} \, .
\end{align}

\subsubsection*{$\boxed{\boldsymbol{\psi^2 X \vp}}$}

\begin{align}
\left[ \beta_{e W} \right]_{rs} & = \frac{1}{12} \left( 3 \gpc - 11 \gc \right) \left[ C_{e W} \right]_{rs} - \frac{1}{2} g g^\prime \left[ C_{e B} \right]_{rs} \nn \\
& - \left[ g \left( C_{\vp W} + i C_{\vp \widetilde W} \right) - \frac{3}{2} g^\prime \left( C_{\vp W B} + i C_{\vp \widetilde W B} \right) \right] \left[ \Gamma_e \right]_{rs} \nn \\
& - 6 g \left[ C_{\ell e q u}^{(3)} \right]_{rspt} \left[ \Gamma_u \right]_{pt}^\ast + \left[ C_{e W} \Gamma_e^\dagger \Gamma_e \right]_{rs} + \gamma_H^{(Y)} \left[ C_{e W} \right]_{rs} + \left[ \gamma_\ell^{(Y)} C_{e W} \right]_{rs} + \left[ C_{eW} \gamma_e^{(Y)} \right]_{rs} \, , \\[4mm]
\left[ \beta_{e B} \right]_{rs} & = \frac{1}{4} \left( \frac{151}{3} \gpc - 9 \gc \right) \left[ C_{e B} \right]_{rs} - \frac{3}{2} g g^\prime \left[ C_{e W} \right]_{rs} \nn \\
& - \left[ \frac{3}{2} g \left( C_{\vp W B} + i C_{\vp \widetilde W B} \right) - 3 g^\prime \left( C_{\vp B} + i C_{\vp \widetilde B} \right) \right] \left[ \Gamma_e \right]_{rs} \nn \\
& + 10 g^\prime \left[ C_{\ell e q u}^{(3)} \right]_{rspt} \left[ \Gamma_u \right]_{pt}^\ast + \left[ C_{e B} \Gamma_e^\dagger \Gamma_e \right]_{rs} + 2 \left[ \Gamma_e \Gamma_e^\dagger C_{eB} \right]_{rs} \nn \\
& + \gamma_H^{(Y)} \left[ C_{e B} \right]_{rs} + \left[ \gamma_\ell^{(Y)} C_{e B} \right]_{rs} + \left[ C_{eB} \gamma_e^{(Y)} \right]_{rs} \, , \\[4mm]
\left[ \beta_{u G} \right]_{rs} & = - \frac{1}{36} \left( 81 \gc + 19 \gpc + 204 \gsc \right) \left[ C_{u G} \right]_{rs} + 6 \, g g_s \left[ C_{u W} \right]_{rs} + \frac{10}{3} g^\prime g_s \left[ C_{u B} \right]_{rs} \nn \\
& - g_s \left[ 4 \left( C_{\vp G} + i C_{\vp \widetilde G} \right) + 9 g_s \left( C_{G} + i C_{\widetilde G} \right) \right] \left[ \Gamma_u \right]_{rs} \nn \\
& - g_s \left( \left[ C_{quqd}^{(1)} \right]_{psrt} - \frac{1}{6} \left[ C_{quqd}^{(8)} \right]_{psrt} \right) \left[ \Gamma_d \right]_{pt}^\ast + 2 \left[ \Gamma_u \Gamma_u^\dagger C_{uG} \right]_{rs} - 2 \left[ \Gamma_d \Gamma_d^\dagger C_{uG} \right]_{rs} \nn \\
& - \left[ C_{dG} \Gamma_d^\dagger \Gamma_u \right]_{rs} + \left[ C_{uG} \Gamma_u^\dagger \Gamma_u \right]_{rs} + \gamma_H^{(Y)} \left[ C_{u G} \right]_{rs} + \left[ \gamma_q^{(Y)} C_{u G} \right]_{rs} + \left[ C_{u G} \gamma_u^{(Y)} \right]_{rs} \, , \\[4mm]
\left[ \beta_{u W} \right]_{rs} & = - \frac{1}{36} \left( 33 \gc + 19 \gpc - 96 \gsc \right) \left[ C_{u W} \right]_{rs} + \frac{8}{3} \, g g_s \left[ C_{u G} \right]_{rs} - \frac{1}{6} g g^\prime \left[ C_{u B} \right]_{rs} \nn \\
& - \left[ g \left( C_{\vp W} + i C_{\vp \widetilde W} \right) - \frac{5}{6} g^\prime \left( C_{\vp W B} + i C_{\vp \widetilde W B} \right) \right] \left[ \Gamma_u \right]_{rs} \nn \\
& + \frac{g}{4} \left( \left[ C_{quqd}^{(1)} \right]_{psrt} + \frac{4}{3} \left[ C_{quqd}^{(8)} \right]_{psrt} \right) \left[ \Gamma_d \right]_{pt}^\ast - 2 g \left[ C_{\ell e q u}^{(3)} \right]_{ptrs} \left[ \Gamma_e \right]_{tp}^\ast + 2 \left[ \Gamma_d \Gamma_d^\dagger C_{u W} \right]_{rs} \nn \\
& - \left[ C_{d W} \Gamma_d^\dagger \Gamma_u \right]_{rs} + \left[ C_{u W} \Gamma_u^\dagger \Gamma_u \right]_{rs} + \gamma_H^{(Y)} \left[ C_{u W} \right]_{rs} + \left[ \gamma_q^{(Y)} C_{u W} \right]_{rs} + \left[ C_{u W} \gamma_u^{(Y)} \right]_{rs} \, , \\[4mm]
\left[ \beta_{u B} \right]_{rs} & = - \frac{1}{36} \left( 81 \gc - 313 \gpc - 96 \gsc \right) \left[ C_{u B} \right]_{rs} + \frac{40}{9} \, g^\prime g_s \left[ C_{u G} \right]_{rs} - \frac{1}{2} g g^\prime \left[ C_{u W} \right]_{rs} \nn \\
& - \left[ - \frac{3}{2} g \left( C_{\vp W B} + i C_{\vp \widetilde W B} \right) + \frac{5}{3} g^\prime \left( C_{\vp B} + i C_{\vp \widetilde B} \right) \right] \left[ \Gamma_u \right]_{rs} \nn \\
& + \frac{g^\prime}{12} \left( \left[ C_{quqd}^{(1)} \right]_{psrt} + \frac{4}{3} \left[ C_{quqd}^{(8)} \right]_{psrt} \right) \left[ \Gamma_d \right]_{pt}^\ast - 6 g^\prime \left[ C_{\ell e q u}^{(3)} \right]_{ptrs} \left[ \Gamma_e \right]_{tp}^\ast + 2 \left[ \Gamma_u \Gamma_u^\dagger C_{u B} \right]_{rs} \nn \\
& - 2 \left[ \Gamma_d \Gamma_d^\dagger C_{u B} \right]_{rs} - \left[ C_{d B} \Gamma_d^\dagger \Gamma_u \right]_{rs} + \left[ C_{u B} \Gamma_u^\dagger \Gamma_u \right]_{rs} + \gamma_H^{(Y)} \left[ C_{u B} \right]_{rs} \nn \\
& + \left[ \gamma_q^{(Y)} C_{u B} \right]_{rs} + \left[ C_{u B} \gamma_u^{(Y)} \right]_{rs} \, ,\\[4mm]
\left[ \beta_{d G} \right]_{rs} & = - \frac{1}{36} \left( 81 \gc + 31 \gpc + 204 \gsc \right) \left[ C_{d G} \right]_{rs} + 6 \, g g_s \left[ C_{d W} \right]_{rs} - \frac{2}{3} g^\prime g_s \left[ C_{d B} \right]_{rs} \nn \\
& - g_s \left[ 4 \left( C_{\vp G} + i C_{\vp \widetilde G} \right) + 9 g_s \left( C_{G} + i C_{\widetilde G} \right) \right] \left[ \Gamma_d \right]_{rs} \nn \\
& - g_s \left( \left[ C_{quqd}^{(1)} \right]_{rtps} - \frac{1}{6} \left[ C_{quqd}^{(8)} \right]_{rtps} \right) \left[ \Gamma_u \right]_{pt}^\ast - 2 \left[ \Gamma_u \Gamma_u^\dagger C_{dG} \right]_{rs} + 2 \left[ \Gamma_d \Gamma_d^\dagger C_{dG} \right]_{rs} \nn \\
& - \left[ C_{uG} \Gamma_u^\dagger \Gamma_d \right]_{rs} + \left[ C_{dG} \Gamma_d^\dagger \Gamma_d \right]_{rs} + \gamma_H^{(Y)} \left[ C_{d G} \right]_{rs} + \left[ \gamma_q^{(Y)} C_{d G} \right]_{rs} + \left[ C_{d G} \gamma_d^{(Y)} \right]_{rs} \, , \\[4mm]
\left[ \beta_{d W} \right]_{rs} & = - \frac{1}{36} \left( 33 \gc + 31 \gpc - 96 \gsc \right) \left[ C_{d W} \right]_{rs} + \frac{8}{3} \, g g_s \left[ C_{d G} \right]_{rs} + \frac{5}{6} g g^\prime \left[ C_{d B} \right]_{rs} \nn \\
& - \left[ g \left( C_{\vp W} + i C_{\vp \widetilde W} \right) - \frac{1}{6} g^\prime \left( C_{\vp W B} + i C_{\vp \widetilde W B} \right) \right] \left[ \Gamma_d \right]_{rs} \nn \\
& + \frac{g}{4} \left( \left[ C_{quqd}^{(1)} \right]_{rtps} + \frac{4}{3} \left[ C_{quqd}^{(8)} \right]_{rtps} \right) \left[ \Gamma_u \right]_{pt}^\ast + 2 \left[ \Gamma_u \Gamma_u^\dagger C_{d W} \right]_{rs} \nn \\
& - \left[ C_{u W} \Gamma_u^\dagger \Gamma_d \right]_{rs} + \left[ C_{d W} \Gamma_d^\dagger \Gamma_d \right]_{rs} + \gamma_H^{(Y)} \left[ C_{d W} \right]_{rs} + \left[ \gamma_q^{(Y)} C_{d W} \right]_{rs} + \left[ C_{d W} \gamma_d^{(Y)} \right]_{rs} \, , \\[4mm]
\left[ \beta_{d B} \right]_{rs} & = - \frac{1}{36} \left( 81 \gc - 253 \gpc - 96 \gsc \right) \left[ C_{d B} \right]_{rs} - \frac{8}{9} \, g^\prime g_s \left[ C_{d G} \right]_{rs} + \frac{5}{2} g g^\prime \left[ C_{d W} \right]_{rs} \nn \\
& - \left[ \frac{3}{2} g \left( C_{\vp W B} + i C_{\vp \widetilde W B} \right) - \frac{1}{3} g^\prime \left( C_{\vp B} + i C_{\vp \widetilde B} \right) \right] \left[ \Gamma_d \right]_{rs} \nn \\
& - \frac{5 g^\prime}{12} \left( \left[ C_{quqd}^{(1)} \right]_{rtps} + \frac{4}{3} \left[ C_{quqd}^{(8)} \right]_{rtps} \right) \left[ \Gamma_u \right]_{pt}^\ast - 2 \left[ \Gamma_u \Gamma_u^\dagger C_{d B} \right]_{rs} + 2 \left[ \Gamma_d \Gamma_d^\dagger C_{d B} \right]_{rs} \nn \\
& - \left[ C_{u B} \Gamma_u^\dagger \Gamma_d \right]_{rs} + \left[ C_{d B} \Gamma_d^\dagger \Gamma_d \right]_{rs} + \gamma_H^{(Y)} \left[ C_{d B} \right]_{rs}  + \left[ \gamma_q^{(Y)} C_{d B} \right]_{rs} + \left[ C_{d B} \gamma_d^{(Y)} \right]_{rs} \, .
\end{align}

\subsubsection*{$\boxed{\boldsymbol{\psi^2 \vp^2 D}}$}

\begin{align}
\left[ \beta_{\vp \ell}^{(1)} \right]_{rs} & = - \frac{1}{4} \xi_B \gpc \delta_{rs} + \frac{1}{3} \gpc \left[ C_{\vp \ell}^{(1)} \right]_{rs} - \frac{2}{3} \gpc \Big( \left[ C_{\ell d} \right]_{rstt} + \left[ C_{\ell e} \right]_{rstt} + \left[ C_{\ell \ell} \right]_{rstt} + \frac{1}{2} \left[ C_{\ell \ell} \right]_{rtts} \nn \\
& + \frac{1}{2} \left[ C_{\ell \ell} \right]_{tsrt} + \left[ C_{\ell \ell} \right]_{ttrs} - \left[ C_{\ell q}^{(1)} \right]_{rstt} - 2 \left[ C_{\ell u} \right]_{rstt} \Big) - \frac{1}{2} \left( C_{\vp \Box} + C_{\vp D} \right) \left[ \Gamma_e \Gamma_e^\dagger \right]_{rs} \nn \\
& - \left[ \Gamma_e C_{\vp e} \Gamma_e^\dagger \right]_{rs} + \frac{3}{2} \left( \left[ \Gamma_e \Gamma_e^\dagger C_{\vp \ell}^{(1)} \right]_{rs} + \left[ C_{\vp \ell}^{(1)} \Gamma_e \Gamma_e^\dagger \right]_{rs} + 3 \left[ \Gamma_e \Gamma_e^\dagger C_{\vp \ell}^{(3)} \right]_{rs} + 3 \left[ C_{\vp \ell}^{(3)} \Gamma_e \Gamma_e^\dagger \right]_{rs} \right) \nn \\
& + 2 \left[ C_{\ell e} \right]_{rspt} \left[ \Gamma_e^\dagger \Gamma_e \right]_{tp} - \left( 2 \left[ C_{\ell \ell} \right]_{rspt} + 2 \left[ C_{\ell \ell} \right]_{ptrs} + \left[ C_{\ell \ell} \right]_{rtps} + \left[ C_{\ell \ell} \right]_{psrt} \right) \left[ \Gamma_e \Gamma_e^\dagger \right]_{tp} \nn \\
& - 6 \left[ C_{\ell q}^{(1)} \right]_{rspt} \left[ \Gamma_d \Gamma_d^\dagger \right]_{tp} + 6 \left[ C_{\ell q}^{(1)} \right]_{rspt} \left[ \Gamma_u \Gamma_u^\dagger \right]_{tp} - 6 \left[ C_{\ell u} \right]_{rspt} \left[ \Gamma_u^\dagger \Gamma_u \right]_{tp} \nn \\
& + 6 \left[ C_{\ell d} \right]_{rspt} \left[ \Gamma_d^\dagger \Gamma_d \right]_{tp} + 2 \gamma_H^{(Y)} \left[ C_{\vp \ell}^{(1)} \right]_{rs} + \left[ \gamma_\ell^{(Y)} C_{\vp \ell}^{(1)} \right]_{rs} + \left[ C_{\vp \ell}^{(1)} \gamma_\ell^{(Y)} \right]_{rs} \, , \\[4mm]
\left[ \beta_{\vp \ell}^{(3)} \right]_{rs} & = \frac{2}{3} \gc \left( \frac{1}{4} C_{\vp \Box} + \Tr \, C_{\vp \ell}^{(3)} + 3\, \Tr \, C_{\vp q}^{(3)} \right) \delta_{rs} - \frac{17}{3} \gc \left[ C_{\vp \ell}^{(3)} \right]_{rs} + \frac{1}{3} \gc \Big( \left[ C_{\ell \ell} \right]_{rtts} + \left[ C_{\ell \ell} \right]_{tsrt} \Big) \nn \\
& + 2 \gc \left[ C_{\ell q}^{(3)} \right]_{rstt} - \frac{1}{2} C_{\vp \Box} \left[ \Gamma_e \Gamma_e^\dagger \right]_{rs} + \frac{1}{2} \Big( 3 \left[ \Gamma_e \Gamma_e^\dagger C_{\vp \ell}^{(1)} \right]_{rs} + 3 \left[ C_{\vp \ell}^{(1)} \Gamma_e \Gamma_e^\dagger \right]_{rs} + \left[ \Gamma_e \Gamma_e^\dagger C_{\vp \ell}^{(3)} \right]_{rs} \nn \\
& + \left[ C_{\vp \ell}^{(3)} \Gamma_e \Gamma_e^\dagger \right]_{rs} \Big) - \left( \left[ C_{\ell \ell} \right]_{rtps} + \left[ C_{\ell \ell} \right]_{psrt} \right) \left[ \Gamma_e \Gamma_e^\dagger \right]_{tp} - 6 \left[ C_{\ell q}^{(3)} \right]_{rspt} \left[  \Gamma_u \Gamma_u^\dagger + \Gamma_d \Gamma_d^\dagger \right]_{tp} \nn \\
& + 2 \gamma_H^{(Y)} \left[ C_{\vp \ell}^{(3)} \right]_{rs} + \left[ \gamma_\ell^{(Y)} C_{\vp \ell}^{(3)} \right]_{rs} + \left[ C_{\vp \ell}^{(3)} \gamma_\ell^{(Y)} \right]_{rs} \, , \\[4mm]
\left[ \beta_{\vp e} \right]_{rs} & = - \frac{1}{2} \xi_B \gpc \delta_{rs} + \frac{1}{3} \gpc \left[ C_{\vp e} \right]_{rs} - \frac{2}{3} \gpc \Big( \left[ C_{e d} \right]_{rstt} + \left[ C_{e e} \right]_{rstt} + \left[ C_{e e} \right]_{rtts} + \left[ C_{e e} \right]_{tsrt} + \left[ C_{e e} \right]_{ttrs} \nn \\
& - 2 \left[ C_{e u} \right]_{rstt} + \left[ C_{\ell e} \right]_{ttrs} - \left[ C_{q e} \right]_{ttrs} \Big) + \left( C_{\vp \Box} + C_{\vp D} \right) \left[ \Gamma_e^\dagger \Gamma_e \right]_{rs} - 2 \left[ \Gamma_e^\dagger C_{\vp \ell}^{(1)} \Gamma_e \right]_{rs} \nn \\
& + 3 \left( \left[ \Gamma_e^\dagger \Gamma_e C_{\vp e} \right]_{rs} + \left[ C_{\vp e} \Gamma_e^\dagger \Gamma_e \right]_{rs} \right) - 2 \left[ C_{\ell e} \right]_{ptrs} \left[ \Gamma_e \Gamma_e^\dagger \right]_{tp} - 6 \left[ C_{e u} \right]_{rspt} \left[ \Gamma_u^\dagger \Gamma_u \right]_{tp} \\
& + 2 \left( \left[ C_{e e} \right]_{rspt} + \left[ C_{e e} \right]_{ptrs} + \left[ C_{e e} \right]_{rtps} + \left[ C_{e e} \right]_{psrt} \right) \left[ \Gamma_e^\dagger \Gamma_e \right]_{tp} + 6 \left[ C_{e d} \right]_{rspt} \left[ \Gamma_d^\dagger \Gamma_d \right]_{tp} \nn \\
& - 6 \left[ C_{q e} \right]_{ptrs} \left[ \Gamma_d \Gamma_d^\dagger \right]_{tp} \!+ 6 \left[ C_{q e} \right]_{ptrs} \left[ \Gamma_u \Gamma_u^\dagger \right]_{tp} \!+ 2 \gamma_H^{(Y)} \left[ C_{\vp e} \right]_{rs} \!+ \left[ \gamma_e^{(Y)} C_{\vp e} \right]_{rs} \!+ \left[ C_{\vp e} \gamma_e^{(Y)} \right]_{rs} \, ,\nn\\[4mm]
\left[ \beta_{\vp q}^{(1)} \right]_{rs} & = \frac{1}{12} \xi_B \gpc \delta_{rs} + \frac{1}{3} \gpc \left[ C_{\vp q}^{(1)} \right]_{rs} - \frac{2}{3} \gpc \Big( \left[ C_{\ell q}^{(1)} \right]_{ttrs} + \left[ C_{q d}^{(1)} \right]_{rstt} - 2 \left[ C_{q u}^{(1)} \right]_{rstt} + \left[ C_{q e} \right]_{rstt} \nn \\
& - \left[ C_{q q}^{(1)} \right]_{rstt} - \frac{1}{6} \left[ C_{q q}^{(1)} \right]_{rtts} - \frac{1}{6} \left[ C_{q q}^{(1)} \right]_{tsrt} - \left[ C_{q q}^{(1)} \right]_{ttrs} - \frac{1}{2} \left[ C_{q q}^{(3)} \right]_{rtts} - \frac{1}{2} \left[ C_{q q}^{(3)} \right]_{tsrt} \Big) \nn \\
& + \frac{1}{2} \left( C_{\vp \Box} + C_{\vp D} \right) \left( \left[ \Gamma_u \Gamma_u^\dagger \right]_{rs} - \left[ \Gamma_d \Gamma_d^\dagger \right]_{rs} \right) - \left[ \Gamma_u C_{\vp u} \Gamma_u^\dagger \right]_{rs} - \left[ \Gamma_d C_{\vp d} \Gamma_d^\dagger \right]_{rs} \nn \\
& + 2 \left[ C_{q e} \right]_{rspt} \left[ \Gamma_e^\dagger \Gamma_e \right]_{tp} - 2 \left[ C_{\ell q}^{(1)} \right]_{ptrs} \left[ \Gamma_e \Gamma_e^\dagger \right]_{tp} + \frac{3}{2} \Big( \left[ \Gamma_d \Gamma_d^\dagger C_{\vp q}^{(1)} \right]_{rs} + \left[ \Gamma_u \Gamma_u^\dagger C_{\vp q}^{(1)} \right]_{rs} \nn \\
& + \left[ C_{\vp q}^{(1)} \Gamma_d \Gamma_d^\dagger \right]_{rs} + \left[ C_{\vp q}^{(1)} \Gamma_u \Gamma_u^\dagger \right]_{rs} + 3 \left[ \Gamma_d \Gamma_d^\dagger C_{\vp q}^{(3)} \right]_{rs} - 3 \left[ \Gamma_u \Gamma_u^\dagger C_{\vp q}^{(3)} \right]_{rs} + 3 \left[ C_{\vp q}^{(3)} \Gamma_d \Gamma_d^\dagger \right]_{rs}\nn \\
& - 3 \left[ C_{\vp q}^{(3)} \Gamma_u \Gamma_u^\dagger \right]_{rs} \Big) - \Big( 6 \left[ C_{q q}^{(1)} \right]_{rspt} + 6 \left[ C_{q q}^{(1)} \right]_{ptrs} + \left[ C_{q q}^{(1)} \right]_{rtps}+ \left[ C_{q q}^{(1)} \right]_{psrt} \nn \\
& + 3 \left[ C_{q q}^{(3)} \right]_{rtps} + 3 \left[ C_{q q}^{(3)} \right]_{psrt} \Big) \left( \left[ \Gamma_d \Gamma_d^\dagger \right]_{tp} - \left[ \Gamma_u \Gamma_u^\dagger \right]_{tp} \right) - 6 \left[ C_{q u}^{(1)} \right]_{rspt} \left[ \Gamma_u^\dagger \Gamma_u \right]_{tp} \nn \\
& + 6 \left[ C_{q d}^{(1)} \right]_{rspt} \left[ \Gamma_d^\dagger \Gamma_d \right]_{tp} + 2 \gamma_H^{(Y)} \left[ C_{\vp q}^{(1)} \right]_{rs} + \left[ \gamma_q^{(Y)} C_{\vp q}^{(1)} \right]_{rs} + \left[ C_{\vp q}^{(1)} \gamma_q^{(Y)} \right]_{rs} \, , \\[4mm]
\left[ \beta_{\vp q}^{(3)} \right]_{rs} & = \frac{2}{3} \gc \left( \frac{1}{4} C_{\vp \Box} + \Tr \, C_{\vp \ell}^{(3)} + 3\,\Tr \, C_{\vp q}^{(3)} \right) \delta_{rs} - \frac{17}{3} \gc \left[ C_{\vp q}^{(3)} \right]_{rs} + \frac{1}{3} \gc \Big( 2 \left[ C_{\ell q}^{(3)} \right]_{ttrs} \nn \\
& + \left[ C_{q q}^{(1)} \right]_{rtts} + \left[ C_{q q}^{(1)} \right]_{tsrt} + 6 \left[ C_{q q}^{(3)} \right]_{rstt} - \left[ C_{q q}^{(3)} \right]_{rtts} - \left[ C_{q q}^{(3)} \right]_{tsrt} + 6 \left[ C_{q q}^{(3)} \right]_{ttrs} \Big) \nn \\
& - \frac{1}{2} C_{\vp \Box} \left( \left[ \Gamma_u \Gamma_u^\dagger \right]_{rs} + \left[ \Gamma_d \Gamma_d^\dagger \right]_{rs} \right) + \frac{1}{2} \Big( 3 \left[ \Gamma_d \Gamma_d^\dagger C_{\vp q}^{(1)} \right]_{rs} - 3 \left[ \Gamma_u \Gamma_u^\dagger C_{\vp q}^{(1)} \right]_{rs} \nn \\
& + 3 \left[ C_{\vp q}^{(1)} \Gamma_d \Gamma_d^\dagger \right]_{rs} - 3 \left[ C_{\vp q}^{(1)} \Gamma_u \Gamma_u^\dagger \right]_{rs} + \left[ \Gamma_d \Gamma_d^\dagger C_{\vp q}^{(3)} \right]_{rs} + \left[ \Gamma_u \Gamma_u^\dagger C_{\vp q}^{(3)} \right]_{rs} \nn \\
& + \left[ C_{\vp q}^{(3)} \Gamma_d \Gamma_d^\dagger \right]_{rs} + \left[ C_{\vp q}^{(3)} \Gamma_u \Gamma_u^\dagger \right]_{rs} \Big) 
- \Big( 6 \left[ C_{q q}^{(3)} \right]_{rspt} + 6 \left[ C_{q q}^{(3)} \right]_{ptrs} + \left[ C_{q q}^{(1)} \right]_{rtps} \nn \\
& + \left[ C_{q q}^{(1)} \right]_{psrt} - \left[ C_{q q}^{(3)} \right]_{rtps} - \left[ C_{q q}^{(3)} \right]_{psrt} \Big) \left( \left[ \Gamma_d \Gamma_d^\dagger \right]_{tp} + \left[ \Gamma_u \Gamma_u^\dagger \right]_{tp} \right) \nn \\
& - 2 \left[ C_{\ell q}^{(3)} \right]_{ptrs} \left[ \Gamma_e \Gamma_e^\dagger \right]_{tp} + 2 \gamma_H^{(Y)} \left[ C_{\vp q}^{(3)} \right]_{rs} + \left[ \gamma_q^{(Y)} C_{\vp q}^{(3)} \right]_{rs} + \left[ C_{\vp q}^{(3)} \gamma_q^{(Y)} \right]_{rs} \, , \\[4mm]
\left[ \beta_{\vp u} \right]_{rs} & = \frac{1}{3} \xi_B \gpc \delta_{rs} + \frac{1}{3} \gpc \left[ C_{\vp u} \right]_{rs} - \frac{2}{3} \gpc \Big( \left[ C_{e u} \right]_{ttrs} + \left[ C_{\ell u} \right]_{ttrs} - \left[ C_{q u}^{(1)} \right]_{ttrs} + \left[ C_{u d}^{(1)} \right]_{rstt} \nn \\
& - 2 \left[ C_{u u} \right]_{rstt} - \frac{2}{3} \left[ C_{u u} \right]_{rtts} - \frac{2}{3} \left[ C_{u u} \right]_{tsrt} - 2 \left[ C_{u u} \right]_{ttrs} \Big) - \left( C_{\vp \Box} + C_{\vp D} \right) \left[ \Gamma_u^\dagger \Gamma_u \right]_{rs} \nn \\
& - 2 \left[ \Gamma_u^\dagger C_{\vp q}^{(1)} \Gamma_u \right]_{rs} + 3 \left( \left[ \Gamma_u^\dagger \Gamma_u C_{\vp u} \right]_{rs} + \left[ C_{\vp u} \Gamma_u^\dagger \Gamma_u \right]_{rs} \right) + \left[ \Gamma_u^\dagger \Gamma_d C_{\vp u d}^\dagger \right]_{rs} + \left[ C_{\vp u d} \Gamma_d^\dagger \Gamma_u \right]_{rs} \nn \\
& - 2 \Big( 3 \left[ C_{u u} \right]_{rspt} + 3 \left[ C_{u u} \right]_{ptrs} + \left[ C_{u u} \right]_{rtps} + \left[ C_{u u} \right]_{psrt} \Big) \left[ \Gamma_u^\dagger \Gamma_u \right]_{tp} + 2\left[ C_{eu} \right]_{ptrs} \left[\Gamma_e^\dagger\Gamma_e\right]_{tp} \nn \\
& - 2 \left[ C_{\ell u} \right]_{ptrs} \left[ \Gamma_e \Gamma_e^\dagger \right]_{tp} + 6 \left[ C_{u d}^{(1)} \right]_{rspt} \left[ \Gamma_d^\dagger \Gamma_d \right]_{tp} - 6 \left[ C_{q u}^{(1)} \right]_{ptrs} \left[ \Gamma_d \Gamma_d^\dagger \right]_{tp}  \nn \\
& + 6 \left[ C_{q u}^{(1)} \right]_{ptrs} \left[ \Gamma_u \Gamma_u^\dagger \right]_{tp} + 2 \gamma_H^{(Y)} \left[ C_{\vp u} \right]_{rs} + \left[ \gamma_u^{(Y)} C_{\vp u} \right]_{rs} + \left[ C_{\vp u} \gamma_u^{(Y)} \right]_{rs} \, ,\\[4mm]
\left[ \beta_{\vp d} \right]_{rs} & = - \frac{1}{6} \xi_B \gpc \delta_{rs} + \frac{1}{3} \gpc \left[ C_{\vp d} \right]_{rs} - \frac{2}{3} \gpc \Big( \left[ C_{d d} \right]_{rstt} + \frac{1}{3} \left[ C_{d d} \right]_{rtts} + \frac{1}{3} \left[ C_{d d} \right]_{tsrt} + \left[ C_{d d} \right]_{ttrs} \nn \\
& + \left[ C_{e d} \right]_{ttrs} + \left[ C_{\ell d} \right]_{ttrs} - \left[ C_{q d}^{(1)} \right]_{ttrs} - 2 \left[ C_{u d}^{(1)} \right]_{ttrs} \Big) + \left( C_{\vp \Box} + C_{\vp D} \right) \left[ \Gamma_d^\dagger \Gamma_d \right]_{rs} \nn \\
& - 2 \left[ \Gamma_d^\dagger C_{\vp q}^{(1)} \Gamma_d \right]_{rs} + 3 \left( \left[ \Gamma_d^\dagger \Gamma_d C_{\vp d} \right]_{rs} + \left[ C_{\vp d} \Gamma_d^\dagger \Gamma_d \right]_{rs} \right) - \left[ \Gamma_d^\dagger \Gamma_u C_{\vp u d} \right]_{rs} - \left[ C_{\vp u d}^\dagger \Gamma_u^\dagger \Gamma_d \right]_{rs} \nn \\
& + 2 \Big( 3 \left[ C_{d d} \right]_{rspt} + 3 \left[ C_{d d} \right]_{ptrs} + \left[ C_{d d} \right]_{rtps} + \left[ C_{d d} \right]_{psrt} \Big) \left[ \Gamma_d^\dagger \Gamma_d \right]_{tp} + 2 \left[ C_{e d} \right]_{ptrs} \left[ \Gamma_e^\dagger \Gamma_e \right]_{tp} \nn \\
& - 2 \left[ C_{\ell d} \right]_{ptrs} \left[ \Gamma_e \Gamma_e^\dagger \right]_{tp} - 6 \left[ C_{u d}^{(1)} \right]_{ptrs} \left[ \Gamma_u^\dagger \Gamma_u \right]_{tp} - 6 \left[ C_{q d}^{(1)} \right]_{ptrs} \left[ \Gamma_d \Gamma_d^\dagger \right]_{tp} \nn \\
& + 6 \left[ C_{q d}^{(1)} \right]_{ptrs} \left[ \Gamma_u \Gamma_u^\dagger \right]_{tp} + 2 \gamma_H^{(Y)} \left[ C_{\vp d} \right]_{rs} + \left[ \gamma_d^{(Y)} C_{\vp d} \right]_{rs} + \left[ C_{\vp d} \gamma_d^{(Y)} \right]_{rs} \, , \\[4mm]
\left[ \beta_{\vp u d} \right]_{rs} & = -3 \gpc \left[ C_{\vp u d} \right]_{rs} + \left( 2 C_{\vp \Box} - C_{\vp D} \right) \left[ \Gamma_u^\dagger \Gamma_d \right]_{rs} - 2 \left[ \Gamma_u^\dagger \Gamma_d C_{\vp d} \right]_{rs} + 2 \left[ C_{\vp u} \Gamma_u^\dagger \Gamma_d \right]_{rs} \nn \\
& + 4 \left( \left[ C_{ud}^{(1)} \right]_{rtps} + \frac{4}{3} \left[ C_{ud}^{(8)} \right]_{rtps} \right) \left[ \Gamma_u^\dagger \Gamma_d \right]_{tp} + 2 \left[ \Gamma_u^\dagger \Gamma_u C_{\vp u d} \right]_{rs} + 2 \left[ C_{\vp u d} \Gamma_d^\dagger \Gamma_d \right]_{rs} \nn \\
& + 2 \gamma_H^{(Y)} \left[ C_{\vp u d} \right]_{rs} + \left[ \gamma_u^{(Y)} C_{\vp u d} \right]_{rs} + \left[ C_{\vp u d} \gamma_d^{(Y)} \right]_{rs} \, .
\end{align}

\subsubsection*{$\boxed{\boldsymbol{\left( \bar L L \right)\left( \bar L L \right)}}$}

\begin{align}
\left[ \beta_{\ell \ell} \right]_{prst} & = \Bigg[ - \frac{1}{6} \gpc \left[ C_{\vp \ell}^{(1)} \right]_{st} \delta_{pr} - \frac{1}{6} \gc \left( \left[ C_{\vp \ell}^{(3)} \right]_{st} \delta_{pr} - 2 \left[ C_{\vp \ell}^{(3)} \right]_{sr} \delta_{pt} \right) \nn \\
&  + \frac{1}{3} \gpc \left( 2 \left[ C_{\ell \ell} \right]_{prww} + \left[ C_{\ell \ell} \right]_{pwwr} \right) \delta_{st} - \frac{1}{3} \gc \left[ C_{\ell \ell} \right]_{pwwr} \delta_{st} + \frac{2}{3} \gc \left[ C_{\ell \ell} \right]_{swwr} \delta_{pt}\nn \\
& - \frac{1}{3} \gpc \left[ C_{\ell q}^{(1)} \right]_{prww} \delta_{st} - \gc \left[ C_{\ell q}^{(3)} \right]_{prww} \delta_{st} + 2 \gc \left[ C_{\ell q}^{(3)} \right]_{ptww} \delta_{rs} \nn \\
& + \frac{1}{3} \gpc \left( - 2 \left[ C_{\ell u} \right]_{prww} + \left[ C_{\ell d} \right]_{prww} + \left[ C_{\ell e} \right]_{prww} \right) \delta_{st} - \frac{1}{2} \left[ \Gamma_e \Gamma_e^\dagger \right]_{pr} \left[ C_{\vp \ell}^{(1)} - C_{\vp \ell}^{(3)} \right]_{st} \nn \\
& - \left[ \Gamma_e \Gamma_e^\dagger \right]_{pt} \left[ C_{\vp \ell}^{(3)} \right]_{sr}  - \frac{1}{2} \left[ \Gamma_e \right]_{sv} \left[ \Gamma_e \right]_{tw}^\ast \left[ C_{\ell e} \right]_{prvw} + \left[ \gamma_{\ell}^{(Y)} \right]_{pv} \left[ C_{\ell \ell} \right]_{vrst} \nn \\
& + \left[ C_{\ell \ell} \right]_{pvst} \left[ \gamma_{\ell}^{(Y)} \right]_{vr} + \, (pr) \leftrightarrow (st) \Bigg] + 6 \gc \left[ C_{\ell \ell} \right]_{ptsr} + 3 \left( \gpc - \gc \right) \left[ C_{\ell \ell} \right]_{prst} \, ,\\[4mm]
\left[ \beta_{q q}^{(1)} \right]_{prst} & = \Bigg[ \frac{1}{18} \gpc \left[ C_{\vp q}^{(1)} \right]_{st} \delta_{pr} - \frac{1}{9} \gpc \left[ C_{\ell q}^{(1)} \right]_{wwst} \delta_{pr} \nn \\
& + \frac{1}{9} \gpc \left( 2\left[ C_{q q}^{(1)} \right]_{prww} + \frac{1}{3} \left[ C_{q q}^{(1)} + 3 C_{qq}^{(3)}\right]_{pwwr} \right) \delta_{st}  + \frac{1}{3} \gsc \left[ C_{q q}^{(1)} +3 C_{qq}^{(3)}\right]_{swwr} \delta_{pt} \nn \\
&- \frac{2}{9} \gsc \left[ C_{q q}^{(1)} + 3 C_{qq}^{(3)}\right]_{pwwr} \delta_{st} + \frac{2}{9} \gpc \left[ C_{qu}^{(1)} \right]_{prww} \delta_{st} - \frac{1}{9} \gpc \left[ C_{qd}^{(1)} \right]_{prww} \delta_{st} \nn \\
& + \frac{1}{12} \gsc \left( \left[ C_{qu}^{(8)} \right]_{srww} + \left[ C_{qd}^{(8)} \right]_{srww} \right) \delta_{pt} - \frac{1}{18} \gsc \left( \left[ C_{qu}^{(8)} \right]_{prww} + \left[ C_{qd}^{(8)} \right]_{prww} \right) \delta_{st} \nn \\
& - \frac{1}{9} \gpc \left[ C_{qe} \right]_{prww} \delta_{st} + \frac{1}{2} \left[ \Gamma_u \Gamma_u^\dagger - \Gamma_d \Gamma_d^\dagger \right]_{pr} \left[ C_{\vp q}^{(1)} \right]_{st} - \frac{1}{2} \left[ \Gamma_u \right]_{pv} \left[ \Gamma_u \right]_{rw}^\ast \left[ C_{qu}^{(1)} - \frac{1}{6}  C_{qu}^{(8)} \right]_{stvw} \nn \\
& - \frac{1}{2} \left[ \Gamma_d \right]_{pv} \left[ \Gamma_d \right]_{rw}^\ast \left[ C_{qd}^{(1)} - \frac{1}{6} C_{qd}^{(8)} \right]_{stvw} \nn \\
& - \frac{1}{8} \left( \left[ \Gamma_u \right]_{pv} \left[ \Gamma_u \right]_{tw}^\ast \left[ C_{qu}^{(8)} \right]_{srvw} +\left[ \Gamma_d \right]_{pv} \left[ \Gamma_d \right]_{tw}^\ast \left[ C_{qd}^{(8)} \right]_{srvw}  \right) \nn \\
& - \frac{1}{8} \left[ \Gamma_d \right]_{tw}^\ast \left[ \Gamma_u \right]_{rv}^\ast \left[ C_{quqd}^{(1)} - \frac{1}{6}  C_{quqd}^{(8)} \right]_{pvsw} - \frac{1}{8} \left[ \Gamma_d \right]_{sw} \left[ \Gamma_u \right]_{pv} \left[ C_{quqd}^{(1)} - \frac{1}{6}  C_{quqd}^{(8)}  \right]_{rvtw}^\ast \nn \\
& + \frac{1}{16} \left( \left[ \Gamma_d \right]_{tw}^\ast \left[ \Gamma_u \right]_{rv}^\ast \left[ C_{quqd}^{(8)} \right]_{svpw} + \left[ \Gamma_d \right]_{sw} \left[ \Gamma_u \right]_{pv} \left[ C_{quqd}^{(8)} \right]_{tvrw}^\ast \right) \nn \\
& + \left[ \gamma_q^{(Y)} \right]_{pv} \left[ C_{q q}^{(1)} \right]_{vrst} + \left[ C_{q q}^{(1)} \right]_{pvst} \left[ \gamma_q^{(Y)} \right]_{vr} + \, (pr) \leftrightarrow (st) \Bigg]\nn \\
& + 9 \gc \left[ C_{qq}^{(3)} \right]_{prst} - 2 \left( \gsc - \frac{1}{6} \gpc \right) \left[ C_{qq}^{(1)} \right]_{prst}+ 3 \gsc \left[ C_{qq}^{(1)} +3\,C_{qq}^{(3)}\right]_{ptsr}\, ,\\[4mm]
\left[ \beta_{q q}^{(3)} \right]_{prst} & = \Bigg[ \frac{1}{6} \gc \left[ C_{\vp q}^{(3)} \right]_{st} \delta_{pr} + \frac{1}{3} \gc \left[ C_{\ell q}^{(3)} \right]_{wwst} \delta_{pr} + \frac{1}{3} \gc \left[ C_{q q}^{(1)} - C_{q q}^{(3)}\right]_{pwwr} \delta_{st} \nn \\
& + 2\gc \left[ C_{q q}^{(3)} \right]_{prww} \delta_{st}  + \frac{1}{3} \gsc \left[ C_{q q}^{(1)} + 3 C_{q q}^{(3)}\right]_{swwr} \delta_{pt} + \frac{1}{12} \gsc \left[ C_{qu}^{(8)} + C_{qd}^{(8)} \right]_{srww} \delta_{pt} \nn \\
& -\frac{1}{2} \left[ \Gamma_u \Gamma_u^\dagger + \Gamma_d \Gamma_d^\dagger \right]_{pr} \left[ C_{\vp q}^{(3)} \right]_{st} - \frac{1}{8} \left( \left[ \Gamma_u \right]_{pv} \left[ \Gamma_u \right]_{tw}^\ast \left[ C_{qu}^{(8)} \right]_{srvw} +\left[ \Gamma_d \right]_{pv} \left[ \Gamma_d \right]_{tw}^\ast \left[ C_{qd}^{(8)} \right]_{srvw}  \right) \nn \\
& + \frac{1}{8} \left[ \Gamma_d \right]_{tw}^\ast \left[ \Gamma_u \right]_{rv}^\ast \left[ C_{quqd}^{(1)} - \frac{1}{6} C_{quqd}^{(8)}\right]_{pvsw} + \frac{1}{8} \left[ \Gamma_d \right]_{sw} \left[ \Gamma_u \right]_{pv} \left[ C_{quqd}^{(1)} - \frac{1}{6} C_{quqd}^{(8)}\right]_{rvtw}^\ast \nn \\
& - \frac{1}{16} \left( \left[ \Gamma_d \right]_{tw}^\ast \left[ \Gamma_u \right]_{rv}^\ast \left[ C_{quqd}^{(8)} \right]_{svpw} + \left[ \Gamma_d \right]_{sw} \left[ \Gamma_u \right]_{pv} \left[ C_{quqd}^{(8)} \right]_{tvrw}^\ast \right) + \left[ \gamma_q^{(Y)} \right]_{pv} \left[ C_{q q}^{(3)} \right]_{vrst} \nn \\
& + \left[ C_{q q}^{(3)} \right]_{pvst} \left[ \gamma_q^{(Y)} \right]_{vr} + \, (pr) \leftrightarrow (st) \Bigg] + 3 \gsc \left[ C_{qq}^{(1)} - C_{qq}^{(3)} \right]_{ptsr} \nn\\
& - 2 \left( \gsc + 3 \gc - \frac{1}{6} \gpc \right) \left[ C_{qq}^{(3)} \right]_{prst} + 3 \gc \left[ C_{qq}^{(1)} \right]_{prst} \, ,\\[4mm]
\left[ \beta_{\ell q}^{(1)} \right]_{prst} & = - \frac{1}{3} \gpc \left[ C_{\vp q}^{(1)} \right]_{st} \delta_{pr} + \frac{1}{9} \gpc \left[ C_{\vp \ell}^{(1)} \right]_{pr} \delta_{st} - \frac{2}{9} \gpc \left( 2 \left[ C_{\ell \ell} \right]_{prww} + \left[ C_{\ell \ell} \right]_{pwwr} \right) \delta_{st} \nn \\
& + \frac{2}{9} \gpc \left[ C_{\ell q}^{(1)} \right]_{prww} \delta_{st} + \frac{2}{3} \gpc \left[ C_{\ell q}^{(1)} \right]_{wwst} \delta_{pr} - \gpc \left[ C_{\ell q}^{(1)} \right]_{prst} + 9 \gc \left[ C_{\ell q}^{(3)} \right]_{prst} \nn \\
& - \frac{2}{9} \gpc \left( 6 \left[ C_{q q}^{(1)} \right]_{stww} + \left[ C_{q q}^{(1)} + 3\,C_{q q}^{(3)} \right]_{swwt} \right) \delta_{pr} - \left[ \Gamma_u \right]_{sv} \left[ \Gamma_u \right]_{tw}^\ast \left[ C_{\ell u} \right]_{prvw} \nn\\
& - \frac{2}{3} \gpc \left( 2 \left[ C_{q u}^{(1)} \right]_{stww} - \left[ C_{q d}^{(1)} \right]_{stww} - \left[ C_{q e} \right]_{stww} \right) \delta_{pr} - \left[ \Gamma_e \Gamma_e^\dagger \right]_{pr} \left[ C_{\vp q}^{(1)} \right]_{st} \nn \\
& + \frac{2}{9} \gpc \left( 2 \left[ C_{\ell u} \right]_{prww} - \left[ C_{\ell d} \right]_{prww} - \left[ C_{\ell e} \right]_{prww} \right) \delta_{st}  + \left[ \Gamma_u \Gamma_u^\dagger - \Gamma_d \Gamma_d^\dagger \right]_{st} \left[ C_{\vp \ell}^{(1)} \right]_{pr} \nn \\
& + \frac{1}{4} \left( \left[ \Gamma_u \right]_{tw}^\ast \left[ \Gamma_e \right]_{rv}^\ast \left[ C_{\ell e q u}^{(1)} -12\,C_{\ell e q u}^{(3)}\right]_{pvsw} + \left[ \Gamma_u \right]_{sw} \left[ \Gamma_e \right]_{pv} \left[ C_{\ell e q u}^{(1)} -12\,C_{\ell e q u}^{(3)}\right]_{rvtw}^\ast \right) \nn \\
& - \left[ \Gamma_d \right]_{sv} \left[ \Gamma_d \right]_{tw}^\ast \left[ C_{\ell d} \right]_{prvw} - \left[ \Gamma_e \right]_{pv} \left[ \Gamma_e \right]_{rw}^\ast \left[ C_{q e} \right]_{stvw} \nn \\
& + \frac{1}{4} \left( \left[ \Gamma_d \right]_{sw} \left[ \Gamma_e \right]_{rv}^\ast \left[ C_{\ell e d q} \right]_{pvwt} + \left[ \Gamma_e \right]_{pv} \left[ \Gamma_d \right]_{tw}^\ast \left[ C_{\ell e d q} \right]_{rvws}^\ast \right) \nn \\
& + \left[ \gamma_\ell^{(Y)} \right]_{pv} \left[ C_{\ell q}^{(1)} \right]_{vrst} + \left[ \gamma_q^{(Y)} \right]_{sv} \left[ C_{\ell q}^{(1)} \right]_{prvt} + \left[ C_{\ell q}^{(1)} \right]_{pvst} \left[ \gamma_\ell^{(Y)} \right]_{vr} + \left[ C_{\ell q}^{(1)} \right]_{prsv} \left[ \gamma_q^{(Y)} \right]_{vt} \, ,\\[4mm]
\left[ \beta_{\ell q}^{(3)} \right]_{prst} & = \frac{1}{3} \gc \left( \left[ C_{\vp q}^{(3)} \right]_{st} \delta_{pr} + \left[ C_{\vp \ell}^{(3)} \right]_{pr} \delta_{st} \right) + \frac{2}{3} \gc \left( 3 \left[ C_{\ell q}^{(3)} \right]_{prww} \delta_{st} + \left[ C_{\ell q}^{(3)} \right]_{wwst} \delta_{pr} \right) \nn \\
& + \frac{2}{3} \gc \left( 6 \left[ C_{q q}^{(3)} \right]_{stww} + \left[ C_{q q}^{(1)} - C_{q q}^{(3)} \right]_{swwt} \right) \delta_{pr} + \frac{2}{3} \gc \left[ C_{\ell \ell} \right]_{pwwr} \delta_{st} + 3 \gc \left[ C_{\ell q}^{(1)} \right]_{prst}\nn \\
&  - \left( 6 \gc + \gpc \right) \left[ C_{\ell q}^{(3)} \right]_{prst} - \left[ \Gamma_e \Gamma_e^\dagger \right]_{pr} \left[ C_{\vp q}^{(3)} \right]_{st} - \left[ \Gamma_u \Gamma_u^\dagger + \Gamma_d \Gamma_d^\dagger \right]_{st} \left[ C_{\vp \ell}^{(3)} \right]_{pr} \nn \\
& - \frac{1}{4} \left( \left[ \Gamma_u \right]_{tw}^\ast \left[ \Gamma_e \right]_{rv}^\ast \left[ C_{\ell e q u}^{(1)} - 12 C_{\ell e q u}^{(3)}\right]_{pvsw} + \left[ \Gamma_u \right]_{sw} \left[ \Gamma_e \right]_{pv} \left[ C_{\ell e q u}^{(1)} - 12 C_{\ell e q u}^{(3)}\right]_{rvtw}^\ast \right) \nn \\
& + \frac{1}{4} \left( \left[ \Gamma_d \right]_{sw} \left[ \Gamma_e \right]_{rv}^\ast \left[ C_{\ell e d q} \right]_{pvwt} + \left[ \Gamma_e \right]_{pv} \left[ \Gamma_d \right]_{tw}^\ast \left[ C_{\ell e d q} \right]_{rvws}^\ast \right) \nn \\
& + \left[ \gamma_\ell^{(Y)} \right]_{pv} \left[ C_{\ell q}^{(3)} \right]_{vrst} + \left[ \gamma_q^{(Y)} \right]_{sv} \left[ C_{\ell q}^{(3)} \right]_{prvt} + \left[ C_{\ell q}^{(3)} \right]_{pvst} \left[ \gamma_\ell^{(Y)} \right]_{vr} + \left[ C_{\ell q}^{(3)} \right]_{prsv} \left[ \gamma_q^{(Y)} \right]_{vt} \, .
\end{align}

\subsubsection*{$\boxed{\boldsymbol{\left( \bar R R \right)\left( \bar R R \right)}}$}

\begin{align}
\left[ \beta_{ee} \right]_{prst} & = \Bigg[ -\frac{1}{3} \gpc \left[ C_{\vp e} \right]_{st} \delta_{pr} + \frac{2}{3} \gpc \left( \left[ C_{\ell e} \right]_{wwpr} - \left[ C_{qe} \right]_{wwpr} - 2\left[ C_{eu} \right]_{prww}+\left[ C_{ed} \right]_{prww}\right. \nn \\
&\left.+4\left[ C_{ee} \right]_{prww}\right) \delta_{st}+6\,\gpc\left[ C_{ee} \right]_{prst}+ \left[ \Gamma_e^\dagger \Gamma_e \right]_{pr} \left[ C_{\vp e}\right]_{st} - \left[ \Gamma_e \right]_{wr} \left[ \Gamma_e \right]_{vp}^\ast \left[ C_{\ell e} \right]_{vwst}\nn \\
&  + \left[ \gamma_e^{(Y)} \right]_{pv} \left[ C_{ee} \right]_{vrst} + \left[ C_{ee} \right]_{pvst} \left[ \gamma_e^{(Y)} \right]_{vr} + \, (pr) \leftrightarrow (st) \Bigg] \, ,\\[4mm]
\left[ \beta_{uu} \right]_{prst} & = \Bigg[ \frac{2}{9} \gpc \left[ C_{\vp u} \right]_{st} \delta_{pr} - \frac{1}{9} \gsc \left( \left[ C_{q u}^{(8)} \right]_{wwst} \delta_{pr} - 3 \left[ C_{q u}^{(8)} \right]_{wwsr} \delta_{pt} \right) + \frac{2}{3} \gsc \left[ C_{u u} \right]_{pwwt} \delta_{rs} \nn\\
& - \frac{4}{9} \gpc \left( \left[ C_{e u} \right]_{wwst} + \left[ C_{\ell u} \right]_{wwst} - \left[ C_{q u}^{(1)} \right]_{wwst} \!\!\!- 4 \left[ C_{u u} \right]_{wwst} - \frac{4}{3} \left[ C_{u u} \right]_{swwt}\right) \delta_{pr} + 3 \gsc \left[ C_{u u} \right]_{ptsr}  \nn \\
& - \frac{2}{9} \gsc \left[ C_{u u} \right]_{swwt} \delta_{pr} - \frac{4}{9} \gpc \left[ C_{u d}^{(1)} \right]_{stww} \delta_{pr} - \frac{1}{18}\gsc \left( \left[ C_{u d}^{(8)} \right]_{stww} \delta_{pr} - 3 \left[ C_{u d}^{(8)} \right]_{srww} \delta_{pt} \right) \nn \\
& + \left( \frac{8}{3} \gpc - \gsc \right) \left[ C_{u u} \right]_{prst} - \left[ \Gamma_u^\dagger \Gamma_u \right]_{pr} \left[ C_{\vp u}\right]_{st} - \left[ \Gamma_u \right]_{wr} \left[ \Gamma_u \right]_{vp}^\ast \left( \left[ C_{q u}^{(1)} \right]_{vwst} - \frac{1}{6} \left[ C_{q u}^{(8)} \right]_{vwst} \right) \nn \\
& - \frac{1}{2} \left[ \Gamma_u \right]_{wr} \left[ \Gamma_u \right]_{vs}^\ast \left[ C_{q u}^{(8)} \right]_{vwpt} + \left[ \gamma_u^{(Y)} \right]_{pv} \left[ C_{u u} \right]_{vrst} + \left[ C_{u u} \right]_{pvst} \left[ \gamma_u^{(Y)} \right]_{vr}
+ \, (pr) \leftrightarrow (st) \Bigg] \, ,\\[4mm]
\left[ \beta_{dd} \right]_{prst} & = \Bigg[ - \frac{1}{9} \gpc \left[ C_{\vp d} \right]_{st} \delta_{pr} - \frac{1}{9} \gsc \left( \left[ C_{q d}^{(8)} \right]_{wwst} \delta_{pr} - 3 \left[ C_{q d}^{(8)} \right]_{wwsr} \delta_{pt} \right) + \frac{2}{3} \gsc \left[ C_{d d} \right]_{pwwt} \delta_{rs} \nn\\ 
& + \frac{2}{9} \gpc \left( \left[ C_{e d} \right]_{wwst} \!\!+ \left[ C_{\ell d} \right]_{wwst} \!\!- \left[ C_{q d}^{(1)} \right]_{wwst} \!\!\!\!+ 2 \left[ C_{d d} \right]_{wwst} + \frac{2}{3} \left[ C_{d d} \right]_{swwt} \right) \delta_{pr} + 3 \gsc \left[ C_{d d} \right]_{ptsr} \nn \\
& - \frac{2}{9} \gsc \left[ C_{d d} \right]_{swwt} \delta_{pr} - \frac{4}{9} \gpc \left[ C_{u d}^{(1)} \right]_{wwst} \delta_{pr} - \frac{1}{18} \gsc \left( \left[ C_{u d}^{(8)} \right]_{wwst} \delta_{pr} - 3 \left[ C_{u d}^{(8)} \right]_{wwsr} \delta_{pt} \right) \nn \\
& + \left( \frac{2}{3} \gpc - \gsc \right) \left[ C_{d d} \right]_{prst} + \left[ \Gamma_d^\dagger \Gamma_d \right]_{pr} \left[ C_{\vp d}\right]_{st} - \left[ \Gamma_d \right]_{wr} \left[ \Gamma_d \right]_{vp}^\ast \left( \left[ C_{q d}^{(1)} \right]_{vwst} \!\!\!- \frac{1}{6} \left[ C_{q d}^{(8)} \right]_{vwst} \right) \nn \\
& - \frac{1}{2} \left[ \Gamma_d \right]_{wr} \left[ \Gamma_d \right]_{vs}^\ast \left[ C_{q d}^{(8)} \right]_{vwpt} + \left[ \gamma_d^{(Y)} \right]_{pv} \left[ C_{d d} \right]_{vrst} + \left[ C_{d d} \right]_{pvst} \left[ \gamma_d^{(Y)} \right]_{vr}
+ \, (pr) \leftrightarrow (st) \Bigg] \, ,\\[4mm]
\left[ \beta_{e u} \right]_{prst} & = - \frac{2}{3} \gpc \left( \left[ C_{\vp u} \right]_{st} + 2 \left[ C_{q u}^{(1)} - C_{\ell u} + 4 C_{u u} - C_{e u} \right]_{wwst} - 2 \left[ C_{u d}^{(1)} \right]_{stww} + \frac{8}{3} \left[ C_{u u} \right]_{swwt} \right) \delta_{pr} \nn \\
& + \frac{4}{9} \gpc \left( \left[ C_{\vp e} \right]_{pr} + 2 \left[ C_{q e} - C_{\ell e} - 4 C_{e e} \right]_{wwpr} - 2 \left[ C_{ed} - 2 C_{e u} \right]_{prww} \right) \delta_{st} - 8 \gpc \left[ C_{e u} \right]_{prst} \nn \\
& + 2 \left[ \Gamma_e^\dagger \Gamma_e \right]_{pr} \left[ C_{\vp u} \right]_{st} - 2 \left[ \Gamma_u^\dagger \Gamma_u \right]_{st} \left[ C_{\vp e} \right]_{pr} + \left[ \Gamma_e \right]_{vp}^\ast \left[ \Gamma_u \right]_{ws}^\ast \left[ C_{\ell e q u}^{(1)} - 12\, C_{\ell e q u}^{(3)}\right]_{vrwt}\nn \\
&  + \left[ \Gamma_e \right]_{vr} \left[\Gamma_u \right]_{wt} \left[ C_{\ell e q u}^{(1)} - 12\, C_{\ell e q u}^{(3)}\right]_{vpws}^\ast \!\!\!- 2 \left[ \Gamma_e \right]_{vp}^\ast \left[\Gamma_e \right]_{wr} \left[ C_{\ell u} \right]_{vwst} - 2 \left[ \Gamma_u \right]_{vs}^\ast \left[\Gamma_u \right]_{wt} \left[ C_{q e} \right]_{vwpr} \nn \\
& + \left[ \gamma_e^{(Y)} \right]_{pv} \left[ C_{e u} \right]_{vrst} + \left[ \gamma_u^{(Y)} \right]_{sv} \left[ C_{e u} \right]_{prvt} + \left[ C_{e u} \right]_{pvst} \left[ \gamma_e^{(Y)} \right]_{vr} + \left[ C_{e u} \right]_{prsv} \left[ \gamma_u^{(Y)} \right]_{vt} \, ,\\[4mm]
\left[ \beta_{e d} \right]_{prst} & = - \frac{2}{3} \gpc \left( \left[ C_{\vp d} \right]_{st} + 2 \left[ C_{q d}^{(1)} - C_{\ell d} - 2 C_{d d} - C_{e d} + 2 C_{u d}^{(1)} \right]_{wwst} - \frac{4}{3} \left[ C_{d d} \right]_{swwt} \right) \delta_{pr} \nn \\
& - \frac{2}{9} \gpc \left( \left[ C_{\vp e} \right]_{pr} + 2 \left[ C_{q e} - C_{\ell e} - 4 C_{e e} \right]_{wwpr} - 2 \left[ C_{e d} - 2 C_{e u} \right]_{prww} \right) \delta_{st} + 4 \gpc \left[ C_{e d} \right]_{prst} \nn \\
& + 2 \left[ \Gamma_e^\dagger \Gamma_e \right]_{pr} \left[ C_{\vp d} \right]_{st} + 2 \left[ \Gamma_d^\dagger \Gamma_d \right]_{st} \left[ C_{\vp e} \right]_{pr} - 2 \left[ \Gamma_e \right]_{vp}^\ast \left[ \Gamma_e \right]_{wr} \left[ C_{\ell d} \right]_{vwst}  \nn \\
& - 2 \left[ \Gamma_d \right]_{vs}^\ast \left[\Gamma_d \right]_{wt} \left[ C_{q e} \right]_{vwpr} + \left[ \Gamma_e \right]_{vp}^\ast \left[\Gamma_d \right]_{wt} \left[ C_{\ell e d q} \right]_{vrsw} + \left[ \Gamma_e \right]_{vr} \left[\Gamma_d \right]_{ws}^\ast \left[ C_{\ell e d q} \right]_{vptw}^\ast \nn \\
& + \left[ \gamma_e^{(Y)} \right]_{pv} \left[ C_{e d} \right]_{vrst} + \left[ \gamma_d^{(Y)} \right]_{sv} \left[ C_{e d} \right]_{prvt} + \left[ C_{e d} \right]_{pvst} \left[ \gamma_e^{(Y)} \right]_{vr} + \left[ C_{e d} \right]_{prsv} \left[ \gamma_d^{(Y)} \right]_{vt} \, ,\\[4mm]
\left[ \beta_{u d}^{(1)} \right]_{prst} & = \frac{4}{9} \gpc \left( \left[ C_{\vp d} \right]_{st} + 2 \left[ C_{q d}^{(1)} - C_{\ell d} - 2 C_{d d} + 2 C_{u d}^{(1)} - C_{e d} \right]_{wwst} - \frac{4}{3} \left[ C_{d d} \right]_{swwt} \right) \delta_{pr} \nn \\
& - \frac{2}{9} \gpc \left( \left[ C_{\vp u} \right]_{pr} + 2 \left[ C_{q u}^{(1)} - C_{\ell u} + 4 C_{u u} - C_{e u} \right]_{wwpr} - 2 \left[ C_{u d}^{(1)} \right]_{prww} + \frac{8}{3} \left[ C_{u u} \right]_{pwwr} \right) \delta_{st} \nn \\
& - \frac{8}{3} \left( \gpc \left[ C_{u d}^{(1)} \right]_{prst} - \gsc \left[ C_{u d}^{(8)} \right]_{prst} \right) - 2 \left[ \Gamma_u^\dagger \Gamma_u \right]_{pr} \left[ C_{\vp d} \right]_{st} + 2 \left[ \Gamma_d^\dagger \Gamma_d \right]_{st} \left[ C_{\vp u} \right]_{pr} \nn \\
& + \frac{2}{3} \left[ \Gamma_d^\dagger \Gamma_u \right]_{sr} \left[ C_{\vp u d} \right]_{pt} + \frac{2}{3} \left[ \Gamma_u^\dagger \Gamma_d \right]_{pt} \left[ C_{\vp u d} \right]_{rs}^\ast - \left[ \Gamma_d \right]_{ws}^\ast \left[ \Gamma_u \right]_{vp}^\ast \left[ C_{q u q d}^{(1)} \right]_{vrwt} \nn \\
& + \frac{1}{3} \left( \left[ \Gamma_d \right]_{vs}^\ast \left[ \Gamma_u \right]_{wp}^\ast \left[ C_{q u q d}^{(1)} + \frac{4}{3}\, C_{q u q d}^{(8)}\right]_{vrwt} + \left[ \Gamma_d \right]_{vt} \left[ \Gamma_u \right]_{wr} \left[ C_{q u q d}^{(1)} + \frac{4}{3}\, C_{q u q d}^{(8)}\right]_{vpws}^\ast \right) \\
& - \left[ \Gamma_d \right]_{wt} \left[ \Gamma_u \right]_{vr} \left[ C_{q u q d}^{(1)} \right]_{vpws}^\ast - 2 \left[ \Gamma_u \right]_{vp}^\ast \left[ \Gamma_u \right]_{wr} \left[ C_{q d}^{(1)} \right]_{vwst} - 2 \left[ \Gamma_d \right]_{vs}^\ast \left[ \Gamma_d \right]_{wt} \left[ C_{q u}^{(1)} \right]_{vwpr} \nn \\
& + \left[ \gamma_u^{(Y)} \right]_{pv} \left[ C_{u d}^{(1)} \right]_{vrst} + \left[ \gamma_d^{(Y)} \right]_{sv} \left[ C_{u d}^{(1)} \right]_{prvt} + \left[ C_{u d}^{(1)} \right]_{pvst} \left[ \gamma_u^{(Y)} \right]_{vr} + \left[ C_{u d}^{(1)} \right]_{prsv} \left[ \gamma_d^{(Y)} \right]_{vt} \, ,\nn\\[4mm]
\left[ \beta_{u d}^{(8)} \right]_{prst} & = \frac{8}{3} \gsc \left[ C_{u u} \right]_{pwwr} \delta_{st} + \frac{8}{3} \gsc \left[ C_{d d} \right]_{swwt} \delta_{pr} + \frac{4}{3} \gsc \left[ C_{q u}^{(8)} \right]_{wwpr} \delta_{st} + \frac{4}{3} \gsc \left[ C_{q d}^{(8)} \right]_{wwst} \delta_{pr} \nn \\
& + \frac{2}{3} \gsc \left[ C_{u d}^{(8)} \right]_{prww} \delta_{st} + \frac{2}{3} \gsc  \left[ C_{u d}^{(8)} \right]_{wwst} \delta_{pr} - 4 \left( \frac{2}{3} \gpc + \gsc \right) \left[ C_{u d}^{(8)} \right]_{prst} + 12 \gsc \left[ C_{u d}^{(1)} \right]_{prst} \nn \\
& + 4 \left[ \Gamma_d^\dagger \Gamma_u \right]_{sr} \left[ C_{\vp u d} \right]_{pt} + 4 \left[ \Gamma_u^\dagger \Gamma_d \right]_{pt} \left[ C_{\vp u d} \right]_{rs}^\ast - 2 \left[ \Gamma_u \right]_{vp}^\ast \left[ \Gamma_u \right]_{wr} \left[ C_{q d}^{(8)} \right]_{vwst} \nn \\
& + 2 \left( \left[ \Gamma_d \right]_{vs}^\ast \left[ \Gamma_u \right]_{wp}^\ast \left[ C_{q u q d}^{(1)} - \frac{1}{6} C_{q u q d}^{(8)} \right]_{vrwt} + \left[ \Gamma_d \right]_{vt} \left[ \Gamma_u \right]_{wr} \left[ C_{q u q d}^{(1)} - \frac{1}{6} C_{q u q d}^{(8)} \right]_{vpws}^\ast \right)  \\
& - 2 \left[ \Gamma_d \right]_{vs}^\ast \left[ \Gamma_d \right]_{wt} \left[ C_{q u}^{(8)} \right]_{vwpr}  - \left( \left[ \Gamma_d \right]_{ws}^\ast \left[ \Gamma_u \right]_{vp}^\ast \left[ C_{q u q d}^{(8)} \right]_{vrwt} + \left[ \Gamma_d \right]_{wt} \left[ \Gamma_u \right]_{vr} \left[ C_{q u q d}^{(8)} \right]_{vpws}^\ast \right) \nn \\
& + \left[ \gamma_u^{(Y)} \right]_{pv} \left[ C_{u d}^{(8)} \right]_{vrst} + \left[ \gamma_d^{(Y)} \right]_{sv} \left[ C_{u d}^{(8)} \right]_{prvt} + \left[ C_{u d}^{(8)} \right]_{pvst} \left[ \gamma_u^{(Y)} \right]_{vr} + \left[ C_{u d}^{(8)} \right]_{prsv} \left[ \gamma_d^{(Y)} \right]_{vt} \, .\nn
\end{align}

\subsubsection*{$\boxed{\boldsymbol{\left( \bar L L \right)\left( \bar R R \right)}}$}

\begin{align}
\left[ \beta_{\ell e} \right]_{prst} & = -\frac{1}{3}\gpc \left[ C_{\vp e}\right]_{st} \delta_{pr} - \frac{2}{3}\gpc \left[ C_{\vp\ell}^{(1)}\right]_{pr}\delta_{st}+\frac{8}{3}\,\gpc \left[C_{\ell\ell}\right]_{prww}\delta_{st}+\frac{4}{3}\,\gpc \left[C_{\ell\ell}\right]_{pwwr}\delta_{st} \nn\\
&-\frac{4}{3}\gpc\left[C_{\ell q}^{(1)}\right]_{prww}\delta_{st}-\frac{2}{3}\gpc\left[C_{qe}\right]_{wwst}\delta_{pr}+\frac{4}{3}\gpc\left[C_{\ell e}\right]_{prww}\delta_{st}+\frac{2}{3}\gpc\left[C_{\ell e}\right]_{wwst}\delta_{pr}\nn\\
&-\frac{8}{3}\gpc\left[C_{\ell u}\right]_{prww}\delta_{st}+\frac{4}{3}\gpc\left[C_{\ell d}\right]_{prww}\delta_{st}-\frac{4}{3}\gpc\left[C_{eu}\right]_{stww}\delta_{pr}+\frac{2}{3}\gpc\left[C_{ed}\right]_{stww}\delta_{pr}\nn\\
&+\frac{8}{3}\gpc \left[C_{ee}\right]_{wwst}\delta_{pr}-6\gpc\left[C_{\ell e}\right]_{prst}+\left[\Gamma_e\right]^\ast_{rs}\left[\xi_e\right]_{pt}+\left[\Gamma_e\right]_{pt}\left[\xi_e\right]^\ast_{rs}-\left[\Gamma_e\Gamma_e^\dagger\right]_{pr}\left[C_{\vp e}\right]_{st}\nn\\
&+2\left[\Gamma_e^\dagger\Gamma_e\right]_{st}\left[C_{\vp\ell}^{(1)}\right]_{pr}-4\left[\Gamma_e\right]_{pv}\left[\Gamma_e\right]^\ast_{rw}\left[C_{ee}\right]_{vtsw}+\left[\Gamma_e\right]_{pw}\left[\Gamma_e\right]^\ast_{vs}\left[C_{\ell e}\right]_{vrwt}\nn\\
&-2\left[\Gamma_e\right]_{wt}\left[\Gamma_e\right]^\ast_{vs}\left[C_{\ell\ell}\right]_{pwvr}-4\left[\Gamma_e\right]_{wt}\left[\Gamma_e\right]^\ast_{vs}\left[C_{\ell\ell}\right]_{prvw}+\left[\Gamma_e\right]_{vt}\left[\Gamma_e\right]^\ast_{rw}\left[C_{\ell e}\right]_{pvsw}\nn\\
&+\left[\gamma^{(Y)}_\ell\right]_{pv}\left[C_{\ell e}\right]_{vrst}+\left[\gamma^{(Y)}_e\right]_{sv}\left[C_{\ell e}\right]_{prvt}+\left[C_{\ell e}\right]_{pvst}\left[\gamma^{(Y)}_\ell\right]_{vr}+\left[C_{\ell e}\right]_{prsv}\left[\gamma^{(Y)}_e\right]_{vt}\,,\\[4mm]
\left[ \beta_{\ell u} \right]_{prst} & = -\frac{1}{3}\gpc \left[ C_{\vp u}\right]_{st} \delta_{pr} + \frac{4}{9}\gpc \left[ C_{\vp\ell}^{(1)}\right]_{pr}\delta_{st}-\frac{16}{9}\gpc \left[C_{\ell\ell}\right]_{prww}\delta_{st}-\frac{8}{9}\gpc \left[C_{\ell\ell}\right]_{pwwr}\delta_{st}\nn\\
&+\frac{8}{9}\gpc\left[C_{\ell q}^{(1)}\right]_{prww}\delta_{st}-\frac{2}{3}\gpc\left[C_{qu}^{(1)}\right]_{wwst}\delta_{pr}+\frac{16}{9}\gpc\left[C_{\ell u}\right]_{prww}\delta_{st}+\frac{2}{3}\gpc\left[C_{\ell u}\right]_{wwst}\delta_{pr}\nn\\
&-\frac{8}{9}\gpc\left[C_{\ell d}\right]_{prww}\delta_{st}-\frac{8}{9}\gpc\left[C_{\ell e}\right]_{prww}\delta_{st}+\frac{2}{3}\gpc\left[C_{ud}^{(1)}\right]_{stww}\delta_{pr}+\frac{2}{3}\gpc\left[C_{eu}\right]_{wwst}\delta_{pr}\nn\\
&-\frac{8}{3}\gpc\left[C_{uu}\right]_{stww}\delta_{pr}-\frac{8}{9}\gpc\left[C_{uu}\right]_{swwt}\delta_{pr}+4\gpc\left[C_{\ell u}\right]_{prst}-\left[\Gamma_e\Gamma_e^\dagger\right]_{pr}\left[C_{\vp u}\right]_{st}\nn\\
&-2\left[\Gamma_u^\dagger\Gamma_u\right]_{st}\left[C_{\vp\ell}^{(1)}\right]_{pr}-\frac{1}{2}\left[\Gamma_e\right]^\ast_{rv}\left[\Gamma_u\right]^\ast_{ws}\left[C_{\ell equ}^{(1)}+12C_{\ell equ}^{(3)}\right]_{pvwt} \!\!- 2\left[\Gamma_u\right]^\ast_{vs}\left[\Gamma_u\right]_{wt}\left[C_{\ell q}^{(1)}\right]_{prvw} \nn\\
&-\frac{1}{2}\left[\Gamma_e\right]_{pv}\left[\Gamma_u\right]_{wt}\left[C_{\ell equ}^{(1)}+12C_{\ell equ}^{(3)}\right]^\ast_{rvws} - \left[\Gamma_e\right]^\ast_{rw}\left[\Gamma_e\right]_{pv}\left[C_{eu}\right]_{vwst}+\left[\gamma^{(Y)}_\ell\right]_{pv}\left[C_{\ell u}\right]_{vrst} \nn\\
&+\left[\gamma^{(Y)}_u\right]_{sv}\left[C_{\ell u}\right]_{prvt}+\left[C_{\ell u}\right]_{pvst}\left[\gamma^{(Y)}_\ell\right]_{vr} +\left[C_{\ell u}\right]_{prsv}\left[\gamma^{(Y)}_u\right]_{vt}\,,\\[4mm]
\left[ \beta_{\ell d} \right]_{prst} & = -\frac{1}{3}\gpc \left[ C_{\vp d}\right]_{st} \delta_{pr} - \frac{2}{9}\gpc \left[ C_{\vp\ell}^{(1)}\right]_{pr}\delta_{st}+\frac{8}{9}\gpc \left[C_{\ell\ell}\right]_{prww}\delta_{st}+\frac{4}{9}\gpc \left[C_{\ell\ell}\right]_{pwwr}\delta_{st}\nn\\
&-\frac{4}{9}\gpc\left[C_{\ell q}^{(1)}\right]_{prww}\delta_{st}-\frac{2}{3}\gpc\left[C_{qd}^{(1)}\right]_{wwst}\delta_{pr}+\frac{4}{9}\gpc\left[C_{\ell d}\right]_{prww}\delta_{st}+\frac{2}{3}\gpc\left[C_{\ell d}\right]_{wwst}\delta_{pr}\nn\\
&-\frac{8}{9}\gpc\left[C_{\ell u}\right]_{prww}\delta_{st}+\frac{4}{9}\gpc\left[C_{\ell e}\right]_{prww}\delta_{st}-\frac{4}{3}\gpc\left[C_{ud}^{(1)}\right]_{wwst}\delta_{pr}+\frac{2}{3}\gpc\left[C_{ed}\right]_{wwst}\delta_{pr}\nn\\
&+\frac{4}{3}\gpc\left[C_{dd}\right]_{stww}\delta_{pr}+\frac{4}{9}\gpc\left[C_{dd}\right]_{swwt}\delta_{pr}-2\gpc\left[C_{\ell d}\right]_{prst}-\left[\Gamma_e\Gamma_e^\dagger\right]_{pr}\left[C_{\vp d}\right]_{st}\nn\\
&+2\left[\Gamma_d^\dagger\Gamma_d\right]_{st}\left[C_{\vp\ell}^{(1)}\right]_{pr}-\frac{1}{2}\left[\Gamma_e\right]^\ast_{rv}\left[\Gamma_d\right]_{wt}\left[C_{\ell edq}\right]_{pvsw}-\frac{1}{2}\left[\Gamma_e\right]_{pv}\left[\Gamma_d\right]_{ws}^\ast\left[C_{\ell edq}\right]^\ast_{rvtw}\nn\\
&-2\left[\Gamma_d\right]^\ast_{vs}\left[\Gamma_d\right]_{wt}\left[C_{\ell q}^{(1)}\right]_{prvw}-\left[\Gamma_e\right]^\ast_{rw}\left[\Gamma_e\right]_{pv}\left[C_{ed}\right]_{vwst}+\left[\gamma^{(Y)}_\ell\right]_{pv}\left[C_{\ell d}\right]_{vrst}\nn\\
&+\left[\gamma^{(Y)}_d\right]_{sv}\left[C_{\ell d}\right]_{prvt}+\left[C_{\ell d}\right]_{pvst}\left[\gamma^{(Y)}_\ell\right]_{vr}+\left[C_{\ell d}\right]_{prsv}\left[\gamma^{(Y)}_d\right]_{vt}\,,\\[4mm]
\left[ \beta_{qe} \right]_{prst} & = \frac{1}{9}\gpc \left[ C_{\vp e}\right]_{st} \delta_{pr} - \frac{2}{3}\gpc \left[ C_{\vp q}^{(1)}\right]_{pr}\delta_{st}-\frac{8}{3}\gpc \left[C_{qq}^{(1)}\right]_{prww}\delta_{st}\nn\\
&-\frac{4}{9}\gpc \left[C_{qq}^{(1)}+3\,C_{qq}^{(3)}\right]_{pwwr}\delta_{st}+\frac{4}{3}\gpc\left[C_{\ell q}^{(1)}\right]_{wwpr}\delta_{st}-\frac{2}{9}\gpc\left[C_{\ell e}\right]_{wwst}\delta_{pr}\nn\\
&+\frac{4}{3}\gpc\left[C_{qe}\right]_{prww}\delta_{st}+\frac{2}{9}\gpc\left[C_{qe}\right]_{wwst}\delta_{pr}-\frac{8}{3}\gpc\left[C_{qu}^{(1)}\right]_{prww}\delta_{st}+\frac{4}{3}\gpc\left[C_{qd}^{(1)}\right]_{prww}\delta_{st}\nn\\
&+\frac{4}{9}\gpc\left[C_{eu}\right]_{stww}\delta_{pr}-\frac{2}{9}\gpc\left[C_{ed}\right]_{stww}\delta_{pr}-\frac{8}{9}\gpc\left[C_{ee}\right]_{wwst}\delta_{pr}+2\gpc\left[C_{qe}\right]_{prst}\nn\\
&+\left[\Gamma_u\Gamma_u^\dagger\right]_{pr}\left[C_{\vp e}\right]_{st}-\left[\Gamma_d\Gamma_d^\dagger\right]_{pr}\left[C_{\vp e}\right]_{st}+2\left[\Gamma_e^\dagger\Gamma_e\right]_{st}\left[C_{\vp q}^{(1)}\right]_{pr}\nn\\
&-\frac{1}{2}\left[\Gamma_d\right]_{pw}\left[\Gamma_e\right]^\ast_{vs}\left[C_{\ell edq}\right]_{vtwr}-\frac{1}{2}\left[\Gamma_e\right]_{vt}\left[\Gamma_d\right]_{rw}^\ast\left[C_{\ell edq}\right]^\ast_{vswp}-2\left[\Gamma_e\right]^\ast_{vs}\left[\Gamma_e\right]_{wt}\left[C_{\ell q}^{(1)}\right]_{vwpr}\nn\\
&-\frac{1}{2}\left[\Gamma_u\right]^\ast_{rw}\left[\Gamma_e\right]_{vs}^\ast\left[C_{\ell equ}^{(1)}+12\, C_{\ell equ}^{(3)}\right]_{vtpw}-\frac{1}{2}\left[\Gamma_u\right]_{pw}\left[\Gamma_e\right]_{vt}\left[C_{\ell equ}^{(1)}+12\, C_{\ell equ}^{(3)}\right]^\ast_{vsrw}\nn\\
&-\left[\Gamma_d\right]^\ast_{rw}\left[\Gamma_d\right]_{pv}\left[C_{ed}\right]_{stvw}-\left[\Gamma_u\right]^\ast_{rw}\left[\Gamma_u\right]_{pv}\left[C_{eu}\right]_{stvw}+\left[\gamma^{(Y)}_q\right]_{pv}\left[C_{qe}\right]_{vrst}\nn\\
&+\left[\gamma^{(Y)}_e\right]_{sv}\left[C_{qe}\right]_{prvt}+\left[C_{qe}\right]_{pvst}\left[\gamma^{(Y)}_q\right]_{vr}+\left[C_{qe}\right]_{prsv}\left[\gamma^{(Y)}_e\right]_{vt}\,,\\[4mm]
\left[ \beta_{qu}^{(1)} \right]_{prst} & = \frac{1}{9}\gpc\left[C_{\vp u}\right]_{st}\delta_{pr} +\frac{4}{9}\gpc \left[C_{\vp q}^{(1)}\right]_{pr} \delta_{st}+\frac{16}{9}\gpc\left[C_{qq}^{(1)}\right]_{prww}\delta_{st}\nn\\
&+\frac{8}{27}\gpc\left[C_{qq}^{(1)}+3\,C_{qq}^{(3)}\right]_{pwwr}\delta_{st}-\frac{8}{9}\gpc\left[C_{\ell q}^{(1)}\right]_{wwpr}\delta_{st}-\frac{8}{9}\gpc\left[C_{qe}\right]_{prww}\delta_{st}\nn\\
&-\frac{8}{9}\gpc\left[C_{qd}^{(1)}\right]_{prww}\delta_{st}+\frac{16}{9}\gpc\left[C_{qu}^{(1)}\right]_{prww}\delta_{st}+\frac{2}{9}\gpc\left[C_{qu}^{(1)}\right]_{wwst}\delta_{pr}\nn\\
&-\frac{2}{9}\gpc\left[C_{\ell u}\right]_{wwst}\delta_{pr}-\frac{2}{9}\gpc\left[C_{eu}\right]_{wwst}\delta_{pr}-\frac{2}{9}\gpc\left[C_{ud}^{(1)}\right]_{stww}\delta_{pr}+\frac{8}{9}\gpc\left[C_{uu}\right]_{stww}\delta_{pr}\nn\\
&+\frac{8}{27}\,\gpc\left[C_{uu}\right]_{swwt}\delta_{pr}-\frac{4}{3}\gpc\left[C_{qu}^{(1)}\right]_{prst}-\frac{8}{3}\gsc\left[C_{qu}^{(8)}\right]_{prst}+\frac{1}{3}\left[\Gamma_u\right]^\ast_{rs}\left[\xi_u\right]_{pt}\nn\\
&+\frac{1}{3}\left[\Gamma_u\right]_{pt}\left[\xi_u\right]^\ast_{rs}+\left[\Gamma_u\Gamma_u^\dagger-\Gamma_d\Gamma_d^\dagger\right]_{pr}\left[C_{\vp u}\right]_{st}-2\left[\Gamma_u^\dagger\Gamma_u\right]_{st}\left[C_{\vp q}^{(1)}\right]_{pr}\nn\\
&+\frac{1}{3}\left[\Gamma_u\right]_{pw}\left[\Gamma_u\right]^\ast_{vs}\left[C_{qu}^{(1)}+\frac{4}{3}C_{qu}^{(8)}\right]_{vrwt}+\frac{1}{3}\left[\Gamma_u\right]_{vt}\left[\Gamma_u\right]^\ast_{rw}\left[C_{qu}^{(1)}+\frac{4}{3}C_{qu}^{(8)}\right]_{pvsw}\nn\\
&+\frac{1}{3}\left[\Gamma_d\right]^\ast_{rw}\left[\Gamma_u\right]^\ast_{vs}\left[C_{quqd}^{(1)}+\frac{4}{3}C_{quqd}^{(8)}\right]_{ptvw}+\frac{1}{3}\left[\Gamma_d\right]_{pw}\left[\Gamma_u\right]_{vt}\left[C_{quqd}^{(1)}+\frac{4}{3}C_{quqd}^{(8)}\right]^\ast_{rsvw}\nn\\
&+\frac{1}{2}\left[\Gamma_d\right]^\ast_{rw}\left[\Gamma_u\right]^\ast_{vs}\left[C_{quqd}^{(1)}\right]_{vtpw}+\frac{1}{2}\left[\Gamma_d\right]_{pw}\left[\Gamma_u\right]_{vt}\left[C_{quqd}^{(1)}\right]^\ast_{vsrw}\nn\\
&-\frac{2}{3}\left[\Gamma_u\right]_{vt}\left[\Gamma_u\right]^\ast_{ws}\left[C_{qq}^{(1)}+3C_{qq}^{(3)}\right]_{pvwr}-4\left[\Gamma_u\right]_{wt}\left[\Gamma_u\right]^\ast_{vs}\left[C_{qq}^{(1)}\right]_{prvw} \\
&-\frac{2}{3}\left[\Gamma_u\right]_{pv}\left[\Gamma_u\right]^\ast_{rw}\left[C_{uu}\right]_{vtsw}-2\left[\Gamma_u\right]_{pv}\left[\Gamma_u\right]^\ast_{rw}\left[C_{uu}\right]_{vwst}-\left[\Gamma_d\right]_{pv}\left[\Gamma_d\right]^\ast_{rw}\left[C_{ud}^{(1)}\right]_{stvw}\nn\\
&+\left[\gamma^{(Y)}_q\right]_{pv}\left[C_{qu}^{(1)}\right]_{vrst}+\left[\gamma^{(Y)}_u\right]_{sv}\left[C_{qu}^{(1)}\right]_{prvt}+\left[C_{qu}^{(1)}\right]_{pvst}\left[\gamma^{(Y)}_q\right]_{vr}+\left[C_{qu}^{(1)}\right]_{prsv}\left[\gamma^{(Y)}_u\right]_{vt}\,,\nn\\[4mm]
\left[ \beta_{qd}^{(1)} \right]_{prst} & = \frac{1}{9}\gpc\left[C_{\vp d}\right]_{st}\delta_{pr} -\frac{2}{9}\gpc\left[C_{\vp q}^{(1)}\right]_{pr}\delta_{st}-\frac{8}{9}\gpc\left[C_{qq}^{(1)}\right]_{prww}\delta_{st}\nn\\
&-\frac{4}{27}\gpc\left[C_{qq}^{(1)}+3\,C_{qq}^{(3)}\right]_{pwwr}\delta_{st}+\frac{4}{9}\gpc\left[C_{\ell q}^{(1)}\right]_{wwpr}\delta_{st}+\frac{4}{9}\gpc\left[C_{qe}\right]_{prww}\delta_{st}\nn\\
&-\frac{8}{9}\gpc\left[C_{qu}^{(1)}\right]_{prww}\delta_{st}+\frac{4}{9}\gpc\left[C_{qd}^{(1)}\right]_{prww}\delta_{st}+\frac{2}{9}\gpc\left[C_{qd}^{(1)}\right]_{wwst}\delta_{pr}\nn\\
&-\frac{2}{9}\gpc\left[C_{\ell d}\right]_{wwst}\delta_{pr}-\frac{2}{9}\gpc\left[C_{ed}\right]_{wwst}\delta_{pr}+\frac{4}{9}\gpc\left[C_{ud}^{(1)}\right]_{wwst}\delta_{pr}-\frac{4}{9}\gpc\left[C_{dd}\right]_{stww}\delta_{pr}\nn\\
&-\frac{4}{27}\,\gpc\left[C_{dd}\right]_{swwt}\delta_{pr}+\frac{2}{3}\gpc\left[C_{qd}^{(1)}\right]_{prst}-\frac{8}{3}\gsc\left[C_{qd}^{(8)}\right]_{prst}+\frac{1}{3}\left[\Gamma_d\right]^\ast_{rs}\left[\xi_d\right]_{pt}\nn\\
&+\frac{1}{3}\left[\Gamma_d\right]_{pt}\left[\xi_d\right]^\ast_{rs}+\left[\Gamma_u\Gamma_u^\dagger-\Gamma_d\Gamma_d^\dagger\right]_{pr}\left[C_{\vp d}\right]_{st}+2\left[\Gamma_d^\dagger\Gamma_d\right]_{st}\left[C_{\vp q}^{(1)}\right]_{pr}\nn\\
&+\frac{1}{3}\left[\Gamma_d\right]_{pw}\left[\Gamma_d\right]^\ast_{vs}\left[C_{qd}^{(1)}+\frac{4}{3}C_{qd}^{(8)}\right]_{vrwt}+\frac{1}{3}\left[\Gamma_d\right]_{vt}\left[\Gamma_d\right]^\ast_{rw}\left[C_{qd}^{(1)}+\frac{4}{3}C_{qd}^{(8)}\right]_{pvsw}\nn\\
&+\frac{1}{3}\left[\Gamma_u\right]^\ast_{rw}\left[\Gamma_d\right]^\ast_{vs}\left[C_{quqd}^{(1)}+\frac{4}{3}C_{quqd}^{(8)}\right]_{vwpt}+\frac{1}{3}\left[\Gamma_u\right]_{pw}\left[\Gamma_d\right]_{vt}\left[C_{quqd}^{(1)}+\frac{4}{3}C_{quqd}^{(8)}\right]^\ast_{vwrs}\nn\\
&+\frac{1}{2}\left[\Gamma_d\right]^\ast_{ws}\left[\Gamma_u\right]^\ast_{rv}\left[C_{quqd}^{(1)}\right]_{pvwt}+\frac{1}{2}\left[\Gamma_u\right]_{pv}\left[\Gamma_d\right]_{wt}\left[C_{quqd}^{(1)}\right]^\ast_{rvws}\nn\\
&-\frac{2}{3}\left[\Gamma_d\right]_{vt}\left[\Gamma_d\right]^\ast_{ws}\left[C_{qq}^{(1)}+3C_{qq}^{(3)}\right]_{pvwr}-4\left[\Gamma_d\right]_{wt}\left[\Gamma_d\right]^\ast_{vs}\left[C_{qq}^{(1)}\right]_{prvw} \\
&-\frac{2}{3}\left[\Gamma_d\right]_{pv}\left[\Gamma_d\right]^\ast_{rw}\left[C_{dd}\right]_{vtsw}-2\left[\Gamma_d\right]_{pv}\left[\Gamma_d\right]^\ast_{rw}\left[C_{dd}\right]_{vwst}-\left[\Gamma_u\right]_{pv}\left[\Gamma_u\right]^\ast_{rw}\left[C_{ud}^{(1)}\right]_{vwst}\nn\\
&+\left[\gamma^{(Y)}_q\right]_{pv}\left[C_{qd}^{(1)}\right]_{vrst}+\left[\gamma^{(Y)}_d\right]_{sv}\left[C_{qd}^{(1)}\right]_{prvt}+\left[C_{qd}^{(1)}\right]_{pvst}\left[\gamma^{(Y)}_q\right]_{vr}+\left[C_{qd}^{(1)}\right]_{prsv}\left[\gamma^{(Y)}_d\right]_{vt}\,,\nn\\[4mm]
\left[ \beta_{qu}^{(8)} \right]_{prst} & = \frac{8}{3}\gsc\left[C_{qq}^{(1)}+3C_{qq}^{(3)}\right]_{pwwr}\delta_{st}+\frac{2}{3}\gsc\left[C_{qu}^{(8)}\right]_{prww}\delta_{st}+\frac{2}{3}\gsc\left[C_{qd}^{(8)}\right]_{prww}\delta_{st}\nn\\
&+\frac{4}{3}\gsc\left[C_{qu}^{(8)}\right]_{wwst}\delta_{pr}+\frac{2}{3}\gsc\left[C_{ud}^{(8)}\right]_{stww}\delta_{pr}+\frac{8}{3}\gsc\left[C_{uu}\right]_{swwt}\delta_{pr}\nn\\
&-\left(\frac{4}{3}\gpc+14\gsc\right)\left[C_{qu}^{(8)}\right]_{prst}-12\gsc\left[C_{qu}^{(1)}\right]_{prst}+2\left[\Gamma_u\right]^\ast_{rs}\left[\xi_u\right]_{pt}+2\left[\Gamma_u\right]_{pt}\left[\xi_u\right]^\ast_{rs}\nn\\
&+2\left[\Gamma_u\right]_{pw}\left[\Gamma_u\right]^\ast_{vs}\left[C_{qu}^{(1)}-\frac{1}{6}C_{qu}^{(8)}\right]_{vrwt}+2\left[\Gamma_u\right]_{vt}\left[\Gamma_u\right]^\ast_{rw}\left[C_{qu}^{(1)}-\frac{1}{6}C_{qu}^{(8)}\right]_{pvsw}\nn\\
&+2\left[\Gamma_d\right]^\ast_{rw}\left[\Gamma_u\right]^\ast_{vs}\left[C_{quqd}^{(1)}-\frac{1}{6}C_{quqd}^{(8)}\right]_{ptvw}+2\left[\Gamma_d\right]_{pw}\left[\Gamma_u\right]_{vt}\left[C_{quqd}^{(1)}-\frac{1}{6}C_{quqd}^{(8)}\right]^\ast_{rsvw}\nn\\
&+\frac{1}{2}\left[\Gamma_u\right]^\ast_{vs}\left[\Gamma_d\right]^\ast_{rw}\left[C_{quqd}^{(8)}\right]_{vtpw}+\frac{1}{2}\left[\Gamma_u\right]_{vt}\left[\Gamma_d\right]_{pw}\left[C_{quqd}^{(8)}\right]^\ast_{vsrw} - 4\left[\Gamma_u\right]_{vt}\left[\Gamma_u\right]^\ast_{ws}\left[C_{qq}^{(1)}+3C_{qq}^{(3)}\right]_{pvwr}\nn\\
&-4\left[\Gamma_u\right]_{pv}\left[\Gamma_u\right]^\ast_{rw}\left[C_{uu}\right]_{vtsw}-\left[\Gamma_d\right]_{pv}\left[\Gamma_d\right]^\ast_{rw}\left[C_{ud}^{(8)}\right]_{stvw}+\left[\gamma^{(Y)}_q\right]_{pv}\left[C_{qu}^{(8)}\right]_{vrst}\nn\\
&+\left[\gamma^{(Y)}_u\right]_{sv}\left[C_{qu}^{(8)}\right]_{prvt}+\left[C_{qu}^{(8)}\right]_{pvst}\left[\gamma^{(Y)}_q\right]_{vr}+\left[C_{qu}^{(8)}\right]_{prsv}\left[\gamma^{(Y)}_u\right]_{vt}\,,\\[4mm]
\left[ \beta_{qd}^{(8)} \right]_{prst} & = \frac{8}{3}\gsc\left[C_{qq}^{(1)}+3C_{qq}^{(3)}\right]_{pwwr}\delta_{st}+\frac{2}{3}\gsc\left[C_{qu}^{(8)}\right]_{prww}\delta_{st}+\frac{2}{3}\gsc\left[C_{qd}^{(8)}\right]_{prww}\delta_{st}\nn\\
&+\frac{4}{3}\gsc\left[C_{qd}^{(8)}\right]_{wwst}\delta_{pr}+\frac{2}{3}\gsc\left[C_{ud}^{(8)}\right]_{wwst}\delta_{pr}+\frac{8}{3}\gsc\left[C_{dd}\right]_{swwt}\delta_{pr}\nn\\
&-\left(-\frac{2}{3}\gpc+14\gsc\right)\left[C_{qd}^{(8)}\right]_{prst}-12\gsc\left[C_{qd}^{(1)}\right]_{prst}+2\left[\Gamma_d\right]^\ast_{rs}\left[\xi_d\right]_{pt}+2\left[\Gamma_d\right]_{pt}\left[\xi_d\right]^\ast_{rs}\nn\\
&+2\left[\Gamma_d\right]_{pw}\left[\Gamma_d\right]^\ast_{vs}\left[C_{qd}^{(1)}-\frac{1}{6}C_{qd}^{(8)}\right]_{vrwt}+2\left[\Gamma_d\right]_{vt}\left[\Gamma_d\right]^\ast_{rw}\left[C_{qd}^{(1)}-\frac{1}{6}C_{qd}^{(8)}\right]_{pvsw}\nn\\
&+2\left[\Gamma_u\right]^\ast_{rw}\left[\Gamma_d\right]^\ast_{vs}\left[C_{quqd}^{(1)}-\frac{1}{6}C_{quqd}^{(8)}\right]_{vwpt}+2\left[\Gamma_u\right]_{pw}\left[\Gamma_d\right]_{vt}\left[C_{quqd}^{(1)}-\frac{1}{6}C_{quqd}^{(8)}\right]^\ast_{vwrs}\nn\\
&+\frac{1}{2}\left[\Gamma_d\right]^\ast_{vs}\left[\Gamma_u\right]^\ast_{rw}\left[C_{quqd}^{(8)}\right]_{pwvt}+\frac{1}{2}\left[\Gamma_d\right]_{vt}\left[\Gamma_u\right]_{pw}\left[C_{quqd}^{(8)}\right]^\ast_{rwvs}\nn\\
&-4\left[\Gamma_d\right]_{vt}\left[\Gamma_d\right]^\ast_{ws}\left[C_{qq}^{(1)}+3C_{qq}^{(3)}\right]_{pvwr} \nn\\
&-4\left[\Gamma_d\right]_{pv}\left[\Gamma_d\right]^\ast_{rw}\left[C_{dd}\right]_{vtsw}-\left[\Gamma_u\right]_{pv}\left[\Gamma_u\right]^\ast_{rw}\left[C_{ud}^{(8)}\right]_{vwst}+\left[\gamma^{(Y)}_q\right]_{pv}\left[C_{qd}^{(8)}\right]_{vrst}\nn\\
&+\left[\gamma^{(Y)}_d\right]_{sv}\left[C_{qd}^{(8)}\right]_{prvt}+\left[C_{qd}^{(8)}\right]_{pvst}\left[\gamma^{(Y)}_q\right]_{vr}+\left[C_{qd}^{(8)}\right]_{prsv}\left[\gamma^{(Y)}_d\right]_{vt}\,,
\end{align}

\subsubsection*{$\boxed{\boldsymbol{\left( \bar L R \right)\left( \bar R L \right)}}$}

\begin{align}
\left[ \beta_{\ell edq} \right]_{prst} & = -\left(\frac{8}{3}\gpc+8\gsc\right)\left[ C_{\ell edq} \right]_{prst}-2\left[\Gamma_d\right]_{ts}^*\left[\xi_e\right]_{pr}-2\left[\Gamma_e\right]_{rp}\left[\xi_d\right]_{st}^*+2\left[\Gamma_e\right]_{pv}\left[\Gamma_d\right]_{tw}^*\left[C_{ed}\right]_{vrsw}\nn\\
&-2\left[\Gamma_e\right]_{vr}\left[\Gamma_d\right]_{tw}^*\left[C_{\ell d}\right]_{pvsw}+2\left[\Gamma_e\right]_{vr}\left[\Gamma_d\right]_{ws}^*\left[C_{\ell q}^{(1)}\right]_{pvwt}+6\left[\Gamma_e\right]_{vr}\left[\Gamma_d\right]^\ast_{ws}\left[C_{\ell q}^{(3)}\right]_{pvwt}\nn\\
&-2\left[\Gamma_e\right]_{pw}\left[\Gamma_d\right]^\ast_{vs}\left[C_{qe}\right]_{vtwr}+2\left[\Gamma_d\right]^\ast_{vs}\left[\Gamma_u\right]^\ast_{tw}\left[C_{\ell equ}^{(1)}\right]_{prvw}+ \left[\gamma_{\ell}^{(Y)} \right]_{pv} \left[ C_{\ell edq} \right]_{vrst} \nn \\
&  + \left[\gamma_{q}^{(Y)} \right]_{sv} \left[ C_{\ell edq} \right]_{prvt} + \left[ C_{\ell edq} \right]_{pvst} \left[ \gamma_{e}^{(Y)} \right]_{vr} + \left[ C_{\ell edq} \right]_{prsv} \left[ \gamma_{d}^{(Y)} \right]_{vt}\, ,
\end{align}

\subsubsection*{$\boxed{\boldsymbol{\left( \bar L R \right)\left( \bar L R \right)}}$}

\begin{align}
\left[ \beta_{quqd}^{(1)} \right]_{prst} & = \frac{10}{3} g^\prime \left[ C_{dB} \right]_{st} \left[ \Gamma_u \right]_{pr} - 6g \left[ C_{dW} \right]_{st} \left[ \Gamma_u \right]_{pr} - \frac{20}{9} g^\prime \left[ C_{dB} \right]_{pt} \left[ \Gamma_u \right]_{sr} + 4 g \left[ C_{dW} \right]_{pt} \left[ \Gamma_u \right]_{sr} \nn\\
&- \frac{64}{9} g_s \left[ C_{dG} \right]_{pt} \left[ \Gamma_u \right]_{sr} - \frac{2}{3} g^\prime \left[ C_{uB} \right]_{pr} \left[ \Gamma_d \right]_{st} - 6 g \left[ C_{uW} \right]_{pr} \left[ \Gamma_d \right]_{st} + \frac{4}{9} g^\prime \left[ C_{uB} \right]_{sr} \left[ \Gamma_d \right]_{pt} \nn\\
&+ 4 g \left[ C_{uW} \right]_{sr} \left[ \Gamma_d \right]_{pt} - \frac{64}{9} g_s \left[ C_{uG} \right]_{sr} \left[ \Gamma_d \right]_{pt} -\frac{1}{2}\left(\frac{11}{9} \gpc + 3 \gc + 32 \gsc\right) \left[ C_{quqd}^{(1)} \right]_{prst} \nn\\
& - \frac{1}{3}\left( -\frac{5}{9}\gpc-3\gc+\frac{64}{3}\gsc\right)\left[ C_{quqd}^{(1)} \right]_{srpt} - \frac{4}{9}\left( -\frac{5}{9}\gpc-3\gc+\frac{28}{3}\gsc\right)\left[ C_{quqd}^{(8)} \right]_{srpt} \nn \\
& + \frac{16}{9} \gsc \left[ C_{quqd}^{(8)} \right]_{prst} - 2 \left[\Gamma_u \right]_{pr}\left[\xi_d\right]_{st} - 2 \left[\Gamma_d \right]_{st}\left[\xi_u\right]_{pr} + \frac{4}{3}\left( \left[\Gamma_u\right]_{vr}\left[\Gamma_d\right]_{pw} \left[ C_{qd}^{(1)} + \frac{4}{3}\, C_{qd}^{(8)}\right]_{svwt}\right.\nn\\
&\left.+ \left[\Gamma_d\right]_{vt}\left[\Gamma_u\right]_{sw} \left[ C_{qu}^{(1)} + \frac{4}{3}\, C_{qu}^{(8)}\right]_{pvwr} + \left[\Gamma_d\right]_{pw}\left[\Gamma_u\right]_{sv} \left[ C_{ud}^{(1)} + \frac{4}{3}\, C_{ud}^{(8)}\right]_{vrwt}\right) \nn\\
& +\frac{8}{3} \left[\Gamma_d\right]_{wt} \left[\Gamma_u\right]_{vr} \left( \left[ C_{qq}^{(1)} -3\, C_{qq}^{(3)} \right]_{svpw} - 3 \left[ C_{qq}^{(1)} -3 C_{qq}^{(3)}\right]_{swpv} \right) \nn \\
&-4 \left[\Gamma_d\right]_{sw} \left[\Gamma_u\right]_{pv} \left[ C_{ud}^{(1)}\right]_{vrwt} + \left[\gamma_q^{(Y)}\right]_{pv} \left[C_{quqd}^{(1)}\right]_{vrst} + \left[\gamma_q^{(Y)}\right]_{sv} \left[C_{quqd}^{(1)}\right]_{prvt}\nn \\
& + \left[C_{quqd}^{(1)}\right]_{pvst} \left[\gamma_u^{(Y)}\right]_{vr} + \left[C_{quqd}^{(1)}\right]_{prsv} \left[\gamma_d^{(Y)}\right]_{vt}\, ,\\[4mm]
\left[ \beta_{quqd}^{(8)} \right]_{prst} & = 8 g_s \left[ C_{dG} \right]_{st} \left[ \Gamma_u \right]_{pr} - \frac{40}{3} g^\prime \left[ C_{dB} \right]_{pt} \left[ \Gamma_u \right]_{sr} + 24 g \left[ C_{dW} \right]_{pt} \left[ \Gamma_u \right]_{sr} + \frac{16}{3} g_s \left[ C_{dG} \right]_{pt} \left[ \Gamma_u \right]_{sr}\nn\\
&+ 8 g_s \left[ C_{uG} \right]_{pr} \left[ \Gamma_d \right]_{st} + \frac{8}{3} g^\prime \left[ C_{uB} \right]_{sr} \left[ \Gamma_d \right]_{pt} + 24 g \left[ C_{uW} \right]_{sr} \left[ \Gamma_d \right]_{pt} + \frac{16}{3} g_s \left[ C_{uG} \right]_{sr} \left[ \Gamma_d \right]_{pt}\nn\\
&+ 8\gsc\left[C_{quqd}^{(1)}\right]_{prst}+\left(\frac{10}{9} \gpc + 6 \gc +\frac{16}{3} \gsc\right) \left[C_{quqd}^{(1)}\right]_{srpt}\nn\\
&+\left(-\frac{11}{18}\gpc -\frac{3}{2}\gc +\frac{16}{3}\gsc\right)\left[C_{quqd}^{(8)}\right]_{prst}-\frac{1}{3}\left(\frac{5}{9}\gpc+3\gc+\frac{44}{3}\gsc\right)\left[C_{quqd}^{(8)}\right]_{srpt}\nn\\
&+8\left( \left[\Gamma_u\right]_{vr}\left[\Gamma_d\right]_{pw} \left[ C_{qd}^{(1)} -\frac{1}{6}\, C_{qd}^{(8)}\right]_{svwt}+ \left[\Gamma_d\right]_{vt}\left[\Gamma_u\right]_{sw} \left[ C_{qu}^{(1)} - \frac{1}{6}\, C_{qu}^{(8)}\right]_{pvwr} \right.\nn\\
&\left. + \left[\Gamma_d\right]_{pw}\left[\Gamma_u\right]_{sv} \left[ C_{ud}^{(1)} - \frac{1}{6}\, C_{ud}^{(8)}\right]_{vrwt}\right)
+ 16 \left[\Gamma_d\right]_{wt} \left[\Gamma_u\right]_{vr} \left[ C_{qq}^{(1)} -3 C_{qq}^{(3)}\right]_{svpw} \nn\\
& -4 \left[\Gamma_d\right]_{sw} \left[\Gamma_u\right]_{pv} \left[ C_{ud}^{(8)}\right]_{vrwt} + \left[\gamma_q^{(Y)}\right]_{pv} \left[C_{quqd}^{(8)}\right]_{vrst} \nn \\
&+ \left[\gamma_q^{(Y)}\right]_{sv} \left[C_{quqd}^{(8)}\right]_{prvt} + \left[C_{quqd}^{(8)}\right]_{pvst} \left[\gamma_u^{(Y)}\right]_{vr} + \left[C_{quqd}^{(8)}\right]_{prsv} \left[\gamma_d^{(Y)}\right]_{vt}\, ,\\[4mm]
\left[ \beta_{\ell equ}^{(1)} \right]_{prst} & = -\left(\frac{11}{3}\gpc + 8 \gsc\right) \left[ C_{\ell equ}^{(1)} \right]_{prst}+\left(30\gpc + 18 \gc\right) \left[ C_{\ell equ}^{(3)} \right]_{prst}\nn\\
&+2\left[ \Gamma_u\right]_{st} \left[\xi_e\right]_{pr} + 2 \left[ \Gamma_e\right]_{pr} \left[\xi_u\right]_{st} + 2 \left[\Gamma_d\right]_{sv}\left[\Gamma_u\right]_{wt} \left[ C_{\ell edq} \right]_{prvw} + 2 \left[\Gamma_e\right]_{pv}\left[\Gamma_u\right]_{sw} \left[ C_{eu} \right]_{vrwt}\nn \\
& + 2 \left[\Gamma_e\right]_{vr}\left[\Gamma_u\right]_{wt} \left[ C_{\ell q}^{(1)} \right]_{pvsw} - 6 \left[\Gamma_e\right]_{vr}\left[\Gamma_u\right]_{wt} \left[ C_{\ell q}^{(3)} \right]_{pvsw} -2 \left[\Gamma_e\right]_{vr}\left[\Gamma_u\right]_{sw} \left[ C_{\ell u} \right]_{pvwt}\nn\\
&- 2 \left[\Gamma_e\right]_{pw}\left[\Gamma_u\right]_{vt} \left[ C_{qe} \right]_{svwr} + \left[\gamma_\ell^{(Y)}\right]_{pv} \left[C_{\ell equ}^{(1)}\right]_{vrst} + \left[\gamma_q^{(Y)}\right]_{sv} \left[C_{\ell equ}^{(1)}\right]_{prvt} \nn \\
& + \left[C_{\ell equ}^{(1)}\right]_{pvst} \left[\gamma_e^{(Y)}\right]_{vr} + \left[C_{\ell equ}^{(1)}\right]_{prsv} \left[\gamma_u^{(Y)}\right]_{vt}\, ,\\[4mm]
\left[ \beta_{\ell equ}^{(3)} \right]_{prst} & = \frac{5}{6} g^\prime \left[C_{eB}\right]_{pr}\left[\Gamma_u\right]_{st}-\frac{3}{2} g \left[C_{uW}\right]_{st}\left[\Gamma_e\right]_{pr} -\frac{3}{2} g^\prime \left[C_{uB}\right]_{st}\left[\Gamma_e\right]_{pr} - \frac{3}{2} g\left[C_{eW}\right]_{pr}\left[\Gamma_u\right]_{st}\nn\\ç
&+ \left(\frac{2}{9}\gpc - 3\gc +\frac{8}{3}\gsc\right) \left[C_{\ell equ}^{(3)}\right]_{prst}+\frac{1}{8}\left(5\gpc + 3 \gc\right) \left[C_{\ell equ}^{(1)}\right]_{prst}\nn \\
& -\frac{1}{2} \left[\Gamma_u\right]_{sw}\left[\Gamma_e\right]_{pv} \left[ C_{eu} \right]_{vrwt} -\frac{1}{2} \left[\Gamma_e\right]_{vr}\left[\Gamma_u\right]_{wt} \left[ C_{\ell q}^{(1)} \right]_{pvsw} + \frac{3}{2} \left[\Gamma_e\right]_{vr}\left[\Gamma_u\right]_{wt} \left[ C_{\ell q}^{(3)} \right]_{pvsw}\nn\\
&- \frac{1}{2} \left[\Gamma_e\right]_{vr}\left[\Gamma_u\right]_{sw} \left[ C_{\ell u} \right]_{pvwt} - \frac{1}{2} \left[\Gamma_e\right]_{pw}\left[\Gamma_u\right]_{vt} \left[ C_{qe} \right]_{svwr} + \left[\gamma_\ell^{(Y)}\right]_{pv} \left[C_{\ell equ}^{(3)}\right]_{vrst}  \nn \\
& + \left[\gamma_q^{(Y)}\right]_{sv} \left[C_{\ell equ}^{(3)}\right]_{prvt}+ \left[C_{\ell equ}^{(3)}\right]_{pvst} \left[\gamma_e^{(Y)}\right]_{vr} + \left[C_{\ell equ}^{(3)}\right]_{prsv} \left[\gamma_u^{(Y)}\right]_{vt}\, .
\end{align}

\subsubsection*{$\boxed{\text{Baryon-number-violating}}$}

\begin{align}
\left[ \beta_{d u q \ell} \right]_{prst} & = - \left( \frac{9}{2} \gc + \frac{11}{6} \gpc + 4 \gsc \right) \left[C_{duq\ell} \right]_{prst} - \left[C_{duq\ell} \right]_{vrwt} \left[ \Gamma_d \right]_{sv}^\ast \left[ \Gamma_d \right]_{wp} \nn \\
& - \left[C_{duq\ell} \right]_{pvwt} \left[ \Gamma_u \right]_{sv}^\ast \left[ \Gamma_u \right]_{wr} + \left( 2 \left[C_{duue} \right]_{prwv} + \left[C_{duue} \right]_{pwrv} \right) \left[ \Gamma_e \right]_{tv}^\ast \left[ \Gamma_u \right]_{sw}^\ast \nn \\
& + \left( 4 \left[C_{qqq\ell} \right]_{vwst} + 4 \left[C_{qqq\ell} \right]_{wvst} - \left[C_{qqq\ell} \right]_{vswt} - \left[C_{qqq\ell} \right]_{wsvt} \right) \left[ \Gamma_d \right]_{vp} \left[ \Gamma_u \right]_{wr} \nn \\
& + 2 \left[C_{qque} \right]_{wsrv} \left[ \Gamma_d \right]_{wp} \left[ \Gamma_e \right]_{tv}^\ast + \left[C_{duq\ell} \right]_{vrst} \left[ \Gamma_d^\dagger \Gamma_d \right]_{vp} + \left[C_{duq\ell} \right]_{pvst} \left[ \Gamma_u^\dagger \Gamma_u \right]_{vr} \nn \\
& + \frac{1}{2} \left[C_{duq\ell} \right]_{prvt} \left[ \Gamma_u \Gamma_u^\dagger + \Gamma_d \Gamma_d^\dagger \right]_{vs} + \frac{1}{2} \left[C_{duq\ell} \right]_{prsv} \left[ \Gamma_e \Gamma_e^\dagger \right]_{vt} \, ,\\[4mm]
\left[ \beta_{q q u e} \right]_{prst} & = - \left( \frac{9}{2} \gc + \frac{23}{6} \gpc + 4 \gsc \right) \left[C_{qque} \right]_{prst} + \Bigg[ - \left[C_{qque} \right]_{pwvt} \left[ \Gamma_u \right]_{rv}^\ast \left[ \Gamma_u \right]_{ws} \nn \\
& + \frac{1}{2} \left[C_{duq\ell} \right]_{vspw} \left[ \Gamma_e \right]_{wt} \left[ \Gamma_d \right]_{rv}^\ast - \frac{1}{2} \left( 2 \left[C_{duue} \right]_{vwst} + \left[C_{duue} \right]_{vswt} \right) \left[ \Gamma_d \right]_{pv}^\ast \left[ \Gamma_u \right]_{rw}^\ast \nn \\
& + \frac{1}{2} \left( - 2 \left[C_{qqq\ell} \right]_{prwv} + \left[C_{qqq\ell} \right]_{pwrv} - 2 \left[ C_{qqq\ell} \right]_{wprv} \right) \left[ \Gamma_u \right]_{ws} \left[ \Gamma_e \right]_{vt} \nn \\
& + \frac{1}{2} \left[ C_{qque} \right]_{vrst} \left[ \Gamma_u \Gamma_u^\dagger + \Gamma_d \Gamma_d^\dagger \right]_{vp} + \, p \leftrightarrow r \Bigg] \nn \\
& + \left[ C_{qque} \right]_{prvt} \left[ \Gamma_u^\dagger \Gamma_u \right]_{vs} + \left[ C_{qque} \right]_{prsv} \left[ \Gamma_e^\dagger \Gamma_e \right]_{vt} \, ,\\[4mm]
\left[ \beta_{q q q \ell} \right]_{prst} & = - \left( 3 \gc + \frac{1}{3} \gpc + 4 \gsc \right) \left[C_{qqq\ell} \right]_{prst} - 4 \gc \left( \left[C_{qqq\ell} \right]_{rpst} + \left[C_{qqq\ell} \right]_{srpt} + \left[C_{qqq\ell} \right]_{psrt} \right) \nn \\
& - 4 \left[C_{qque} \right]_{prwv} \left[ \Gamma_e \right]_{tv}^\ast \left[ \Gamma_u \right]_{sw}^\ast + 2 \left[C_{duq\ell} \right]_{vwst} \left( \left[ \Gamma_d \right]_{pv}^\ast \left[ \Gamma_u \right]_{rw}^\ast + \left[ \Gamma_d \right]_{rv}^\ast \left[ \Gamma_u \right]_{pw}^\ast \right) \nn \\
& + \frac{1}{2} \left[C_{qqq\ell} \right]_{vrst} \left[ \Gamma_u \Gamma_u^\dagger + \Gamma_d \Gamma_d^\dagger \right]_{vp} + \frac{1}{2} \left[C_{qqq\ell} \right]_{pvst} \left[ \Gamma_u \Gamma_u^\dagger + \Gamma_d \Gamma_d^\dagger \right]_{vr} \nn \\
& + \frac{1}{2} \left[C_{qqq\ell} \right]_{prvt} \left[ \Gamma_u \Gamma_u^\dagger + \Gamma_d \Gamma_d^\dagger \right]_{vs} + \frac{1}{2} \left[C_{qqq\ell} \right]_{prsv} \left[ \Gamma_e \Gamma_e^\dagger \right]_{vt} \, ,\\[4mm]
\left[ \beta_{d u u e} \right]_{prst} & = - \left( 2 \gpc + 4 \gsc \right) \left[C_{duue} \right]_{prst} - \frac{20}{3} \gpc \left[C_{duue} \right]_{psrt} \nn \\
& + 4 \left[ C_{duq\ell} \right]_{prwv} \left[ \Gamma_u \right]_{ws} \left[ \Gamma_e \right]_{vt} - 8 \left[ C_{qque} \right]_{vwst} \left[ \Gamma_d \right]_{vp} \left[ \Gamma_u \right]_{wr} + \left[ C_{duue} \right]_{vrst} \left[ \Gamma_d^\dagger \Gamma_d \right]_{vp} \nn \\
& + \left[ C_{duue} \right]_{pvst} \left[ \Gamma_u^\dagger \Gamma_u \right]_{vr} + \left[ C_{duue} \right]_{prvt} \left[ \Gamma_u^\dagger \Gamma_u \right]_{vs} + \left[ C_{duue} \right]_{prsv} \left[ \Gamma_e^\dagger \Gamma_e \right]_{vt} \, .
\end{align}

\subsubsection*{$\boxed{\text{SM parameters}}$}

The RGEs for the SM parameters get also modified in the presence of
the dim-6 operators. Using an analogous notation for their $\beta$
functions
\begin{equation}
\frac{dX}{dt} \equiv \frac{1}{16 \pi^2} \, \beta_X \, .
\end{equation}
these are given by
\begin{align}
\beta_{g} & = - \frac{19}{6} g^3 - 8 \, g \, \frac{m^2}{\Lambda^2} \, C_{\vp W} \, , \\[2mm]
\beta_{g^\prime} & = \frac{41}{6} g^{\prime 3} - 8 \, g^\prime \, \frac{m^2}{\Lambda^2} \, C_{\vp B} \, , \\[2mm]
\beta_{g_s} & = -7 g_s^3 - 8 \, g_s \, \frac{m^2}{\Lambda^2} \, C_{\vp G} \, , \\[2mm]
\beta_{\lambda} & = 12 \lambda^2 + \frac{3}{4} g^{\prime 4} +\frac{3}{2} \gpc \gc +\frac{9}{4} g^4 - 3 \left( \gpc + 3 \gc \right) \lambda + 4 \, \lambda \, \gamma_H^{(Y)} \nn \\ 
& - 4 \left( 3 \, \Tr \, \Gamma_d  \Gamma_d^\dagger \Gamma_d \Gamma_d^\dagger + 3 \, \Tr \, \Gamma_u \Gamma_u^\dagger \Gamma_u \Gamma_u^\dagger + \Tr \, \Gamma_e \Gamma_e^\dagger \Gamma_e \Gamma_e^\dagger \right) \nn \\
& + 4 \, \frac{m^2}{\Lambda^2} \Big[ 12 C_\vp + \left( - 16 \lambda + \frac{10}{3} \gc \right) C_{\vp \Box} + \left( 6 \lambda + \frac{3}{2} (\gpc - \gc) \right) C_{\vp D} + 2 \left( \eta_1 + \eta_2 \right) \nn \\
& + 9 \gc C_{\vp W} + 3 \gpc C_{\vp B} + 3 g g^\prime C_{\vp W B} + \frac{4}{3} \gc \left( \Tr \, C_{\vp \ell}^{(3)} + 3 \, \Tr \, C_{\vp q}^{(3)} \right) \Big] \, , \\[2mm]
\beta_{m^2} & = m^2 \left[ 6 \lambda - \frac{9}{2} \gc - \frac{3}{2} \gpc + 2 \gamma_H^{(Y)} + 4 \, \frac{m^2}{\Lambda^2} \left( C_{\vp D} - 2 C_{\vp \Box} \right) \right] \, , \\
\left[ \beta_{\Gamma_u} \right]_{rs} & = \frac{3}{2} \left( \left[ \Gamma_u \Gamma_u^\dagger \Gamma_u \right]_{rs} - \left[ \Gamma_d \Gamma_d^\dagger \Gamma_u \right]_{rs}\right) + \left(\gamma_H^{(Y)} - \frac{9}{4} \gc - \frac{17}{12} \gpc - 8 \gsc \right) \left[ \Gamma_u \right]_{rs} \nn \\
& + 2 \, \frac{m^2}{\Lambda^2} \Bigg[ 3 \left[ C_{u \vp} \right]_{rs} + \frac{1}{2} \left( C_{\vp D} - 2 C_{\vp \Box} \right) \left[ \Gamma_u \right]_{rs} - \left[ C_{\vp q}^{(1) \dagger} \Gamma_u \right]_{rs} + 3 \left[ C_{\vp q}^{(3) \dagger} \Gamma_u \right]_{rs} \nn \\
& + \left[ \Gamma_u C_{\vp u}^\dagger \right]_{rs} - \left[ \Gamma_d C_{\vp u d}^\dagger \right]_{rs} - 2 \left( \left[ C_{q u}^{(1)} \right]_{rpts} + \frac{4}{3} \left[ C_{q u}^{(8)} \right]_{rpts} \right) \left[ \Gamma_u \right]_{pt} - \left[ C_{\ell e q u}^{(1)} \right]_{ptrs} \left[ \Gamma_e \right]_{pt}^\ast \nn \\
& + 3 \left[ C_{q u q d}^{(1)} \right]_{rspt} \left[ \Gamma_d \right]_{pt}^\ast + \frac{1}{2} \left( \left[ C_{q u q d}^{(1)} \right]_{psrt} + \frac{4}{3} \left[ C_{q u q d}^{(8)} \right]_{psrt} \right) \left[ \Gamma_d \right]_{pt}^\ast \Bigg] \, , \\
\left[ \beta_{\Gamma_d} \right]_{rs} & = \frac{3}{2} \left( \left[ \Gamma_d \Gamma_d^\dagger \Gamma_d \right]_{rs} - \left[ \Gamma_u \Gamma_u^\dagger \Gamma_d \right]_{rs} \right) + \left( \gamma_H^{(Y)} - \frac{9}{4} \gc - \frac{5}{12} \gpc - 8 \gsc \right)  \left[ \Gamma_d \right]_{rs} \nn \\
& + 2 \, \frac{m^2}{\Lambda^2} \Bigg[ 3 \left[ C_{d \vp} \right]_{rs} + \frac{1}{2} \left( C_{\vp D} - 2 C_{\vp \Box} \right) \left[ \Gamma_d \right]_{rs} + \left[ C_{\vp q}^{(1) \dagger} \Gamma_d \right]_{rs} + 3 \left[ C_{\vp q}^{(3) \dagger} \Gamma_d \right]_{rs} \nn \\
& - \left[ \Gamma_d C_{\vp d}^\dagger \right]_{rs} - \left[ \Gamma_u C_{\vp u d} \right]_{rs} - 2 \left( \left[ C_{q d}^{(1)} \right]_{rpts} + \frac{4}{3} \left[ C_{q d}^{(8)} \right]_{rpts} \right) \left[ \Gamma_d \right]_{pt} + \left[ C_{\ell e q d}^\ast \right]_{ptsr} \left[ \Gamma_e \right]_{tp}^\ast \nn \\
& + 3 \left[ C_{q u q d}^{(1)} \right]_{ptrs} \left[ \Gamma_u \right]_{pt}^\ast + \frac{1}{2} \left( \left[ C_{q u q d}^{(1)} \right]_{rpts} + \frac{4}{3} \left[ C_{q u q d}^{(8)} \right]_{rpts} \right) \left[ \Gamma_u \right]_{pt}^\ast \Bigg] \, , \\
\left[ \beta_{\Gamma_e} \right]_{rs} & = \frac{3}{2} \left[ \Gamma_e \Gamma_e^\dagger \Gamma_e \right]_{rs} + \left( \gamma_H^{(Y)} - \frac{3}{4} (3 \gc + 5 \gpc) \right) \left[ \Gamma_e \right]_{rs} \nn \\
& + 2 \, \frac{m^2}{\Lambda^2} \Bigg[ 3 \left[ C_{e \vp} \right]_{rs} + \frac{1}{2} \left( C_{\vp D} - 2 C_{\vp \Box} \right) \left[ \Gamma_e \right]_{rs} + \left[ C_{\vp \ell}^{(1) \dagger} \Gamma_e \right]_{rs} + 3 \left[ C_{\vp \ell}^{(3) \dagger} \Gamma_e \right]_{rs} \nn \\
& - \left[ \Gamma_e C_{\vp e}^\dagger \right]_{rs} - 2 \left[ C_{\ell e} \right]_{rpts} \left[ \Gamma_e \right]_{pt} + 3 \left[ C_{\ell e d q} \right]_{rspt} \left[ \Gamma_d \right]_{tp} - 3 \left[ C_{\ell e q u}^{(1)} \right]_{rspt} \left[ \Gamma_u \right]_{pt}^\ast \Bigg] \, .
\end{align}
In the absence of contributions from the dim-6 operators one recovers
the well-known SM RGEs \cite{Machacek:1983tz,Machacek:1983fi,Machacek:1984zw,Luo:2002ey}. Notice however that we do
not include an $SU(5)$ normalization factor for $g^\prime$, and hence
the usual expressions are found by replacing $\gpc = 3/5 \, g_1^2$. Finally, we also have in the SM Lagrangian
\begin{align}
\mathcal{L}_{\theta} =  \frac{\theta^{\prime} g^{\prime 2 }}{  32 \pi^2 }  \widetilde B_{\mu \nu}  B^{\mu \nu} + 
 \frac{\theta  g^{2}}{  32 \pi^2 }  \widetilde W_{\mu \nu}^{I}  W^{\mu \nu}_{I}  +  \frac{\theta_s g_s^{2}}{  32 \pi^2 }  \widetilde G_{\mu \nu}^{A}  G^{\mu \nu}_{A}  
\end{align}
with 
\begin{align}
\beta_{\theta^{\prime}} &= - \frac{128 \pi^2}{\gpc} \, \frac{m^2}{\Lambda^2} \, C_{\vp \widetilde B} \,,  \nonumber \\
\beta_{\theta} &= - \frac{128 \pi^2}{\gc} \, \frac{m^2}{\Lambda^2} \, C_{\vp \widetilde W} \,, \nonumber \\
\beta_{\theta_s} &= - \frac{128 \pi^2}{\gsc} \, \frac{m^2}{\Lambda^2} \, C_{\vp \widetilde G} \,. 
\end{align}

\section{Weak Effective Theory for $B$ physics}
\label{sec:WET}

The Hamiltonian for the WET can be written as
\begin{equation}
\mathcal{H}_{\rm WET} = -\mathcal{L}_{\rm QCD+QED}^{(u,d,c,s,b,e,\mu,\tau)} - \mathcal{L}_{\rm SM}^{(6)}
- \frac{4 G_F}{\sqrt2} \sum_i \Big[ \ C_i\,O_i + \hc \Big]\ ,
\label{HWET}
\end{equation}
where $\mathcal{L}_{\rm QCD+QED}^{(u,d,c,s,b,e,\mu,\tau)}$ is the
usual QCD and QED Lagrangian for the light fermions, $\mathcal{L}_{\rm
  SM}^{(6)}$ contains pure-SM dimension-six operators and the $C_i$
coefficients contain all beyond the Standard Model effects. The
sum over the $i$ index runs over all operators $O_i$ defined
below.\footnote{At the moment \dsix only includes a limited subset of
  WET operators: $\Delta B = \Delta S = 2$, $\Delta B = \Delta C = 1$
  and $\Delta B = \Delta S = 1$ operators.} See Ref.~\cite{Aebischer:2017gaw} for definitions
and conventions.

\subsection{$\Delta B = \Delta S = 2$ operators}
\label{subsec:WET-BS2}

In the case of $\Delta B = \Delta S = 2$ operators, the basis is given
by
\begin{align}
O_1^{sbsb}   \,= & \, ({\bar s} \gamma_{\mu} P_{L} b)\, ({\bar s} \gamma^{\mu} P_{L} b)\,, &
O_5^{sbsb}  \,= & \,({\bar s}_{\alpha} P_{L} b_{\beta})\, ({\bar s}_{\beta}  P_{R} b_{\alpha}) \,,   \notag \\[1mm]
O_2^{sbsb}  \,= & \, ({\bar s}P_{L} b)\, ({\bar s}  P_{L} b)  \,,  &
O_{1^\prime}^{sbsb}   \,= & \, ({\bar s} \gamma_{\mu} P_{R} b)\, ({\bar s} \gamma^{\mu} P_{R} b)\,, \notag \\[1mm]
O_3^{sbsb}  \,= &\, ({\bar s}_{\alpha} P_{L} b_{\beta})\, ({\bar s}_{\beta}  P_{L} b_{\alpha}) \,,  &    
O_{2^\prime}^{sbsb}  \,= & \, ({\bar s}P_{R} b)\, ({\bar s}  P_{R} b)  \,, \notag \\[1mm]
O_4^{sbsb}   \,= & \, ({\bar s} P_{L} b)\, ({\bar s}  P_{R} b)  \,,  &
O_{3^\prime}^{sbsb}  \,= &\, ({\bar s}_{\alpha} P_{R} b_{\beta})\, ({\bar s}_{\beta}  P_{R} b_{\alpha})   \,,
&
\label{opbasisDF2}
\end{align}
where we denote with primed indices the operators with opposite
chirality. $\alpha$ and $\beta$ are $SU(3)_c$ indices.

\subsection{$\Delta B = \Delta C = 1$ operators}
\label{subsec:WET-BC1}

The basis for the $\Delta B = \Delta C = 1$ operators is given by
\begin{align}
O_{1}^{cb\ell\ell} &= \left( \bar{c} \,P_R\, \gamma^\mu \, b \right)  \left( \bar{\ell} \, \gamma_\mu \,\nu_{\ell} \right)\,, &
O_{5}^{cb\ell\ell} &= \left( \bar{c}\, P_R\, b \right)  \left( \bar{\ell}  \,\nu_{\ell} \right)\,, &
\\[1mm]
O_{1^\prime}^{cb\ell\ell} &= \left( \bar{c} \, P_L\, \gamma^\mu \, b \right)  \left( \bar{\ell} \, \gamma_\mu \,\nu_{\ell} \right)\,, &
O_{5^\prime}^{cb\ell\ell} &= \left( \bar{c}\, P_L \,b \right)  \left( \bar{\ell} \, \nu_{\ell} \right)\,, &
O_{7^\prime}^{cb\ell\ell} &= \left( \bar{c} \, P_L\, \sigma^{\mu\nu}\,b \right)  \left( \bar{\ell} \, \sigma_{\mu\nu} \,\nu_{\ell} \right)\,,
\notag
\end{align}
with $\ell\in\{e,\mu,\tau\}$.

\subsection{$\Delta B = \Delta S = 1$ operators}
\label{subsec:WET-BS1}

There are three classes of $\Delta B = \Delta S = 1$ operators: Magnetic, hadronic (4-quark) and semileptonic operators. These are chosen as:

\begin{itemize}
\item Magnetic penguins:
\end{itemize}
\begin{align}
O_{7\gamma}^{sb} &= \frac{e}{g_s^2} m_b \,  ( \bar{s}  \, P_R\, \sigma_{\mu\nu}  \,  b ) \; F^{\mu\nu},&
O_{8g}^{sb} &= \frac{1}{g_s} m_b \, (\bar{s} \, P_R\, \sigma_{\mu\nu} \,\, T^{A} \, b ) \; G^{\mu\nu}_A\,,\notag\\[1mm]
O_{7^\prime \gamma}^{sb} &= \frac{e}{g_s^2} m_b \,  ( \bar{s}  \, P_L\, \sigma_{\mu\nu} \,  b ) \; F^{\mu\nu},&
O_{8^\prime g}^{sb} &= \frac{1}{g_s} m_b \, (\bar{s} \, P_L\, \sigma_{\mu\nu} \,\, T^{A} \, b ) \; G^{\mu\nu}_A\,, \label{sbM}
\end{align}
$T^A$ ($A=1,\dots,8$) are the $SU(3)$ generators.

\begin{itemize}
\item Hadronic ($q \neq s$):
\end{itemize}
\begin{align}
O_1^{sbqq} &= (\bar{s}\,P_R\, \gamma_\mu  b)\;   ( \bar{q} \gamma^\mu q )\,, &
O_2^{sbqq} &=(\bar{s}\,P_R\,  \gamma_\mu \,T^{\mysmall A}b)\; (\bar{q} \gamma^\mu\, T^{\mysmall A} q )\,, \notag\\[1mm]
O_3^{sbqq}&=(\bar{s}\,P_R\,\gamma_{\mu\nu\rho} \, b)\;( \bar{q} \gamma^{\mu\nu\rho} q)\,, &
O_4^{sbqq} &=(\bar{s}\,P_R\, \gamma_{\mu\nu\rho}  T^{\mysmall A}b)\;(\bar{q} \gamma^{\mu\nu\rho}\, T^{\mysmall A} q)\,, \notag\\[1mm]
O_5^{sbqq} &= (\bar{s}\,P_R\, b) (\bar{q}\, q)\,, &
O_6^{sbqq} &=(\bar{s}\,P_R\,T^{\mysmall A} b) (\bar{q}\, T^{\mysmall A}q)\; ,  \notag\\[1mm]
O_7^{sbqq}&=(\bar{s}\,P_R\, \sigma^{\mu\nu} \, b) (\bar{q} \,\sigma_{\mu\nu} \,q)\,, &
O_8^{sbqq} &=(\bar{s}\,P_R\,\sigma^{\mu\nu} \,T^{\mysmall A} b) (\bar{q} \sigma_{\mu\nu}\,T^{\mysmall A} q)\,,\label{sbqq}\\[1mm]
O_9^{sbqq}&=(\bar{s}\,P_R\,\gamma_{\mu\nu\rho\sigma}\,b)\;( \bar{q} \gamma^{\mu\nu\rho\sigma} q)\,, &
O_{10}^{sbqq} &=(\bar{s}\,P_R\,\gamma_{\mu\nu\rho\sigma} \, T^{\mysmall A} b)\;(\bar{q} \gamma^{\mu\nu\rho\sigma}\, T^{\mysmall A}  q)\,, \notag
\end{align}
where $q=\{ u, d, c\}$. In the case of $q=b$, the color-octet
operators $O^{sbbb}_{2,4,6,8,10}$ are Fierz-equivalent to the
color-singlet ones and are not included in the basis.  In addition,
the analogous set with opposite chirality is needed:
\begin{equation}
O_{i^\prime}^{sbqq} = O_{i}^{sbqq} \Big|_{P_{L,R}\to P_{R,L}}\,,
\end{equation}

\begin{itemize}
\item Hadronic ($q = s$):
\end{itemize}
\begin{align}
O_1^{sbss} &= (\bar{s}\, \gamma_\mu\,P_L \, b)\;   ( \bar{s} \gamma^\mu \,s )\,, &
O_{1^\prime}^{sbss} &=(\bar{s}\, \gamma_\mu\,P_R \,b)\; (\bar{s} \gamma^\mu\, s )\,, \notag\\[1mm]
O_3^{sbss}&=(\bar{s}\,\gamma_{\mu\nu\rho}\,P_L \, b)\;( \bar{s} \gamma^{\mu\nu\rho} s)\,, &
O_{3^\prime}^{sbss} &=(\bar{s}\,\, \gamma_{\mu\nu\rho}\,P_R \, b)\;(\bar{s} \gamma^{\mu\nu\rho}\, s)\,, \notag\\[1mm]
O_5^{sbss} &= (\bar{s}\,P_L\, b) (\bar{s}\, s)\,, &
O_{5^\prime}^{sbss} &=(\bar{s}\,P_R\, b) (\bar{s}\, s)\; ,  \notag\\[1mm]
O_7^{sbss}&=(\bar{s}\, \sigma^{\mu\nu}\,P_L\, b) (\bar{s} \,\sigma_{\mu\nu} \,s)\,, &
O_{7^\prime}^{sbss} &=(\bar{s}\,\sigma^{\mu\nu}\,P_R\, b) (\bar{s} \,\sigma_{\mu\nu} \,s)\,,\notag \\[1mm]
O_9^{sbss}&=(\bar{s}\,\gamma_{\mu\nu\rho\sigma}\,P_L \, b)\;( \bar{s} \gamma^{\mu\nu\rho\sigma} s)\,, &
O_{9^\prime}^{sbss} &=(\bar{s}\,\gamma_{\mu\nu\rho\sigma}\,P_R \, b)\;( \bar{s} \gamma^{\mu\nu\rho\sigma} s)\,. \label{sbss}
\end{align}
Again, the color-octet operators are Fierz-redundant and have been ommited.

\begin{itemize}
\item Semileptonic:
\end{itemize}
\begin{align}
O_1^{sb\ell\ell} &= (\bar{s}\, P_R\, \gamma_\mu \, b)\;   ( \bar{\ell} \gamma^\mu \ell )\,, &
O_{1^\prime}^{sb\ell\ell} &=(\bar{s}\, P_L \gamma_\mu \,b)\; (\bar{\ell} \gamma^\mu\, \ell )\,, \notag\\[1mm]
O_3^{sb\ell\ell}&=(\bar{s}\,P_R\,\gamma_{\mu\nu\rho}  \,b)\;( \bar{\ell} \gamma^{\mu\nu\rho} \ell)\,, &
O_{3^\prime}^{sb\ell\ell} &=(\bar{s}\,P_L\, \gamma_{\mu\nu\rho}  \,b)\;(\bar{\ell} \gamma^{\mu\nu\rho}\, \ell)\,, \notag\\[1mm]
O_5^{sb\ell\ell} &= (\bar{s}\,P_R \,b) (\bar{\ell}\, \ell)\,, &
O_{5^\prime}^{sb\ell\ell} &=(\bar{s}\,P_L \,b) (\bar{\ell}\, \ell)\; ,  \notag\\[1mm]
O_7^{sb\ell\ell}&=(\bar{s}\, P_R\, \sigma^{\mu\nu} \,b) (\bar{\ell} \,\sigma_{\mu\nu} \,\ell')\,, &
O_{7^\prime}^{sb\ell\ell} &=(\bar{s}\, P_L\,\sigma^{\mu\nu}  \,b) (\bar{\ell} \,\sigma_{\mu\nu} \,\ell)\,, \notag\\[1mm]
O_9^{sb\ell\ell}&=(\bar{s}\, P_R\,\gamma_{\mu\nu\rho\sigma}  \,b)\;( \bar{\ell} \gamma^{\mu\nu\rho\sigma} \ell)\,, &
O_{9^\prime}^{sb\ell\ell} &=(\bar{s}\,P_L\,\gamma_{\mu\nu\rho\sigma}  \,b)\;( \bar{\ell} \gamma^{\mu\nu\rho\sigma} \ell)\,,
\label{sbll}
\end{align}
with $\ell \in \{ e,\mu,\tau \}$.

\newpage
\section{\dsixbf routines}
\label{ap:routines}

In this Appendix we list the public routines implemented in \dsix. The
user can take advantage of them when writing a project with \dsix.

\subsection{General \dsix routines}
\label{ap:routines-general}

\noindent {\bf LoadModule[moduleName]}

\begin{myroutine}
Loads the \dsix module {\tt moduleName}. Once this command is
evaluated, the module is initialized and all its routines are
available to the user. \\

\underline{Requires}: Nothing. This routine can be used as soon as
\dsix is loaded. \\

\underline{Arguments}: In the current version of \dsix, {\tt
  moduleName} can either be {\tt "SMEFTrunner"}, {\tt "EWmatcher"} or {\tt
  "WETrunner"}. \\

\underline{Example}: {\tt LoadModule["SMEFTrunner"]} loads the {\tt
  SMEFTrunner} module.
\end{myroutine}

\noindent {\bf MyPrint[string]}

\begin{myroutine}
Prints the message {\tt string}. It can be switched on and off by using
the \dsix routines {\tt TurnOnMessages} and {\tt TurnOffMessages}. \\

\underline{Requires}: Nothing. This routine can be used as soon as
\dsix is loaded. \\

\underline{Arguments}: {\tt string} must be a valid \mathe string. \\

\underline{Example}: {\tt MyPrint["I am using DsixTools"]} prints a
simple message.
\end{myroutine}

\noindent {\bf TurnOnMessages}

\begin{myroutine}
Turns on the messages written by the \dsix routines. \\

\underline{Requires}: Nothing. This routine can be used as soon as
\dsix is loaded.
\end{myroutine}

\noindent {\bf TurnOffMessages}

\begin{myroutine}
Turns off the messages written by the \dsix routines. The only
exception are the messages written when a new module is loaded, which
cannot be switched off. \\

\underline{Requires}: Nothing. This routine can be used as soon as
\dsix is loaded.
\end{myroutine}

\newpage

\noindent {\bf ReadInputFiles[options\_file,WCsInput\_file,\{SMInput\_file\}]}

\begin{myroutine}
Reads all input files. \\

\underline{Requires}: Nothing. This routine can be used as soon as
\dsix is loaded. \\

\underline{Arguments}: {\tt options\_file} and {\tt WCsInput\_file}
are the paths to the options and WCs input files. {\tt WCsInput\_file}
can contain either SMEFT or WET WCs. The {\tt SMInput\_file} argument
is the path to the input card with the values of the SM parameters and
is optional: it should be absent when this routine is used to read WET
input. \\

\underline{Example}: {\tt ReadInputFiles["Options.dat","WCsInput.dat","SMInput.dat"]} reads the content of three SMEFT
input files. {\tt ReadInputFiles["Options.dat","WCsInput.dat"]} reads the content of two WET input files.
\end{myroutine}

\noindent {\bf WriteInputFiles[options\_file,WCsInput\_file,\{SMInput\_file\},data]}

\begin{myroutine}
Creates input files with the parameter values in {\tt data}. This
routine can be used to export the input values defined by the user in
a project to external text files, later to be read by \dsix. \\

\underline{Requires}: Nothing. This routine can be used as soon as
\dsix is loaded. \\

\underline{Arguments}: {\tt options\_file} and {\tt WCsInput\_file}
are the paths to the options and WCs input files created by this
routine. {\tt data} is an array where the input values have a precise
ordering, defined by the ordering in the {\tt Parameters} (for SMEFT
input) and {\tt WETParameters} (for WET input) global arrays, see
\ref{ap:parameters}. The {\tt SMInput\_file} argument is the path to
the input card with the values of the SM parameters and is optional:
it should be absent when this routine is used to write WET input. \\

\underline{Example}: {\tt WriteInputFiles["Options.dat","WCsInput.dat","SMInput.dat", data]} exports the SMEFT values in
          {\tt data} to three input files. {\tt
            WriteInputFiles["Options.dat", "WCsInput.dat", data]}
          exports the WET values in {\tt data} to two input
          files. These input files can later be read by \dsix.
\end{myroutine}

\noindent {\bf WriteAndReadInputFiles[options\_file,WCsInput\_file,\{SMInput\_file\}]}

\begin{myroutine}
Writes data into new input files and then reads them. In essence,
running this routine is equivalent to running {\tt WriteInputFiles}
and {\tt ReadInputFiles}. \\

\underline{Requires}: Before using the routine the data to be exported
(and later read) must be declared. In case of SMEFT input, all SM
parameters must be set to their input values by giving values to
variables of the form {\tt Init[SM\_parameter]} and all WC values will
be assumed to vanish unless they are declared in a similar way. In
case of WET input, only the WC values must be declared. Finally, the
options will be assumed to take default values unless set to other
choices. \\

\underline{Arguments}: {\tt options\_file} and {\tt WCsInput\_file}
are the paths to the options and WCs input files created by this
routine. The {\tt SMInput\_file} argument is the path to the input
card with the values of the SM parameters and is optional: it should
be absent when this routine is used to write and read WET input.\\

\underline{Example}: {\tt WriteAndReadInputFiles["Options.dat","WCsInput.dat","SMInput.dat"]} exports the previously declared
values to three SMEFT input files and then reads them.\\
{\tt WriteAndReadInputFiles["Options.dat", "WCsInput.dat"]} exports the
previously declared values to two WET input files and then reads them.
\end{myroutine}

\noindent {\bf NewInput[parameter,newvalue,dispatch]}

\begin{myroutine}
Replaces the current input (contained in the \mathe dispatch {\tt
  dispatch}) by a new one in which {\tt parameter} takes the value
{\tt newvalue}. This routine is particularly useful for projects
involving loops with varying parameters. \\

\underline{Requires}: Input values must have been introduced before
using this routine, for instance using the {\tt ReadInputFiles}
routine. Only in this way the dispatch {\tt dispatch} would have been
defined. \\

\underline{Arguments}: {\tt parameter} must be a valid name for a
\dsix parameter. These can be found in the global arrays {\tt
  Parameters} (in case of SMEFT input) and {\tt WETparameters} (in
case of WET input), see \ref{ap:parameters}. {\tt newvalue} must be a
valid number (in general complex). {\tt dispatch} is the name of the
\mathe dispatch where the input values are saved. By default, this is
       {\tt input}. In the {\tt SMEFTrunner} module there is also the
       internal dispatch {\tt inputSMEFTrunner}. \\

\underline{Example}: {\tt NewInput[LL[1,1,1,2],1.0,input]} changes the
input value (at the high-energy scale $\Lambda$) of $\left[ C_{\ell
    \ell} \right]_{1112}$ to $1$.
\end{myroutine}

\noindent {\bf NewScale[scale,newvalue]}

\begin{myroutine}
Replaces the current value of {\tt scale} by {\tt newvalue}. This
routine is particularly useful for projects involving loops with
varying energy scales. \\

\underline{Requires}: Nothing. This routine can be used as soon as
\dsix is loaded. \\

\underline{Arguments}: {\tt scale} can be either {\tt "high"} or {\tt
  "low"}. {\tt newvalue} must be a valid positive real number. As with
other dimensionful parameters, the scale should be given in GeV.\\

\underline{Example}: {\tt NewScale["high",1000]} changes the high
energy scale $\Lambda$ to $1$ TeV.
\end{myroutine}

\noindent {\bf H[mat]}

\begin{myroutine}
Returns the Hermitian conjugate of the matrix {\tt mat}. If the option {\tt CPV} has been set to {\tt 0} then it returns the transpose. \\

\underline{Requires}: The global variable {\tt CPV} must have been set
to either {\tt 0} (for real parameters) or {\tt 1} (for complex
parameters) before using this function. \\

\underline{Arguments}: {\tt mat} must be a valid matrix. \\

\underline{Example}: {\tt H[Gu]} returns the Hermitian conjugate of
the up-quarks Yukawa matrix $\Gamma_u$.
\end{myroutine}

\noindent {\bf CC[x]}

\begin{myroutine}
Returns the complex conjugate of {\tt x}. If the option {\tt CPV} has been set to {\tt 0} then it returns {\tt x}. \\

\underline{Requires}: The global variable {\tt CPV} must have been set
to either {\tt 0} (for real parameters) or {\tt 1} (for complex
parameters) before using this function. \\

\underline{Arguments}: {\tt x} must be a valid complex number. \\

\underline{Example}: {\tt CC[GU[1,2]/.input]} returns $\left[ \Gamma_u
  \right]_{12}^\ast$ if {\tt CPV} has been set to {\tt 1} and $\left[
  \Gamma_u \right]_{12}$ if {\tt CPV} has been set to {\tt 0}, using
for $\left[ \Gamma_u \right]_{12}$ its input value.
\end{myroutine}

\subsection{\SMEFTrunner routines}
\label{ap:routines-SMEFTrunner}

\noindent {\bf InitializeSMEFTrunnerInput}

\begin{myroutine}
Initializes the input for the \SMEFTrunner module. This initialization creates the
dispatches {\tt inputSMEFTrunner} and {\tt inputSMEFTrunnerSM}. The
user does not need to use this routine, as it is automatically
executed when the \SMEFTrunner module is loaded. \\

\underline{Requires}: The \SMEFTrunner module must have been loaded in
order to use this routine.
\end{myroutine}


\noindent {\bf FindParameterSMEFT[parameter]}

\begin{myroutine}
Returns the position of {\tt parameter} in the {\tt Parameters}
list. If one uses a generic element name as argument, {\tt
  FindParameterSMEFT} will return a list with all positions where it
appears in {\tt Parameters}. \\

\underline{Requires}: Nothing. This routine can be used as soon as
\dsix is loaded. \\

\underline{Arguments}: {\tt parameter} can be either the name of a SM
parameter or SMEFT WC, or the generic name of an element of one of the
2- or 4-fermion parameters.  \\

\underline{Example}: {\tt FindParameterSMEFT[LQ3[2, 2, 2, 3]]} returns
          {\tt \{355\}}, the position of $\left[ C_{\ell q}^{(3)}
            \right]_{2223}$ in the {\tt Parameters} list. {\tt
            FindParameterSMEFT[LQ3]} returns {\tt \{327, 328, \dots,
            371\}}, since the $C_{\ell q}^{(3)}$ WCs are saved in
          these positions of the {\tt Parameters} list.
\end{myroutine}

\noindent {\bf GetBeta}

\begin{myroutine}
Computes the SMEFT $\beta$ functions analytically. After running this
routine they can be read by evaluating quantities of the form {\tt
  $\beta$[parameter]}. For instance, the $C_{\ell q}^{(1)}$ $\beta$
functions can be obtained by evaluating {\tt $\beta$[lq1]}. Table
\ref{tab:SMEFTparameters} contains the names used for all SMEFT
$\beta$ functions. \\

\underline{Requires}: The \SMEFTrunner module must have been loaded in
order to use this routine. Moreover, the global variable {\tt CPV}
must have been set to either {\tt 0} (for real parameters) or {\tt 1}
(for complex parameters) before using this function.
\end{myroutine}

\noindent {\bf LoadBetaFunctions}

\begin{myroutine}
Constructs the SMEFT $\beta$ functions (using {\tt GetBeta}
internally) or reads them from a file. \\

\underline{Requires}: The \SMEFTrunner module must have been loaded
and the option {\tt ReadRGEs} set in order to use this routine.
\end{myroutine}

\noindent {\bf RunRGEsSMEFT}

\begin{myroutine}
Runs the SMEFT RGEs. \\

\underline{Requires}: The \SMEFTrunner module must have been loaded
and the option {\tt RGEsMethod} set in order to use this routine.
\end{myroutine}

\noindent {\bf ExportSMEFTrunner}

\begin{myroutine}
Exports the \SMEFTrunner results to the output file {\tt Output\_SMEFTrunner.dat}. \\

\underline{Requires}: The \SMEFTrunner module must have been loaded
and the option {\tt exportSMEFTrunner} set in order to use this routine.
\end{myroutine}

\newpage

\noindent {\bf WriteSMEFTrunnerOutputFile[Output\_file,data]}

\begin{myroutine}
Exports the \SMEFTrunner results in {\tt data} to {\tt Output\_file}. \\

\underline{Requires}: The \SMEFTrunner module must have been loaded in
order to use this routine. \\

\underline{Arguments}: {\tt Output\_file} is the path to the
\SMEFTrunner output file. {\tt data} is an array containing the
results obtained with \SMEFTrunner. It must contain the same of
elements and ordered in exactly the same way as {\tt outSMEFTrunner}. \\

\underline{Example}: \\[1mm]
{\tt WriteSMEFTrunnerOutputFile["Output\_SMEFTrunner.dat",outSMEFTrunner/.t->tLOW]}
exports the data saved in the {\tt outSMEFTrunner} array, evaluated at
$\mu = \mu_{\rm EW}$, into the text file {\tt Output\_SMEFTrunner.dat}.
\end{myroutine}

\subsection{\EWmatcher routines}
\label{ap:routines-EWmatcher}

\noindent {\bf InitializeEWmatcherInput}

\begin{myroutine}
Initializes the input for the \EWmatcher module. The user does not
need to use this routine, as it is automatically executed when the
\EWmatcher module is loaded. \\

\underline{Requires}: The \EWmatcher module must have been loaded in
order to use this routine.
\end{myroutine}

\noindent {\bf FindParameterWET[parameter]}

\begin{myroutine}
Returns the position of {\tt parameter} in the {\tt WETParameters}
list. If one uses a generic element name as argument, {\tt
  FindParameterWET} will return a list with all positions where it
appears in {\tt WETParameters}. \\

\underline{Requires}: Nothing. This routine can be used as soon as
\dsix is loaded. \\

\underline{Arguments}: {\tt parameter} must be the name of a WET WC.  \\

\underline{Example}: {\tt FindParameterWET[CBS1[$\mu$][1]]} returns
          {\tt \{71\}}, the position of $C_1^{sb\mu\mu}$ within the {\tt
            WETParameters} list. {\tt FindParameterWET[CBS1[$\mu$]]} returns 
            {\tt \{71, 72, 73, 74, 75\}}, since the $C_i^{sb\mu\mu}$ WCs are saved in
          these positions of the {\tt WETParameters} list.
\end{myroutine}

\noindent {\bf RotateToMassBasis}

\begin{myroutine}
Transforms the SMEFT WCs to the fermion mass basis. It also creates
the replacements array {\tt ToMassBasis}, which can be used to print
the numerical values of the SMEFT WCs in the fermion mass basis.\\

\underline{Requires}: The \EWmatcher module must have been loaded in
order to use this routine.
\end{myroutine}

\noindent {\bf Biunitary[mat,dim]}

\begin{myroutine}
Applies a biunitary transformation that diagonalizes the {\tt dim} $\times$ {\tt dim}
matrix {\tt mat}. It returns three outputs: the square root of the
eigenvalues of {\tt mat}$^\dagger$ {\tt mat} (or, equivalently, {\tt
  mat} {\tt mat}$^\dagger$) and the unitary matrices $U_L$ and $U_R$
that diagonalize {\tt mat} as $U_L^\dagger$ {\tt mat} $U_R = \big(
${\tt mat}$ \big)_{\rm diag}$. \\

\underline{Requires}: The \EWmatcher module must have been loaded in order to use this function. \\

\underline{Arguments}: {\tt mat} must be a valid {\tt dim} $\times$
          {\tt dim} matrix. \\

\underline{Example}: {\tt Biunitary[Gu/.input,3]} returns the square
root of the eigenvalues of $\Gamma_u^\dagger \Gamma_u$ (or,
equivalently, $\Gamma_u \Gamma_u^\dagger$), as well as the matrices
$U_L$ and $U_R$ that diagonalize $\Gamma_u$ as $U_L^\dagger \Gamma_u
U_R = \left( \Gamma_u \right)_{\rm diag}$. Here $\Gamma_u$ is the
input value for the up-quarks Yukawa matrix.
\end{myroutine}

\noindent {\bf ApplyEWmatching}

\begin{myroutine}
Matches the SMEFT WCs onto the WET WCs. After using this routine
several arrays containing the numerical values of the WET WCs at the
electroweak scale are created. The result of this step can also be accessed by using the {\tt Match} and {\tt MatchAnalytical} functions. \\

\underline{Requires}: The \EWmatcher module must have been loaded in order to use this function.
\end{myroutine}

\noindent {\bf Match[WC]}

\begin{myroutine}
Returns the value of the WET Wilson coefficient {\tt WC} after matching it to the SMEFT. \\

\underline{Requires}: The \EWmatcher module must have been loaded in order to use this function. Furthermore, the {\tt ApplyEWmatching} routines must have been run before using this function. \\

\underline{Arguments}: {\tt WC} must be the name of a valid SMEFT WC. \\

\underline{Example}: {\tt Match[CBS1[d][1]]} prints the numerical value of the $C_1^{sbdd}$ WC.
\end{myroutine}

\noindent {\bf MatchAnalytical[WC]}

\begin{myroutine}
Returns the analytical expression of the WET Wilson coefficient {\tt WC} after matching it to the SMEFT. Only the energy scales and the SM parameters are replaced by numerical values in this function. \\

\underline{Requires}: The \EWmatcher module must have been loaded in order to use this function. Furthermore, the {\tt ApplyEWmatching} routines must have been run before using this function. \\

\underline{Arguments}: {\tt WC} must be the name of a valid SMEFT WC. \\

\underline{Example}: {\tt Match[CBS1[d][1]]} prints the analytical expression of the $C_1^{sbdd}$ WC.
\end{myroutine}

\noindent {\bf WriteWCsMassBasisOutputFile[Output\_file]}

\begin{myroutine}
Exports the SMEFT 2- and 4-fermion WCs in the fermion mass basis to {\tt Output\_file}. \\

\underline{Requires}: The \EWmatcher module must have been loaded in order to use this function. Furthermore, the {\tt ApplyEWmatching} routines must have been run before using this function. \\

\underline{Arguments}: {\tt Output\_file} is the path to the text file
where the SMEFT WCs in the fermion mass basis will be exported. \\

\underline{Example}:
{\tt WriteWCsMassBasisOutputFile["Output\_WCsMassBasis.dat"]} will export the SMEFT
WCs in the fermion mass basis to the file {\tt "Output\_WCsMassBasis.dat"}.
\end{myroutine}

\noindent {\bf ExportEWmatcher}

\begin{myroutine}
Exports the \EWmatcher results to the output file {\tt Output\_EWmatcher.dat}. \\

\underline{Requires}: The \EWmatcher module must have been loaded
and the option {\tt exportEWmatcher} set in order to use this routine.
\end{myroutine}

\noindent {\bf WriteEWmatcherOutputFile[Output\_file,data]}

\begin{myroutine}
Exports the \EWmatcher results in {\tt data} to {\tt Output\_file}. \\

\underline{Requires}: The \EWmatcher module must have been loaded in
order to use this routine. \\

\underline{Arguments}: {\tt Output\_file} is the path to the
\EWmatcher output file. {\tt data} is an array containing the results
obtained with \EWmatcher. The precise ordering of the WET WCs in {\tt
  data} is given by that in {\tt WETParameters}, see
\ref{ap:parameters}. \\

\underline{Example}: {\tt
  WriteEWmatcherOutputFile["Output\_EWmatcher.dat",dataOutput]}
exports the data saved in the {\tt dataOutput} array to the {\tt
  Output\_EWmatcher.dat} text file.
\end{myroutine}

\subsection{\WETrunner routines}
\label{ap:routines-WETrunner}

\noindent {\bf InitializeWETrunnerInput}

\begin{myroutine}
Initializes the input for the \WETrunner module. The user does not
need to use this routine, as it is automatically executed when the
\WETrunner module is loaded. \\

\underline{Requires}: The \WETrunner module must have been loaded in
order to use this routine.
\end{myroutine}

\noindent {\bf RunRGEsWET}

\begin{myroutine}
Runs the WET RGEs. \\

\underline{Requires}: The \WETrunner module must have been loaded in
order to use this routine.
\end{myroutine}

\noindent {\bf ExportWETrunner}

\begin{myroutine}
Exports the \WETrunner results to the output file {\tt
  Output\_WETrunner.dat}. \\

\underline{Requires}: The \WETrunner module must have been loaded
and the option {\tt exportWETrunner} set in order to use this routine.
\end{myroutine}

\noindent {\bf WriteWETrunnerOutputFile[Output\_file,data,scale]}

\begin{myroutine}
Exports the \WETrunner results in {\tt data} to {\tt Output\_file}  after evaluating them at $\mu =$ {\tt scale}. \\

\underline{Requires}: The \WETrunner module must have been loaded in
order to use this routine. \\

\underline{Arguments}: {\tt Output\_file} is the path to the
\WETrunner output file. {\tt data} is an array containing the results
obtained with \WETrunner. The precise ordering of the WET WCs in {\tt
  data} is given by {\tt BS2Low}-{\tt BC1Low}-{\tt
  BS1unprimedLow}-{\tt BS1primedLow}, following exactly the ordering
in {\tt WETParameters} (see \ref{ap:parameters}), and they must be
  evaluated at $t = \log_{10}$ {\tt scale}. \\

\underline{Example}:\\[1mm]
{\tt WriteWETrunnerOutputFile["Output\_WETrunner.dat",dataOutput,mb]}
will export the data saved in the {\tt dataOutput} array, evaluated at
$\mu = m_b$, into the {\tt Output\_WETrunner.dat}  text file.
\end{myroutine}

\newpage

\section{\dsixbf parameters}
\label{ap:parameters}

In this Appendix we provide additional details about the SMEFT and WET
parameters used in \dsix. These can be useful to properly read or
write some variables in a \mathe session using \dsix.

\subsection{SMEFT parameters}
\label{ap:parameters-SMEFT}

\parskip2ex plus1ex 

Table \ref{tab:SMEFTparameters} provides a complete list of the SMEFT
  parameters used in \dsix. In addition to the SMEFT WCs, this
  includes the SM parameters (gauge couplings, Yukawa matrices and
  scalar and $\theta$ parameters). This table is particularly useful
  to identify the names given to the elements of 2- and 4-fermion WCs,
  as well as the $\beta$ functions used in the \SMEFTrunner module.

\bigskip

{
\small
\begin{longtable}{|ccccccc|}
\caption{
SMEFT parameters. \emph{Position} denotes the position of the
  parameter (or parameters for 2- and 4-fermion objects) in the {\tt
    Parameters} global array. The column \emph{$\beta$ function} gives
  the name of the $\beta$ function (obtained with {\tt
    $\beta$[parameter]} after using the {\tt GetBeta}
  routine). \emph{Type} indicates the type of parameter (with nF
  standing for n-fermion) and \emph{Category}  denotes the index
  symmetry category of the coefficient, being relevant for 2- and
  4-fermion WCs.
}
\label{tab:SMEFTparameters}\\
\hline
Position & Parameter(s) & \dsix name & Elements & $\beta$ function & Type & Category \\
\hline
1 & $g$ & {\tt g} & - & {\tt $\beta$[g]} & 0F & 0 \\
2 & $g^\prime$ & {\tt gp} & - & {\tt $\beta$[gp]} & 0F & 0 \\
3 & $g_s$ & {\tt gs} & - & {\tt $\beta$[gs]} & 0F & 0 \\
4 & $\lambda$ & {\tt $\lambda$} & - & {\tt $\beta$[$\lambda$]} & 0F & 0 \\
5 & $m^2$ [GeV$^2$] & {\tt m2} & - & {\tt $\beta$[m2]} & 0F & 0 \\
6-14 & $\Gamma_u$ & {\tt Gu} & {\tt GU[i,j]} & {\tt $\beta$[GuX]} & 2F & 1 \\
15-23 & $\Gamma_d$ & {\tt Gd} & {\tt GD[i,j]} & {\tt $\beta$[GdX]} & 2F & 1 \\
24-32 & $\Gamma_e$ & {\tt Ge} & {\tt GE[i,j]} & {\tt $\beta$[GeX]} & 2F & 1 \\
33 & $\theta$ & {\tt $\theta$} & - & {\tt $\beta$[$\theta$]} & 0F & 0 \\
34 & $\theta^\prime$ & {\tt $\theta$p} & - & {\tt $\beta$[$\theta$p]} & 0F & 0 \\
35 & $\theta_s$ & {\tt $\theta$s} & - & {\tt $\beta$[$\theta$s]} & 0F & 0  \\
36 & $C_G$ & {\tt G} & - & {\tt $\beta$[G]} & 0F & 0 \\
37 & $C_{\widetilde G}$ & {\tt Gtilde} & - & {\tt $\beta$[Gtilde]} & 0F & 0 \\
38 & $C_W$ & {\tt W} & - & {\tt $\beta$[W]} & 0F & 0 \\
39 & $C_{\widetilde W}$ & {\tt Wtilde} & - & {\tt $\beta$[Wtilde]} & 0F & 0 \\
40 & $C_\vp$ & {\tt $\vp$} & - & {\tt $\beta$[$\vp$]} & 0F & 0 \\
41 & $C_{\vp\Box}$ & {\tt $\vp\Box$} & - & {\tt $\beta$[$\vp\Box$]} & 0F & 0 \\
42 & $C_{\vp D}$ & {\tt $\vp$DD} & - & {\tt $\beta$[$\vp$D]} & 0F & 0 \\
43 & $C_{\vp G}$ & {\tt $\vp$G} & - & {\tt $\beta$[$\vp$G]} & 0F & 0 \\
44 & $C_{\vp B}$ & {\tt $\vp$B} & - & {\tt $\beta$[$\vp$B]} & 0F & 0 \\
45 & $C_{\vp W}$ & {\tt $\vp$W} & - & {\tt $\beta$[$\vp$W]} & 0F & 0 \\
46 & $C_{\vp W B}$ & {\tt $\vp$WB} & - & {\tt $\beta$[$\vp$WB]} & 0F & 0 \\
47 & $C_{\vp \widetilde G}$ & {\tt $\vp$Gtilde} & - & {\tt $\beta$[$\vp$Gtilde]} & 0F & 0 \\
48 & $C_{\vp \widetilde B}$ & {\tt $\vp$Btilde} & - & {\tt $\beta$[$\vp$Btilde]} & 0F & 0 \\
49 & $C_{\vp \widetilde W}$ & {\tt $\vp$Wtilde} & - & {\tt $\beta$[$\vp$Wtilde]} & 0F & 0 \\
50 & $C_{\vp \widetilde W B}$ & {\tt $\vp$WtildeB} & - & {\tt $\beta$[$\vp$WtildeB]} & 0F & 0 \\
51-59 & $C_{u \vp}$ & {\tt WC[u$\vp$]} & {\tt U$\vp$[i,j]} & {\tt $\beta$[u$\vp$]} & 2F & 1 \\
60-68 & $C_{d \vp}$ & {\tt WC[d$\vp$]} & {\tt D$\vp$[i,j]} & {\tt $\beta$[d$\vp$]} & 2F & 1 \\
69-77 & $C_{e \vp}$ & {\tt WC[e$\vp$]} & {\tt E$\vp$[i,j]} & {\tt $\beta$[e$\vp$]} & 2F & 1 \\
78-86 & $C_{eW}$ & {\tt WC[eW]} & {\tt EW[i,j]} & {\tt $\beta$[eW]} & 2F & 1 \\
87-95 & $C_{eB}$ & {\tt WC[eB]} & {\tt EB[i,j]} & {\tt $\beta$[eB]} & 2F & 1 \\
96-104 & $C_{uG}$ & {\tt WC[uG]} & {\tt UG[i,j]} & {\tt $\beta$[uG]} & 2F & 1 \\
105-113 & $C_{uW}$ & {\tt WC[uW]} & {\tt UW[i,j]} & {\tt $\beta$[uW]} & 2F & 1 \\
114-122 & $C_{uB}$ & {\tt WC[uB]} & {\tt UB[i,j]} & {\tt $\beta$[uB]} & 2F & 1 \\
123-131 & $C_{dG}$ & {\tt WC[dG]} & {\tt DG[i,j]} & {\tt $\beta$[dG]} & 2F & 1 \\
132-140 & $C_{dW}$ & {\tt WC[dW]} & {\tt DW[i,j]} & {\tt $\beta$[dW]} & 2F & 1 \\
141-149 & $C_{dB}$ & {\tt WC[dB]} & {\tt DB[i,j]} & {\tt $\beta$[dB]} & 2F & 1 \\
150-155 & $C_{\vp \ell}^{(1)}$ & {\tt WC[$\vp$l1]} & {\tt $\vp$L1[i,j]} & {\tt $\beta$[$\vp$l1]} & 2F & 2 \\
156-161 & $C_{\vp \ell}^{(3)}$ & {\tt WC[$\vp$l3]} & {\tt $\vp$L3[i,j]} & {\tt $\beta$[$\vp$l3]} & 2F & 2 \\
162-167 & $C_{\vp e}$ & {\tt WC[$\vp$e]} & {\tt $\vp$E[i,j]} & {\tt $\beta$[$\vp$e]} & 2F & 2 \\
168-173 & $C_{\vp q}^{(1)}$ & {\tt WC[$\vp$q1]} & {\tt $\vp$Q1[i,j]} & {\tt $\beta$[$\vp$q1]} & 2F & 2 \\
174-179 & $C_{\vp q}^{(3)}$ & {\tt WC[$\vp$q3]} & {\tt $\vp$Q3[i,j]} & {\tt $\beta$[$\vp$q3]} & 2F & 2 \\
180-185 & $C_{\vp u}$ & {\tt WC[$\vp$u]} & {\tt $\vp$U[i,j]} & {\tt $\beta$[$\vp$u]} & 2F & 2 \\
186-191 & $C_{\vp d}$ & {\tt WC[$\vp$d]} & {\tt $\vp$D[i,j]} & {\tt $\beta$[$\vp$d]} & 2F & 2 \\
192-200 & $C_{\vp u d}$ & {\tt WC[$\vp$ud]} & {\tt $\vp$UD[i,j]} & {\tt $\beta$[$\vp$ud]} & 2F & 1 \\
201-227 & $C_{\ell \ell}$ & {\tt WC[ll]} & {\tt LL[i,j,k,l]} & {\tt $\beta$[ll]} & 4F & 4 \\
228-254 & $C_{qq}^{(1)}$ & {\tt WC[qq1]} & {\tt QQ1[i,j,k,l]} & {\tt $\beta$[qq1]} & 4F & 4 \\
255-281 & $C_{qq}^{(3)}$ & {\tt WC[qq3]} & {\tt QQ3[i,j,k,l]} & {\tt $\beta$[qq3]} & 4F & 4 \\
282-326 & $C_{\ell q}^{(1)}$ & {\tt WC[lq1]} & {\tt LQ1[i,j,k,l]} & {\tt $\beta$[lq1]} & 4F & 5 \\
327-371 & $C_{\ell q}^{(3)}$ & {\tt WC[lq3]} & {\tt LQ3[i,j,k,l]} & {\tt $\beta$[lq3]} & 4F & 5 \\
372-392 & $C_{ee}$ & {\tt WC[ee]} & {\tt EE[i,j,k,l]} & {\tt $\beta$[ee]} & 4F & 6 \\
393-419 & $C_{uu}$ & {\tt WC[uu]} & {\tt UU[i,j,k,l]} & {\tt $\beta$[uu]} & 4F & 4 \\
420-446 & $C_{dd}$ & {\tt WC[dd]} & {\tt DD[i,j,k,l]} & {\tt $\beta$[dd]} & 4F & 4 \\
447-491 & $C_{eu}$ & {\tt WC[eu]} & {\tt EU[i,j,k,l]} & {\tt $\beta$[eu]} & 4F & 5 \\
492-536 & $C_{ed}$ & {\tt WC[ed]} & {\tt ED[i,j,k,l]} & {\tt $\beta$[ed]} & 4F & 5 \\
537-581 & $C_{ud}^{(1)}$ & {\tt WC[ud1]} & {\tt UD1[i,j,k,l]} & {\tt $\beta$[ud1]} & 4F & 5 \\
582-626 & $C_{ud}^{(8)}$ & {\tt WC[ud8]} & {\tt UD8[i,j,k,l]} & {\tt $\beta$[ud8]} & 4F & 5 \\
627-671 & $C_{\ell e}$ & {\tt WC[le]} & {\tt LE[i,j,k,l]} & {\tt $\beta$[le]} & 4F & 5 \\
672-716 & $C_{\ell u}$ & {\tt WC[lu]} & {\tt LU[i,j,k,l]} & {\tt $\beta$[lu]} & 4F & 5 \\
717-761 & $C_{\ell d}$ & {\tt WC[ld]} & {\tt LD[i,j,k,l]} & {\tt $\beta$[ld]} & 4F & 5 \\
762-806 & $C_{q e}$ & {\tt WC[qe]} & {\tt QE[i,j,k,l]} & {\tt $\beta$[qe]} & 4F & 5 \\
807-851 & $C_{q u}^{(1)}$ & {\tt WC[qu1]} & {\tt QU1[i,j,k,l]} & {\tt $\beta$[qu1]} & 4F & 5 \\
852-896 & $C_{q u}^{(8)}$ & {\tt WC[qu8]} & {\tt QU8[i,j,k,l]} & {\tt $\beta$[qu8]} & 4F & 5 \\
897-941 & $C_{q d}^{(1)}$ & {\tt WC[qd1]} & {\tt QD1[i,j,k,l]} & {\tt $\beta$[qd1]} & 4F & 5 \\
942-986 & $C_{q d}^{(8)}$ & {\tt WC[qd8]} & {\tt QD8[i,j,k,l]} & {\tt $\beta$[qd8]} & 4F & 5 \\
987-1067 & $C_{\ell e d q}$ & {\tt WC[ledq]} & {\tt LEDQ[i,j,k,l]} & {\tt $\beta$[ledq]} & 4F & 3 \\
1068-1148 & $C_{quqd}^{(1)}$ & {\tt WC[quqd1]} & {\tt QUQD1[i,j,k,l]} & {\tt $\beta$[quqd1]} & 4F & 3 \\
1149-1229 & $C_{quqd}^{(8)}$ & {\tt WC[quqd8]} & {\tt QUQD8[i,j,k,l]} & {\tt $\beta$[quqd8]} & 4F & 3 \\
1230-1310 & $C_{\ell e q u}^{(1)}$ & {\tt WC[lequ1]} & {\tt LEQU1[i,j,k,l]} & {\tt $\beta$[lequ1]} & 4F & 3 \\
1311-1391 & $C_{\ell e q u}^{(3)}$ & {\tt WC[lequ3]} & {\tt LEQU3[i,j,k,l]} & {\tt $\beta$[lequ3]} & 4F & 3 \\
1392-1472 & $C_{duq\ell}$ & {\tt WC[duql]} & {\tt DUQL[i,j,k,l]} & {\tt $\beta$[duql]} & 4F & 3 \\
1473-1526 & $C_{qque}$ & {\tt WC[qque]} & {\tt QQUE[i,j,k,l]} & {\tt $\beta$[qque]} & 4F & 7 \\
1527-1583 & $C_{qqq\ell}$ & {\tt WC[qqql]} & {\tt QQQL[i,j,k,l]} & {\tt $\beta$[qqql]} & 4F & 8 \\
1584-1664 & $C_{duue}$ & {\tt WC[duue]} & {\tt DUUE[i,j,k,l]} & {\tt $\beta$[duue]} & 4F & 3 \\
\hline
\end{longtable}
}

It is well known that some of the 2- and 4-fermion operators in the
SMEFT posess specific symmetries under the exchange of flavor
indices. This translates into an index symmetry for the corresponding
WCs. For instance, the Wilson coefficient $C_{\vp e}$ is a Hermitian
matrix, hence following the symmetry relation $\left[ C_{\vp e}
  \right]_{ij} = \left[ C_{\vp e} \right]_{ji}^\ast$. More complicated
index symmetries exist for some of the 4-fermion WCs. In all these
cases, the number of independent WCs gets reduced. For example, the
$C_{ee}$ 4-fermion WC does not contain $81$ ($=3^4$) independent
complex WCs, but just $21$ real and $15$ imaginary independent
components. Therefore, it is convenient to restrict the number of
parameters considered in SMEFT calculations to just the independent
ones. In \dsix we have followed this approach, dropping redudant WCs
in all calculations. This is what motivates the introduction of an
index symmetry category column in Table \ref{tab:SMEFTparameters}. The
meaning of the different categories is given by:

\begin{center}
\begin{tabular}{cl}
\hline
Category & Meaning \\
\hline
0 & 0F scalar object \\
1 & 2F general $3 \times 3$ matrix \\
2 & 2F Hermitian matrix \\
3 & 4F general $3 \times 3 \times 3 \times 3$ object \\
4 & 4F two identical $\bar \psi \psi$ currents \\
5 & 4F two independent $\bar \psi \psi$ currents \\
6 & 4F two identical $\bar \psi \psi$ currents - special case $C_{ee}$ \\
7 & 4F Baryon-number-violating - special case $C_{qque}$ \\
8 & 4F Baryon-number-violating - special case $C_{qqql}$ \\
\hline
\end{tabular}
\end{center}

We see that, apart from the WCs in categories 0, 1 and 3, all other
WCs have index symmetries. Furthermore, there are three WCs with
special symmetries, not shared by any other WC: $C_{ee}$, $C_{qque}$
and $C_{qqql}$. In Table \ref{tab:SMEFTnonredundant} we list the
independent WCs contained in each category. This, combined with Table
\ref{tab:SMEFTparameters}, completely allows the user to determine the
position of a given parameter in the {\tt Parameters} array. In any
case, we remind the reader that the function {\tt FindParametersSMEFT}
can also be used for this purpose.

{\small
\renewcommand{\arraystretch}{1.06}
\begin{longtable}{|ccccccccc|}
\caption{Independent SMEFT WCs in each category. Elements in red
  denote real WCs.
\label{tab:SMEFTnonredundant}} \\
\hline
 & 1 & 2 & 3 & 4 & 5 & 6 & 7 & 8 \\
\hline
1 & \{1, 1\} & \real{\{1, 1\}} & \{1, 1, 1, 1\} & \real{\{1, 1, 1, 1\}} & \real{\{1, 1, 1, 1\}} & \real{\{1, 1, 1, 1\}} & \{1, 1, 1, 1\} & \{1, 1, 1, 1\} \\ 
2 & \{1, 2\} & \{1, 2\} & \{1, 1, 1, 2\} & \{1, 1, 1, 2\} & \{1, 1, 1, 2\} & \{1, 1, 1, 2\} & \{1, 1, 1, 2\} & \{1, 1, 1, 2\} \\ 
3 & \{1, 3\} & \{1, 3\} & \{1, 1, 1, 3\} & \{1, 1, 1, 3\} & \{1, 1, 1, 3\} & \{1, 1, 1, 3\} & \{1, 1, 1, 3\} & \{1, 1, 1, 3\} \\ 
4 & \{2, 1\} & \real{\{2, 2\}} & \{1, 1, 2, 1\} & \real{\{1, 1, 2, 2\}} & \real{\{1, 1, 2, 2\}} & \real{\{1, 1, 2, 2\}} & \{1, 1, 2, 1\} & \{1, 1, 2, 1\} \\ 
5 & \{2, 2\} & \{2, 3\} & \{1, 1, 2, 2\} & \{1, 1, 2, 3\} & \{1, 1, 2, 3\} & \{1, 1, 2, 3\} & \{1, 1, 2, 2\} & \{1, 1, 2, 2\} \\ 
6 & \{2, 3\} & \real{\{3, 3\}} & \{1, 1, 2, 3\} & \real{\{1, 1, 3, 3\}} & \real{\{1, 1, 3, 3\}} & \real{\{1, 1, 3, 3\}} & \{1, 1, 2, 3\} & \{1, 1, 2, 3\} \\ 
7 & \{3, 1\} &  & \{1, 1, 3, 1\} & \{1, 2, 1, 2\} & \{1, 2, 1, 1\} & \{1, 2, 1, 2\} & \{1, 1, 3, 1\} & \{1, 1, 3, 1\} \\ 
8 & \{3, 2\} &  & \{1, 1, 3, 2\} & \{1, 2, 1, 3\} & \{1, 2, 1, 2\} & \{1, 2, 1, 3\} & \{1, 1, 3, 2\} & \{1, 1, 3, 2\} \\ 
9 & \{3, 3\} &  & \{1, 1, 3, 3\} & \real{\{1, 2, 2, 1\}} & \{1, 2, 1, 3\} & \{1, 2, 2, 2\} & \{1, 1, 3, 3\} & \{1, 1, 3, 3\} \\ 
10 &  &  & \{1, 2, 1, 1\} & \{1, 2, 2, 2\} & \{1, 2, 2, 1\} & \{1, 2, 2, 3\} & \{1, 2, 1, 1\} & \{1, 2, 1, 1\} \\ 
11 &  &  & \{1, 2, 1, 2\} & \{1, 2, 2, 3\} & \{1, 2, 2, 2\} & \{1, 2, 3, 2\} & \{1, 2, 1, 2\} & \{1, 2, 1, 2\} \\ 
12 &  &  & \{1, 2, 1, 3\} & \{1, 2, 3, 1\} & \{1, 2, 2, 3\} & \{1, 2, 3, 3\} & \{1, 2, 1, 3\} & \{1, 2, 1, 3\} \\ 
13 &  &  & \{1, 2, 2, 1\} & \{1, 2, 3, 2\} & \{1, 2, 3, 1\} & \{1, 3, 1, 3\} & \{1, 2, 2, 1\} & \{1, 2, 2, 1\} \\ 
14 &  &  & \{1, 2, 2, 2\} & \{1, 2, 3, 3\} & \{1, 2, 3, 2\} & \{1, 3, 2, 3\} & \{1, 2, 2, 2\} & \{1, 2, 2, 2\} \\ 
15 &  &  & \{1, 2, 2, 3\} & \{1, 3, 1, 3\} & \{1, 2, 3, 3\} & \{1, 3, 3, 3\} & \{1, 2, 2, 3\} & \{1, 2, 2, 3\} \\ 
16 &  &  & \{1, 2, 3, 1\} & \{1, 3, 2, 2\} & \{1, 3, 1, 1\} & \real{\{2, 2, 2, 2\}} & \{1, 2, 3, 1\} & \{1, 2, 3, 1\} \\ 
17 &  &  & \{1, 2, 3, 2\} & \{1, 3, 2, 3\} & \{1, 3, 1, 2\} & \{2, 2, 2, 3\} & \{1, 2, 3, 2\} & \{1, 2, 3, 2\} \\ 
18 &  &  & \{1, 2, 3, 3\} & \real{\{1, 3, 3, 1\}} & \{1, 3, 1, 3\} & \real{\{2, 2, 3, 3\}} & \{1, 2, 3, 3\} & \{1, 2, 3, 3\} \\ 
19 &  &  & \{1, 3, 1, 1\} & \{1, 3, 3, 2\} & \{1, 3, 2, 1\} & \{2, 3, 2, 3\} & \{1, 3, 1, 1\} & \{1, 3, 1, 1\} \\ 
20 &  &  & \{1, 3, 1, 2\} & \{1, 3, 3, 3\} & \{1, 3, 2, 2\} & \{2, 3, 3, 3\} & \{1, 3, 1, 2\} & \{1, 3, 1, 2\} \\ 
21 &  &  & \{1, 3, 1, 3\} & \real{\{2, 2, 2, 2\}} & \{1, 3, 2, 3\} & \real{\{3, 3, 3, 3\}} & \{1, 3, 1, 3\} & \{1, 3, 1, 3\} \\ 
22 &  &  & \{1, 3, 2, 1\} & \{2, 2, 2, 3\} & \{1, 3, 3, 1\} &  & \{1, 3, 2, 1\} & \{1, 3, 2, 1\} \\ 
23 &  &  & \{1, 3, 2, 2\} & \real{\{2, 2, 3, 3\}} & \{1, 3, 3, 2\} &  & \{1, 3, 2, 2\} & \{1, 3, 2, 2\} \\ 
24 &  &  & \{1, 3, 2, 3\} & \{2, 3, 2, 3\} & \{1, 3, 3, 3\} &  & \{1, 3, 2, 3\} & \{1, 3, 2, 3\} \\ 
25 &  &  & \{1, 3, 3, 1\} & \real{\{2, 3, 3, 2\}} & \real{\{2, 2, 1, 1\}} &  & \{1, 3, 3, 1\} & \{1, 3, 3, 1\} \\ 
26 &  &  & \{1, 3, 3, 2\} & \{2, 3, 3, 3\} & \{2, 2, 1, 2\} &  & \{1, 3, 3, 2\} & \{1, 3, 3, 2\} \\ 
27 &  &  & \{1, 3, 3, 3\} & \real{\{3, 3, 3, 3\}} & \{2, 2, 1, 3\} &  & \{1, 3, 3, 3\} & \{1, 3, 3, 3\} \\ 
28 &  &  & \{2, 1, 1, 1\} &  & \real{\{2, 2, 2, 2\}} &  & \{2, 2, 1, 1\} & \{2, 1, 2, 1\} \\ 
29 &  &  & \{2, 1, 1, 2\} &  & \{2, 2, 2, 3\} &  & \{2, 2, 1, 2\} & \{2, 1, 2, 2\} \\ 
30 &  &  & \{2, 1, 1, 3\} &  & \real{\{2, 2, 3, 3\}} &  & \{2, 2, 1, 3\} & \{2, 1, 2, 3\} \\ 
31 &  &  & \{2, 1, 2, 1\} &  & \{2, 3, 1, 1\} &  & \{2, 2, 2, 1\} & \{2, 1, 3, 1\} \\ 
32 &  &  & \{2, 1, 2, 2\} &  & \{2, 3, 1, 2\} &  & \{2, 2, 2, 2\} & \{2, 1, 3, 2\} \\ 
33 &  &  & \{2, 1, 2, 3\} &  & \{2, 3, 1, 3\} &  & \{2, 2, 2, 3\} & \{2, 1, 3, 3\} \\ 
34 &  &  & \{2, 1, 3, 1\} &  & \{2, 3, 2, 1\} &  & \{2, 2, 3, 1\} & \{2, 2, 2, 1\} \\ 
35 &  &  & \{2, 1, 3, 2\} &  & \{2, 3, 2, 2\} &  & \{2, 2, 3, 2\} & \{2, 2, 2, 2\} \\ 
36 &  &  & \{2, 1, 3, 3\} &  & \{2, 3, 2, 3\} &  & \{2, 2, 3, 3\} & \{2, 2, 2, 3\} \\ 
37 &  &  & \{2, 2, 1, 1\} &  & \{2, 3, 3, 1\} &  & \{2, 3, 1, 1\} & \{2, 2, 3, 1\} \\ 
38 &  &  & \{2, 2, 1, 2\} &  & \{2, 3, 3, 2\} &  & \{2, 3, 1, 2\} & \{2, 2, 3, 2\} \\ 
39 &  &  & \{2, 2, 1, 3\} &  & \{2, 3, 3, 3\} &  & \{2, 3, 1, 3\} & \{2, 2, 3, 3\} \\ 
40 &  &  & \{2, 2, 2, 1\} &  & \real{\{3, 3, 1, 1\}} &  & \{2, 3, 2, 1\} & \{2, 3, 1, 1\} \\ 
41 &  &  & \{2, 2, 2, 2\} &  & \{3, 3, 1, 2\} &  & \{2, 3, 2, 2\} & \{2, 3, 1, 2\} \\ 
42 &  &  & \{2, 2, 2, 3\} &  & \{3, 3, 1, 3\} &  & \{2, 3, 2, 3\} & \{2, 3, 1, 3\} \\ 
43 &  &  & \{2, 2, 3, 1\} &  & \real{\{3, 3, 2, 2\}} &  & \{2, 3, 3, 1\} & \{2, 3, 2, 1\} \\ 
44 &  &  & \{2, 2, 3, 2\} &  & \{3, 3, 2, 3\} &  & \{2, 3, 3, 2\} & \{2, 3, 2, 2\} \\ 
45 &  &  & \{2, 2, 3, 3\} &  & \real{\{3, 3, 3, 3\}} &  & \{2, 3, 3, 3\} & \{2, 3, 2, 3\} \\ 
46 &  &  & \{2, 3, 1, 1\} &  &  &  & \{3, 3, 1, 1\} & \{2, 3, 3, 1\} \\ 
47 &  &  & \{2, 3, 1, 2\} &  &  &  & \{3, 3, 1, 2\} & \{2, 3, 3, 2\} \\ 
48 &  &  & \{2, 3, 1, 3\} &  &  &  & \{3, 3, 1, 3\} & \{2, 3, 3, 3\} \\ 
49 &  &  & \{2, 3, 2, 1\} &  &  &  & \{3, 3, 2, 1\} & \{3, 1, 3, 1\} \\ 
50 &  &  & \{2, 3, 2, 2\} &  &  &  & \{3, 3, 2, 2\} & \{3, 1, 3, 2\} \\ 
51 &  &  & \{2, 3, 2, 3\} &  &  &  & \{3, 3, 2, 3\} & \{3, 1, 3, 3\} \\ 
52 &  &  & \{2, 3, 3, 1\} &  &  &  & \{3, 3, 3, 1\} & \{3, 2, 3, 1\} \\ 
53 &  &  & \{2, 3, 3, 2\} &  &  &  & \{3, 3, 3, 2\} & \{3, 2, 3, 2\} \\ 
54 &  &  & \{2, 3, 3, 3\} &  &  &  & \{3, 3, 3, 3\} & \{3, 2, 3, 3\} \\ 
55 &  &  & \{3, 1, 1, 1\} &  &  &  &  & \{3, 3, 3, 1\} \\ 
56 &  &  & \{3, 1, 1, 2\} &  &  &  &  & \{3, 3, 3, 2\} \\ 
57 &  &  & \{3, 1, 1, 3\} &  &  &  &  & \{3, 3, 3, 3\} \\ 
58 &  &  & \{3, 1, 2, 1\} &  &  &  &  &  \\ 
59 &  &  & \{3, 1, 2, 2\} &  &  &  &  &  \\ 
60 &  &  & \{3, 1, 2, 3\} &  &  &  &  &  \\ 
61 &  &  & \{3, 1, 3, 1\} &  &  &  &  &  \\ 
62 &  &  & \{3, 1, 3, 2\} &  &  &  &  &  \\ 
63 &  &  & \{3, 1, 3, 3\} &  &  &  &  &  \\ 
64 &  &  & \{3, 2, 1, 1\} &  &  &  &  &  \\ 
65 &  &  & \{3, 2, 1, 2\} &  &  &  &  &  \\ 
66 &  &  & \{3, 2, 1, 3\} &  &  &  &  &  \\ 
67 &  &  & \{3, 2, 2, 1\} &  &  &  &  &  \\ 
68 &  &  & \{3, 2, 2, 2\} &  &  &  &  &  \\ 
69 &  &  & \{3, 2, 2, 3\} &  &  &  &  &  \\ 
70 &  &  & \{3, 2, 3, 1\} &  &  &  &  &  \\ 
71 &  &  & \{3, 2, 3, 2\} &  &  &  &  &  \\ 
72 &  &  & \{3, 2, 3, 3\} &  &  &  &  &  \\ 
73 &  &  & \{3, 3, 1, 1\} &  &  &  &  &  \\ 
74 &  &  & \{3, 3, 1, 2\} &  &  &  &  &  \\ 
75 &  &  & \{3, 3, 1, 3\} &  &  &  &  &  \\ 
76 &  &  & \{3, 3, 2, 1\} &  &  &  &  &  \\ 
77 &  &  & \{3, 3, 2, 2\} &  &  &  &  &  \\ 
78 &  &  & \{3, 3, 2, 3\} &  &  &  &  &  \\ 
79 &  &  & \{3, 3, 3, 1\} &  &  &  &  &  \\ 
80 &  &  & \{3, 3, 3, 2\} &  &  &  &  &  \\ 
81 &  &  & \{3, 3, 3, 3\} &  &  &  &  &  \\
\hline
\end{longtable}
}

\subsection{WET parameters}
\label{ap:parameters-WET}

Table \ref{tab:WETparameters} provides a complete list of the WET
  parameters considered in \dsix.

\begin{center}
\renewcommand{\arraystretch}{1.06}
\setlength{\tabcolsep}{3.4mm}
\begin{longtable}{|ccc||ccc|}
\caption{WET parameters. \emph{Position} denotes the position of the
  parameter in the {\tt
    WETParameters} global array.
\label{tab:WETparameters}} \\
\hline
Position & Parameter & \dsix name & Position & Parameter & \dsix name \\
\hline
1 & $C_1^{sbsb}$   & {\tt CBS2[1]} & 70 & $C_9^{sbee}$   & {\tt CBS1[e][9]} \\ 
2 & $C_2^{sbsb}$   & {\tt CBS2[2]} & 71 & $C_1^{sb\mu\mu}$   & {\tt CBS1[$\mu$][1]} \\ 
3 & $C_3^{sbsb}$   & {\tt CBS2[3]} & 72 & $C_3^{sb\mu\mu}$   & {\tt CBS1[$\mu$][3]} \\ 
4 & $C_4^{sbsb}$   & {\tt CBS2[4]} & 73 & $C_5^{sb\mu\mu}$   & {\tt CBS1[$\mu$][5]} \\ 
5 & $C_5^{sbsb}$   & {\tt CBS2[5]} & 74 & $C_7^{sb\mu\mu}$   & {\tt CBS1[$\mu$][7]} \\ 
6 & $C_{1^\prime}^{sbsb}$   & {\tt CBS2p[1]} & 75 & $C_9^{sb\mu\mu}$   & {\tt CBS1[$\mu$][9]} \\ 
7 & $C_{2^\prime}^{sbsb}$   & {\tt CBS2p[2]} & 76 & $C_1^{sb\tau\tau}$   & {\tt CBS1[$\tau$][1]} \\ 
8 & $C_{3^\prime}^{sbsb}$   & {\tt CBS2p[3]} & 77 & $C_3^{sb\tau\tau}$   & {\tt CBS1[$\tau$][3]} \\ 
9 &  $C_1^{cbee}$  & {\tt CBC1[e][1]} & 78 & $C_5^{sb\tau\tau}$   & {\tt CBS1[$\tau$][5]} \\ 
10 & $C_5^{cbee}$   & {\tt CBC1[e][5]} & 79 & $C_7^{sb\tau\tau}$   & {\tt CBS1[$\tau$][7]} \\ 
11 & $C_{1^\prime}^{cbee}$   & {\tt CBC1p[e][1]} & 80 & $C_9^{sb\tau\tau}$   & {\tt CBS1[$\tau$][9]} \\ 
12 & $C_{5^\prime}^{cbee}$   & {\tt CBC1p[e][5]} & 81 & $C_{1^\prime}^{sbuu}$   & {\tt CBS1p[u][1]} \\ 
13 & $C_{7^\prime}^{cbee}$   & {\tt CBC1p[e][7]} & 82 & $C_{2^\prime}^{sbuu}$   & {\tt CBS1p[u][2]} \\ 
14 & $C_1^{cb\mu\mu}$   & {\tt CBC1[$\mu$][1]} & 83 & $C_{3^\prime}^{sbuu}$   & {\tt CBS1p[u][3]} \\ 
15 & $C_5^{cb\mu\mu}$   & {\tt CBC1[$\mu$][5]} & 84 & $C_{4^\prime}^{sbuu}$   & {\tt CBS1p[u][4]} \\ 
16 & $C_{1^\prime}^{cb\mu\mu}$   & {\tt CBC1p[$\mu$][1]} & 85 & $C_{5^\prime}^{sbuu}$   & {\tt CBS1p[u][5]} \\ 
17 & $C_{5^\prime}^{cb\mu\mu}$   & {\tt CBC1p[$\mu$][5]} & 86 & $C_{6^\prime}^{sbuu}$   & {\tt CBS1p[u][6]} \\ 
18 & $C_{7^\prime}^{cb\mu\mu}$   & {\tt CBC1p[$\mu$][7]} & 87 & $C_{7^\prime}^{sbuu}$   & {\tt CBS1p[u][7]} \\ 
19 & $C_1^{cb\tau\tau}$   & {\tt CBC1[$\tau$][1]} & 88 & $C_{8^\prime}^{sbuu}$   & {\tt CBS1p[u][8]} \\ 
20 & $C_5^{cb\tau\tau}$   & {\tt CBC1[$\tau$][5]} & 89 & $C_{9^\prime}^{sbuu}$   & {\tt CBS1p[u][9]} \\ 
21 & $C_{1^\prime}^{cb\tau\tau}$   & {\tt CBC1p[$\tau$][1]} & 90 & $C_{10^\prime}^{sbuu}$    & {\tt CBS1p[u][10]} \\ 
22 & $C_{5^\prime}^{cb\tau\tau}$   & {\tt CBC1p[$\tau$][5]} & 91 & $C_{1^\prime}^{sbdd}$   & {\tt CBS1p[d][1]} \\ 
23 & $C_{7^\prime}^{cb\tau\tau}$   & {\tt CBC1p[$\tau$][7]} & 92 & $C_{2^\prime}^{sbdd}$   & {\tt CBS1p[d][2]} \\ 
24 & $C_1^{sbuu}$   & {\tt CBS1[u][1]} & 93 & $C_{3^\prime}^{sbdd}$   & {\tt CBS1p[d][3]} \\ 
25 & $C_2^{sbuu}$   & {\tt CBS1[u][2]} & 94 & $C_{4^\prime}^{sbdd}$   & {\tt CBS1p[d][4]} \\ 
26 & $C_3^{sbuu}$   & {\tt CBS1[u][3]} & 95 & $C_{5^\prime}^{sbdd}$   & {\tt CBS1p[d][5]} \\ 
27 & $C_4^{sbuu}$   & {\tt CBS1[u][4]} & 96 & $C_{6^\prime}^{sbdd}$   & {\tt CBS1p[d][6]} \\ 
28 & $C_5^{sbuu}$   & {\tt CBS1[u][5]} & 97 & $C_{7^\prime}^{sbdd}$   & {\tt CBS1p[d][7]} \\ 
29 & $C_6^{sbuu}$   & {\tt CBS1[u][6]} & 98 & $C_{8^\prime}^{sbdd}$   & {\tt CBS1p[d][8]} \\ 
30 & $C_7^{sbuu}$   & {\tt CBS1[u][7]} & 99 & $C_{9^\prime}^{sbdd}$   & {\tt CBS1p[d][9]} \\ 
31 & $C_8^{sbuu}$   & {\tt CBS1[u][8]} & 100 & $C_{10^\prime}^{sbdd}$   & {\tt CBS1p[d][10]} \\ 
32 & $C_9^{sbuu}$   & {\tt CBS1[u][9]} & 101 & $C_{1^\prime}^{sbcc}$   & {\tt CBS1p[c][1]} \\ 
33 & $C_{10}^{sbuu}$   & {\tt CBS1[u][10]} & 102 & $C_{2^\prime}^{sbcc}$   & {\tt CBS1p[c][2]} \\ 
34 & $C_1^{sbdd}$   & {\tt CBS1[d][1]} & 103 & $C_{3^\prime}^{sbcc}$   & {\tt CBS1p[c][3]} \\ 
35 & $C_2^{sbdd}$   & {\tt CBS1[d][2]} & 104 & $C_{4^\prime}^{sbcc}$   & {\tt CBS1p[c][4]} \\ 
36 & $C_3^{sbdd}$   & {\tt CBS1[d][3]} & 105 & $C_{5^\prime}^{sbcc}$   & {\tt CBS1p[c][5]} \\ 
37 & $C_4^{sbdd}$   & {\tt CBS1[d][4]} & 106 & $C_{6^\prime}^{sbcc}$   & {\tt CBS1p[c][6]} \\ 
38 & $C_5^{sbdd}$   & {\tt CBS1[d][5]} & 107 & $C_{7^\prime}^{sbcc}$   & {\tt CBS1p[c][7]} \\ 
39 & $C_6^{sbdd}$   & {\tt CBS1[d][6]} & 108 & $C_{8^\prime}^{sbcc}$   & {\tt CBS1p[c][8]} \\ 
40 & $C_7^{sbdd}$   & {\tt CBS1[d][7]} & 109 & $C_{9^\prime}^{sbcc}$   & {\tt CBS1p[c][9]} \\ 
41 & $C_8^{sbdd}$   & {\tt CBS1[d][8]} & 110 & $C_{10^\prime}^{sbcc}$   & {\tt CBS1p[c][10]} \\ 
42 & $C_9^{sbdd}$   & {\tt CBS1[d][9]} & 111 & $C_{1^\prime}^{sbss}$   & {\tt CBS1p[s][1]} \\ 
43 & $C_{10}^{sbdd}$   & {\tt CBS1[d][10]} & 112 & $C_{3^\prime}^{sbss}$   & {\tt CBS1p[s][3]} \\ 
44 & $C_1^{sbcc}$   & {\tt CBS1[c][1]} & 113 & $C_{5^\prime}^{sbss}$   & {\tt CBS1p[s][5]} \\ 
45 & $C_2^{sbcc}$   & {\tt CBS1[c][2]} & 114 & $C_{7^\prime}^{sbss}$   & {\tt CBS1p[s][7]} \\ 
46 & $C_3^{sbcc}$   & {\tt CBS1[c][3]} & 115 & $C_{9^\prime}^{sbss}$   & {\tt CBS1p[s][9]} \\ 
47 & $C_4^{sbcc}$   & {\tt CBS1[c][4]} & 116 & $C_{1^\prime}^{sbbb}$   & {\tt CBS1p[b][1]} \\ 
48 & $C_5^{sbcc}$   & {\tt CBS1[c][5]} & 117 & $C_{3^\prime}^{sbbb}$   & {\tt CBS1p[b][3]} \\ 
49 & $C_6^{sbcc}$   & {\tt CBS1[c][6]} & 118 & $C_{5^\prime}^{sbbb}$   & {\tt CBS1p[b][5]} \\ 
50 & $C_7^{sbcc}$   & {\tt CBS1[c][7]} & 119 & $C_{7^\prime}^{sbbb}$   & {\tt CBS1p[b][7]} \\ 
51 & $C_8^{sbcc}$   & {\tt CBS1[c][8]} & 120 & $C_{9^\prime}^{sbbb}$   & {\tt CBS1p[b][9]} \\ 
52 & $C_9^{sbcc}$   & {\tt CBS1[c][9]} & 121 & $C_{7^\prime \gamma}$   & {\tt CBS1p[M][7]} \\ 
53 & $C_{10}^{sbcc}$   & {\tt CBS1[c][10]} & 122 & $C_{8^\prime g}$   & {\tt CBS1p[M][8]} \\ 
54 & $C_1^{sbss}$   & {\tt CBS1[s][1]} & 123 & $C_{1^\prime}^{sbee}$   & {\tt CBS1p[e][1]} \\ 
55 & $C_3^{sbss}$   & {\tt CBS1[s][3]} & 124 & $C_{3^\prime}^{sbee}$   & {\tt CBS1p[e][3]} \\ 
56 & $C_5^{sbss}$   & {\tt CBS1[s][5]} & 125 & $C_{5^\prime}^{sbee}$   & {\tt CBS1p[e][5]} \\ 
57 & $C_7^{sbss}$   & {\tt CBS1[s][7]} & 126 & $C_{7^\prime}^{sbee}$   & {\tt CBS1p[e][7]} \\ 
58 & $C_9^{sbss}$   & {\tt CBS1[s][9]} & 127 & $C_{9^\prime}^{sbee}$   & {\tt CBS1p[e][9]} \\ 
59 & $C_1^{sbbb}$   & {\tt CBS1[b][1]} & 128 & $C_{1^\prime}^{sb\mu\mu}$   & {\tt CBS1p[$\mu$][1]} \\ 
60 & $C_3^{sbbb}$   & {\tt CBS1[b][3]} & 129 & $C_{3^\prime}^{sb\mu\mu}$   & {\tt CBS1p[$\mu$][3]} \\ 
61 & $C_5^{sbbb}$   & {\tt CBS1[b][5]} & 130 & $C_{5^\prime}^{sb\mu\mu}$   & {\tt CBS1p[$\mu$][5]} \\ 
62 & $C_7^{sbbb}$   & {\tt CBS1[b][7]} & 131 & $C_{7^\prime}^{sb\mu\mu}$   & {\tt CBS1p[$\mu$][7]} \\ 
63 & $C_9^{sbbb}$   & {\tt CBS1[b][9]} & 132 & $C_{9^\prime}^{sb\mu\mu}$   & {\tt CBS1p[$\mu$][9]} \\ 
64 & $C_{7 \gamma}$   & {\tt CBS1[M][7]} & 133 & $C_{1^\prime}^{sb\tau\tau}$   & {\tt CBS1p[$\tau$][1]} \\ 
65 & $C_{8 g}$   & {\tt CBS1[M][8]} & 134 & $C_{3^\prime}^{sb\tau\tau}$   & {\tt CBS1p[$\tau$][3]} \\ 
66 & $C_1^{sbee}$   & {\tt CBS1[e][1]} & 135 & $C_{5^\prime}^{sb\tau\tau}$   & {\tt CBS1p[$\tau$][5]} \\ 
67 & $C_3^{sbee}$   & {\tt CBS1[e][3]} & 136 & $C_{7^\prime}^{sb\tau\tau}$   & {\tt CBS1p[$\tau$][7]} \\ 
68 & $C_5^{sbee}$   & {\tt CBS1[e][5]} & 137 & $C_{9^\prime}^{sb\tau\tau}$   & {\tt CBS1p[$\tau$][9]} \\ 
69 & $C_7^{sbee}$   & {\tt CBS1[e][7]} &  &  & \\ 
\hline
\end{longtable}
\end{center}

\section{Baryon-number-violating operators in the mass basis}
\label{ap:BVinMB}

After electroweak symmetry breaking, the Warsaw basis operators can be
rotated to the fermion mass basis. This is achieved by performing
unitary transformations of the fermion fields in order to diagonalize
the fermion mass matrices,
\begin{align} \label{eq:rotations}
u_L \to& V_{u_L} u_L \,, \qquad  d_L \to V_{d_L} d_L \,, \qquad   u_R \to V_{u_R} u_R \,, \qquad   d_R \to V_{d_R} d_R  \,, \nonumber \\
e_L \to& V_{e_L} e_L \,, \qquad  e_R \to V_{e_R} e_R \,.
\end{align}
In this way
\begin{align}
m_{\psi}^{\rm{diag}}  \equiv V_{\psi_L}^{\dag} \, m_{\psi}  \, V_{\psi_R} 
\end{align}
is a diagonal and positive matrix corresponding to the physical
fermion masses. We note that these definitions imply that the CKM
matrix is given by $V = V_{u_L}^\dagger V_{d_L}$. The resulting
operators after applying these unitary transformations to the
Baryon-number-violating operators are shown in Table \ref{tab:BVinMB}. For the
rest of Wilson coefficients see \cite{Aebischer:2015fzz}.

\begin{table}
  \centering
  \begin{tabular}{cl}
    \hline
    \hline
    Operator & \multicolumn{1}{c}{Definition in the mass basis} \\
    \hline
    \multirow{2}{*}{$Q_{duq\ell}$} & $\left[ \widetilde{C}_{duq\ell} \right]_{ijkl} \left( d_R^i C u_R^j \right) \, \left[ V_{mk}^\ast \left( u_L^m C e_L^l \right) - \left( d_L^k C \nu_L^l \right) \right]$ \\
    & $\left[ \widetilde{C}_{duq\ell}\right]_{ijkl} = \left[ C_{duq\ell} \right]_{pqrl} \left[ V_{d_R} \right]_{pi} \left[ V_{u_R} \right]_{qj} \left[ V_{d_L} \right]_{rk}$ \\
    \hline
    \multirow{2}{*}{$Q_{qque}$} & $\left[ \widetilde{C}_{qque} \right]_{ijkl} \left[ V_{mi}^\ast V_{jn} \left( u_L^m C d_L^n \right) - \left( u_L^i C d_L^i \right) \right] \left(u_R^k C e_R^l \right)$ \\
    & $\left[ \widetilde{C}_{qque} \right]_{ijkl} = \left[ C_{qque} \right]_{pqrl} \left[ V_{d_L} \right]_{pi} \left[ V_{u_L} \right]_{qj} \left[ V_{u_R} \right]_{rk}$ \\
    \hline
    \multirow{2}{*}{$Q_{qqq\ell}$} & $\begin{aligned} \left[ \widetilde{C}_{qqq\ell} \right]_{ijkl} & \left[ \left( d_L^i C d_L^j \right) \left( u_L^k C \nu_L^l \right) + V_{mi}^\ast V_{nj}^\ast V_{kt} \left( u_L^m C u_L^n \right) \left( d_L^t C e_L^l \right) \right. \\ & \left. - V_{mi}^\ast \left( u_L^m C d_L^j \right) \left( u_L^k C e_L^l \right) - V_{nj}^\ast V_{kt} \left( d_L^i C u_L^n \right) \left( d_L^t C \nu_L^l \right) \right] \end{aligned}$ \\
    & $\left[ \widetilde{C}_{qqq\ell} \right]_{ijkl} = \left[ C_{qqq\ell} \right]_{pqrl} \left[ V_{d_L} \right]_{pi} \left[ V_{d_L} \right]_{qj} \left[ V_{u_L} \right]_{rk}$ \\
    \hline
    \multirow{2}{*}{$Q_{duue}$} & $\left[ \widetilde{C}_{duue} \right]_{ijkl} \left(d_R^i C u_R^j \right) \left( u_R^k C e_R^l \right)$ \\
    & $\left[ \widetilde{C}_{duue} \right]_{ijkl} = \left[ C_{duue} \right]_{pqrl} \left[ V_{d_R} \right]_{pi} \left[ V_{u_R} \right]_{qj} \left[ V_{u_R} \right]_{rk}$ \\
    \hline
  \end{tabular}
  \caption{Baryon-number-violating operators in the mass basis.
  \label{tab:BVinMB}}
\end{table}

\newpage

\bibliographystyle{JHEP}

\end{document}